\documentclass[preprint2,usenatbib]{mnras}
\usepackage[pdftex]{graphicx}
\usepackage {amsmath,amssymb}
\usepackage{color}
\usepackage{aas_macros}
\usepackage{array}
\newcommand{\hi}{{\rm H}{\textsc i}}

\newcommand{\degree}{\ensuremath{^\circ}}
\def\h2{\rm{H_2}}
\def\fHI{f_{\rm{HI}}}
\def\fh2{f_{\rm{H_2}}}
\def\mHI{M_{\rm HI}}

\def\mh2{M_{\rm{H2}}}
\def\SHI{\Sigma_{\rm{HI}}}

\def\SH2{\Sigma_{\rm H2}}

\def\ms{M_{\odot}}

\def\mspc{M_{\odot}~\rm{pc^{-2}}}

\begin{document}
\title{The Local Volume HI Survey: star formation properties}

\author[Jing Wang et al.]{Jing Wang$^{1,2}$\thanks{Email: hyacinthwj@gmail.com}, B\"arbel S. Koribalski$^{1}$\thanks{Email:Baerbel.Koribalski@csiro.au}, Tom H.  Jarrett$^3$, Peter Kamphuis$^4$, Zhao-Yu Li$^{5,6}$, \\
\newauthor Luis C. Ho$^{2,7}$, Tobias Westmeier$^8$, Li Shao$^1$, Claudia del P. Lagos$^{8,9}$, O. Ivy Wong$^{8,9}$, \\
\newauthor Paolo Serra$^{10}$, Lister Staveley-Smith$^{8,9}$, Gyula J\'ozsa$^{11,12,13}$, Thijs van der Hulst$^{14}$ \\
\newauthor \'A.R. L\'opez-S\'anchez$^{15,16}$\\
$^1$Australia Telescope National Facility, CSIRO Astronomy and Space Science, PO box 76, Epping, NSW 1710, Australia\\
$^2$Kavli Institute for Astronomy and Astrophysics, Peking University, Beijing 100871, China\\
$^3$Department of Astronomy, University of Cape Town, Private Bag X3, Rondebosch, 7701, South Africa\\
$^4$National Centre for Radio Astrophysics, TIFR, Ganeshkhind, Pune 411007, India\\
$^5$Key Laboratory for Research in Galaxies and Cosmology, Shanghai Astronomical Observatory, Chinese Academy of Sciences, \\
        80 Nandan Road, Shanghai 200030, China\\
$^6$College of Astronomy and Space Sciences, University of Chinese Academy of Sciences, 19A Yuquan Road, Beijing 100049, China\\
$^7$Department of Astronomy, School of Physics, Peking University, Beijing 100871, China\\
$^8$International Centre for Radio Astronomy Research (ICRAR), M468, University of Western Australia, 35 Stirling Hwy, \\
        Crawley, WA 6009, Australia\\
$^9$ARC Centre of Excellence for All-Sky Astrophysics (CAASTRO), Australia\\
$^{10}$INAF - Osservatorio Astronomico di Cagliari, Via della Scienza 5, I-09047 Selargius (CA), Italy\\
$^{11}$SKA South Africa Radio Astronomy Research Group, 3rd Floor, The Park, Park Road, Pinelands, 7405, South Africa\\
$^{12}$Rhodes University, Department of Physics and Electronics, Rhodes Centre for Radio Astronomy Techniques \& Technologies,\\
          PO Box 94, Grahamstown, 6140, South Africa\\
$^{13}$Argelander-Institut f\"ur Astronomie, Auf dem H\"ugel 71, D-53121 Bonn, Germany\\
$^{14}$University of Groningen,  Kapteyn Astronomical Institute, Landleven 12,  9747 AD, Groningen, The Netherlands\\
$^{15}$Australian Astronomical Observatory, PO Box 915, North Ryde, NSW 1670, Australia\\
$^{16}$Department of Physics and Astronomy, Macquarie University, NSW 2109, Australia\\
}
\date{Accepted 2017 ???? ??
      Received 2017 ???? ??;
      in original form 2017 January}
\pubyear{2017}
\maketitle

\begin{abstract}
We built a multi-wavelength dataset for galaxies from the Local Volume \hi\ Survey (LVHIS), which comprises 82 galaxies. We also select a sub-sample of ten large galaxies for investigating properties in the galactic outskirts. The LVHIS sample covers nearly four orders of magnitude in stellar mass and two orders of magnitude in \hi\ mass fraction ($\fHI$). The radial distribution of \hi\ gas with respect to the stellar disc is correlated with $\fHI$ but with a large scatter. We confirm the previously found correlations between the total \hi\ mass and star formation rate (SFR), and between \hi\ surface densities and SFR surface densities beyond R$_{25}$. However, the former correlation becomes much weaker when the average surface densities rather than total mass or rate are considered, and the latter correlation also becomes much weaker when the effect of stellar mass is removed or controlled. Hence the link between SFR and \hi\ is intrinsically weak in these regions, consistent with what was found on kpc scales in the galactic inner regions. We find a strong correlation between the SFR surface density and the stellar mass surface density, which is consistent with the star formation models where the gas is in quasi-equilibrium with the mid-plane pressure. We find no evidence for \hi\ warps to be linked with decreasing star forming efficiencies.
 
\end{abstract}


\section{Introduction}
\label{sec:introduction}
How galaxies evolve in their stellar population is a key question in galactic astronomy, and the properties of atomic hydrogen \hi\ gas contain key information to that question. The star formation rates (SFR) and stellar masses of star forming galaxies are observed to correlate with each other and this relationship is often referred to as the star formation `main sequence' \citep[MS,][]{Noeske07, Elbaz07, Dutton10}. The star formation quenching or quenched galaxies lie below the sequence with a wide range of star formation rates \citep{Whitaker12}. The deviation to the median SFR of the MS at a fixed stellar mass strongly depends on the \hi\  gas fraction of galaxies \citep{Saintonge16}. Hence the quenching of star formation is accompanied by removal or exhaustion of \hi\ gas. This is because \hi\ forms the initial reservoir from which molecular gas and subsequently, stars are formed. In order to better understand galaxy quenching, we first need to understand how \hi\ fuels star formation, i.e. the relation between \hi\ and SFR in star-forming galaxies. 

Several large single-dish \hi\ surveys have been carried out in the past decade. The shallow blind surveys, ALFALFA \citep[Arecibo Legacy Fast ALFA,][]{Giovanelli05} and HIPASS \citep[HI Parkes All-Sky Survey,][]{Meyer04, Koribalski04}, and the targeted, optically selected, GASS survey \citep[GALEX Arecibo SDSS Survey,][]{Catinella10}  have significantly advanced our understanding of the SFR-\hi\ relations on the global scales. 
Among the most interesting findings are, 1) SFR and \hi\ mass are tightly correlated for the ALFALFA sample whose galaxies are mostly on the star-forming sequence \citep{Huang12}; 2) the star forming efficiencies (SFE$=$SFR$/\mHI$) in the \hi-selected ALFALFA sample are on average three times lower than those in the optically selected GASS sample, which includes both star-forming and transitional galaxies \citep{Huang12, Schiminovich10}; 3) there is a weak correlation between SFE and stellar mass in the ALFALFA sample \citep{Huang12}, which was not found within the GASS sample  \citep{Schiminovich10}, possibly due to the small range in stellar masses for GASS and$\slash$or the different sample selections; 4) there is a weak correlation between SFE and optical surface brightness for galaxies selected from the HIPASS sample \citep{Wong16}.

The strong correlation between SFR and \hi\ mass found on global scales seems to be at odds with the lack of correlation between surface densities of  SFR and \hi\ found on sub-kpc scales in the inner parts of galaxies \citep{Bigiel08}. The latter is contrasted by the tight scaling relation between surface densities of SFR and molecular gas \citep{Bigiel08}, and the explanation is that the molecular gas clouds rather than the \hi\ gas directly form stars. This difference in \hi-SFR relations on kpc and global scales might be caused by the difference in scales, but can also possibly indicate different physics working on these two scales. Especially we need to keep in mind that a large fraction of the \hi\ gas in \hi-rich galaxies is distributed beyond the stellar disc where the SFE is so much lower than that in the inner discs that it would take several times the Hubble time to deplete the \hi\ gas there \citep{Bigiel10}.

On the other hand, the global \hi-SFR correlation is present on kpc-scales in the \hi\ dominated regions, i.e., in irregular galaxies \citep{Roychowdhury14} and the outskirts of spiral and early-type galaxies \citep{Koribalski09, Bigiel10, Yim16, Yildiz16}. However there is large scatter ($\sim$1.5 dex) in these relations \citep{Roychowdhury14, Bigiel10}.  \citet{Hunter98} showed that in irregular galaxies the SFR radial profiles actually follow the shape of stellar mass profiles better than that of \hi\ profiles. This makes us wonder whether the same behaviour can be found in the outskirts of spiral galaxies. A possibility that could explain these trends is that both the SFR and the \hi\ surface densities are correlated with the stellar mass surface densities in the outer regions of galaxies.
The mid-plane pressure model has been commonly used to explain the link between mass surface densities and SFR \citep[e.g.]{Leroy08, Ostriker10}, however one significant limitation is that it doesn't explain the stochastic behaviour of star formation in these low gas density regions \citep{Gerola80,Teich16}.


Mass surface density is not the only possible parameter that regulates SFR. Yim \& van der Hulst (2016) found that disturbances from the external environment seem to hardly affect the radial \hi\ SFE of galaxies. On the other hand, the \hi\ disc internal structures may affect SFE in the galactic outskirt. As demonstrated nicely in \citet{vanderKruit07}, \hi\ warps usually start near the optical truncation radius where the optical discs abruptly and steeply turn fainter \citep[also see][]{Jozsa07b, GarciaRuiz02, Briggs90}. One possible explanation is that the \hi\ gas in a warp has an external origin and was accreted with an angular momentum mis-aligned with that of the main disc. Conversely, the existence of an \hi\ warp may affect the radial distribution of SFR, which is adopted in some theoretical models to explain the optical truncation radius.  For example, in the model of \citet{SanchezBlazquez09} the gas near a warp would have low volume densities due to the warp, and hence could not efficiently form stars. As a result a large fraction of stars beyond the warp radius were not formed in situ but rather migrated from the inner part of the galaxy disc. These theories add complexities to the \hi-SFR relationship in the galactic outskirts.  However, the influence of \hi\ warps on localised SF has never been investigated statistically with observational data. 

 We have reviewed the apparently different behaviours of \hi-SFR relations on different scales and in different regions. However, to understand them in a consistent picture has been difficult, largely due to the lack of proper datasets. The single-dish \hi\ data spatially resolve very few galaxies while \hi\ interferometric samples usually lack homogeneity in galaxy selection. Another common problem with HI interferometry is that extended emission of objects with large angular scales (typically $>$ 15 arcmin) is not properly measured because there are no baselines short enough to sample such
angular scales. This affects the surface densities of all extended emission, especially on the outer parts. The Local Volume \hi\ Survey (LVHIS\footnote{http://www.atnf.csiro.au/people/Baerbel.Koribalski/LVHIS/LVHIS-galaxies.html}, Koribalski et al. in prep, Paper I hereafter) tends to overcome these difficulties. It is close to a volume limited \hi\ sample of galaxies beyond the Milky Way, built on the single-dish blind survey HIPASS \citep{Meyer04}, by targeting all detections with a declination less than or similar to $-30$ degree and a distance less than 8 Mpc from the HIPASS catalogs \citep{Meyer04, Koribalski04}. Special care has been taken to address the missing flux problem by using compact array configurations. We also combine the interferometric data with single-dish imaging mosaics \citep[from the Parkes Galactic All Sky Survey, GASS,][]{McClure-Griffiths09} to produce the HI image for the galaxy NGC 300, for which ATCA data alone is not sufficient to sample the extended HI structures \citep[see Paper I and][for a description of such techniques]{Stanimirovic02}. LVHIS hence provides us a suitable laboratory for coherently understanding the different \hi-SFR relations. 
 
 This paper presents the multi-wavelength dataset for the LVHIS sample. As a first application of the dataset it investigates \hi-SFR relations on global scales and in the outskirts, with the goal of reconciling the seeming contradictions of these relations from the relation observed in galactic inner regions.

 The paper is organised as follows.
 In Section~\ref{sec:sample} we introduce the data used in this paper, including a brief description about the LVHIS sample and \hi\ data. In Section~\ref{sec:analysis} we analyse the multi-wavelength data for the whole LVHIS sample while in Section~\ref{sec:analysis2} we focus on a sub-sample of ten large LVHIS galaxies. 
 A comparison of LVHIS with other \hi\ surveys can be found in Section~\ref{sec:scale_relation} compares LVHIS. In Section \ref{sec:dist_HI} we study the dependence of \hi\ radial distributions on stellar properties.  
In Section \ref{sec:glob_SFE} we use resolved \hi\ properties to explore causes for the global star formation relations. In Section~\ref{sec:loc_SFE} we investigate relations between SFR, \hi\ and stellar surface densities in the galactic outskirts. In Section~\ref{sec:warp_SFR} we search for possible links between the onset of an \hi\ warp and the radial change of SFE. 
In Section~\ref{sec:summary} we summarise the paper. 

Throughout this paper, we assume a Kroupa IMF \citep[initial mass function,][]{Kroupa01}, and a $\Lambda$CDM cosmology with $\Omega_{m}=0.3$, $\Omega_{lambda}=0.7$ and $h=0.7$. Any quantities and trends taken from the literature are converted to be consistent with these assumptions before used in this paper. Because this paper extensively discusses correlations, we define here that a strong correlation should have Pearson correlation coefficient $\rho>$0.8, a significant correlation should have $\rho$ between 0.6 and 0.8, a weak$\slash$moderate correlation should have $\rho$  between 0.2 and 0.6, while no correlation corresponds to $\rho<$0.2. Errors for correlation coefficients are calculated through bootstrapping.

\section{Data}
\label{sec:data}

\subsection{Sample and ATCA HI data}
\label{sec:sample}
The Local Volume \hi\ Survey (LVHIS, Koribalski et al. in prep, Paper I) consists of a complete sample of 82 nearby galaxies ($D < 10$ Mpc) with a declination $\delta \lesssim -30\degree$ selected from HIPASS \citep{Meyer04, Koribalski04}. 

Using the Australia Telescope Compact Array (ATCA), sensitive \hi\ spectral 
line and 20-cm radio continuum data were obtained for all LVHIS galaxies.  Large galaxies like M\,83, NGC~3621,
Circinus and others were mosaiced to ensure that the extended \hi\ emission of the outer 
disc was not missed. Paper I presents the LVHIS \hi\ galaxy atlas (cubes and moment maps), based on the 
`natural' weighted {\em uv}-data, and focused on the large-scale \hi\ 
emission in and around the selected galaxies. Typical angular resolution is B$_{\rm maj} \ga$40\arcsec\ (ie. $\sim$1~kpc at $D$ = 5 Mpc), where B$_{\rm maj}$ is the major axis of the synthesized beam. The typical r.m.s. sensitivity is 
$\sim$1.5 mJy\,beam$^{-1}$ per 4 km$/$s channel, which corresponds to a 5 r.m.s column density limit of  0.46$\times 10^{20}$ cm$^{-2}$ assuming a 20 km$/$s line width. This paper makes use of the primary beam corrected cubes and moment 0 images to derive \hi\ related properties.

Results based on the LVHIS dataset have been presented in a series of publications, including \citet{Koribalski09, vanEymeren10, LopezSanchez12, Kirby12, Kamphuis15, Johnson15} and \citet{Wang16}. A parallel study of the 20-cm radio continuum emission of the sample will be presented in Shao et al. (in prep). We take R$_{25}$ (the semi-major axis of the 25 mag arcsec$^{-2}$ isophote in the $B$-band) from the surface photometry catalogue of the ESO-Uppsala galaxies \citep{Lauberts89}, and the \hi\ surface density radial profiles and R$_{\rm HI}$ (the semi-major axis for the \hi\ 1 $\mspc$ isophote) from \citet{Wang16}. 

\begin{figure*} 
\includegraphics[width=14cm,angle=270]{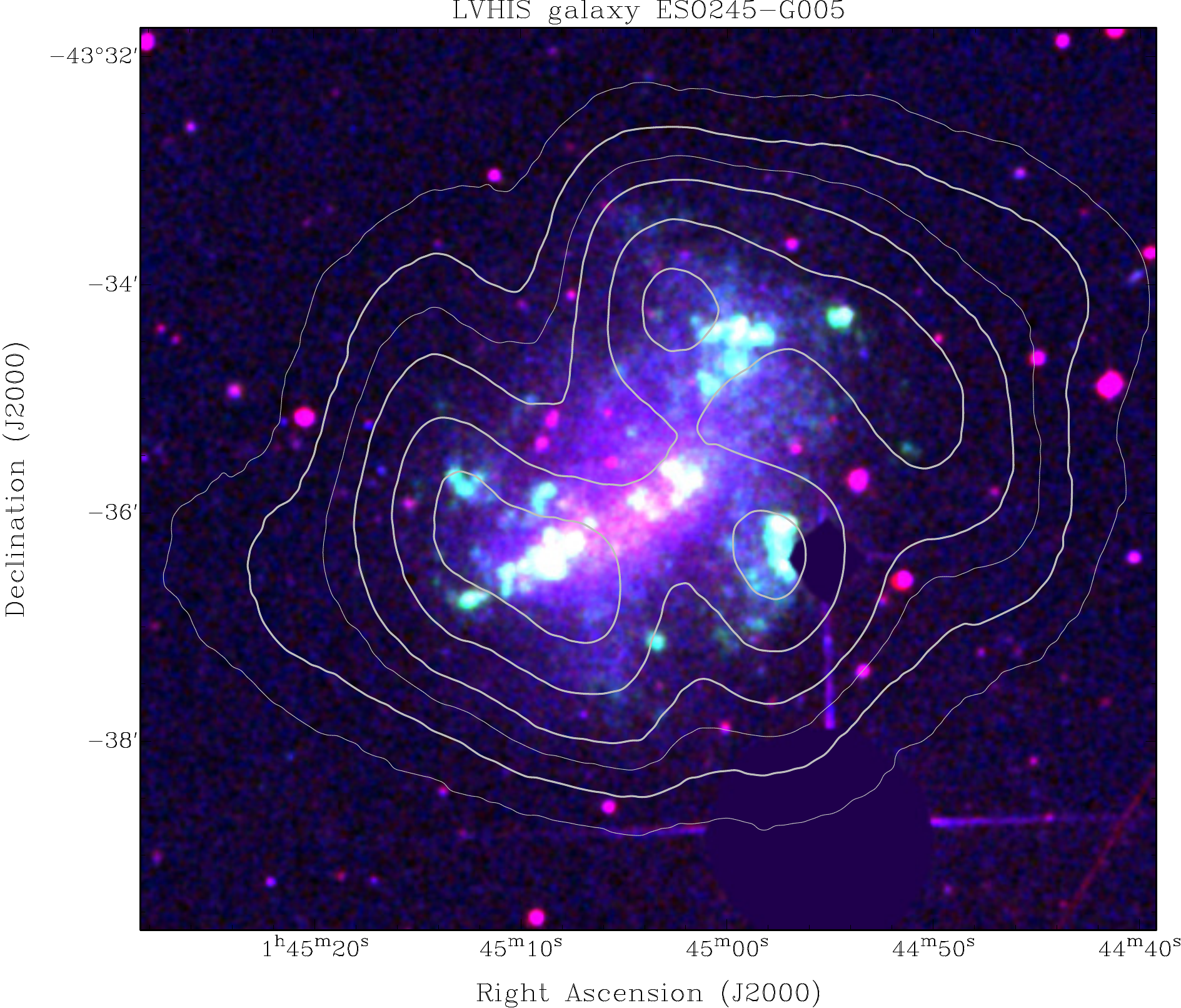}
\caption{Multi-wavelength image of the galaxy ESO245-G005, one of 82 nearby galaxies in the 
  Local Volume HI Survey (LVHIS, Koribalski et al. in prep). The 3-colour composite consists 
  of DSS2 $R$-band (red), GALEX FUV (green) and DSS2 $B$-band (blue) images overlaid 
  with ATCA HI intensity contours at 0.5, 1, 1.5, 2, 3 and 4 Jy/beam km/s. The optical and 
  UV images were smoothed, and we masked two foreground stars.}
\label{fig:HIcontour}
\end{figure*}

 The LVHIS galaxy names along with some of their key properties derived throughout this paper are listed in Table~\ref{tab:galprop}. The table columns are as follows: (1) LVHIS ID, (2) HIPASS name, (3) galaxyÊname, (4) star formation rate (${\rm SFR = SFR_{FUV} + SFR_{W4}}$; Section~\ref{sec:estimate_sfr}),Ê(5) stellar mass ($M_*$) derived from WISE luminosities (Section~\ref{sec:estimate_mst}), (6) HI fraction ($\fHI=\mHI/M_*$), (7) ratio of \hi\ to optical radii, (8) the averaged stellar surface densities within the W1 effective radius ($\Sigma_{\rm *,eff}$, Section~\ref{sec:irphot}), (9) the averaged SFR surface density within R$_{25}$ ($\Sigma_{\rm SFR,r<R25}$, Section~\ref{sec:best_res}), (10) the averaged \hi\ surface densities within R$_{25}$, (11) the averaged \hi\ surface densities within 3.2 r$_{\rm s}$ of the W1 band (Section~\ref{sec:irphot}).ÊProperties in Columns 7, 10 and 11 are obtained only when the relevant radii are resolved in \hi\ (${\rm >b_{maj}}$). 
  The tables presented in this paper have the same order as in Paper I and Shao et al. (in prep) and we also use `ID'  to denote the index (starting from 1). Figure~\ref{fig:HIcontour} shows an example multi-wavelength image of one of the 82 LVHIS galaxies.



From the LVHIS sample we also select a sub-sample of large galaxies for detailed multi-wavelength analysis (marked with {\it l} in Table~\ref{tab:galprop}). These, in total ten galaxies, are all the LVHIS galaxies that have an optical diameter, $D_{25}$, at least four times larger than the \hi\ angular resolution and have GALEX FUV data available (Section~\ref{sec:data_uv}). They include seven galaxies with Hubble types ranging between Sc and Sd, two Sm galaxies and one irregular dwarf galaxy.  An atlas of these galaxies can be found in Appendix~\ref{sec:appendix_figure}. The \hi\ discs are 15 to 73 times larger than the \hi\  angular resolution. We obtain multi-wavelength radial profiles and tilted-ring \hi\ models for this subsample in Section~\ref{sec:analysis2}. 

We note that for analysis based on this sub-sample, we use the natural weighted \hi\ image from THINGS \citep[The HI Nearby Galaxy Survey][]{Walter08} instead of the ATCA \hi\ image for the galaxy NGC 7793. This is because for NGC 7793, the THINGS \hi\ data have much better resolution ($\sim$15.6 acsec) than the ATCA \hi\ data ($\sim$364 arcses) while they do not significantly miss flux (its \hi\ flux 12\% less than the HIPASS \hi\ flux, comparable to the 1$\sigma=15\%$ difference between ATCA and HIPASS fluxes for LVHIS, see Paper I) due to the lack of short baselines. 

\subsection{GALEX}
\label{sec:data_uv}
The Galaxy Evolution Explorer \citep[GALEX]{Martin05} provides near-ultraviolet (NUV, effective wavelength $\lambda\sim1528\AA$) and far-ultraviolet (FUV, $\lambda\sim2271\AA$) images over a large fraction of the sky. Each image has a pixel size of 1\farcs5, and the typical full width half maximum (FWHM) of the PSF (point spread function) is 5\farcs0 (FUV) and 5\farcs5 (NUV). The provided magnitude zero-points are calibrated in the AB mag-system. Of the 82 LVHIS galaxies, 56 are covered by the GALEX imaging survey (data release 6+7) in the NUV band and 46 in the FUV band. The remaining 26 galaxies are missing because GALEX avoided the sky near the Galactic Plane. 

The depth of the available GALEX images varies substantially: 22 galaxies have relatively deep images from the Medium Imaging Survey (MIS), with typical exposure times of $t_{\rm exp}>1500~s$ and a depth of 22.6 mag in FUV, 9 have ultra deep images from the Deep Imaging Survey (DIS), with $t_{\rm exp} > 10,000~s$ and a depth of 24.2 mag in FUV, while 15 have short exposures from the All-Sky Imaging Survey (AIS), with $t_{\rm exp}\sim100~s$ and a depth of 20 mag in FUV. 
\subsection{WISE}
The Wide-field Infrared Survey Explorer \citep[WISE]{Wright10} has mapped the whole sky in the 3.4, 4.6, 12 and 22 $\mu$m MIR (mid-infrared) bands, commonly referred to as  W1, W2, W3 and W4. We make use of specially constructed WISE mosaics produced with ICORE \citet[Image Co-addition with opitional Resolution Enhancement]{Masci13} as described in \citet{Cluver14}. These mosaics notably improve the angular resolution by $\sim$30\% from the ALLWISE images \citep{Cutri12} published in the WISE public release archive, which were optimised for detecting point sources. They have a pixel size of 1\arcsec\ and a resolution of 5\farcs9, 6\farcs5, 7\farcs0 and 12\farcs4 for the four bands, respectively. The typical sky rms (1$\sigma$) is  \mbox{23.0, 21.8, 18.1 and 15.8 mag arcsec$^{-2}$} \citep{Jarrett12}.  The magnitude zero-points are calibrated in the Vega mag system.



\subsection{Other data}
\label{sec:otherdata}
Past studies have provided photometric data for several LVHIS galaxies. A comprehensive overview of the complimentary datasets is given in Table 1 of Paper I. Here we review a few datasets that are used for analysis or comparison in this paper. 


 As part of the  Local Volume Legacy (LVL) project \citep{Kennicutt08, Dale09}, \citet{Lee11} obtained asymptotic fluxes (a measure of the total flux, see Section~\ref{sec:uvphot}) in the GALEX bands for 390 local volume galaxies including 53 LVHIS galaxies.  We will use this dataset to validate our GALEX flux measurements (Section~\ref{sec:uvphot}). 

The Carnegie-Irvine Galaxy Survey \citep[CGS]{Ho11} provides $B$, $V$, $R$ and $I$-band images for the 605 optically brightest galaxies in the southern sky, including 22 LVHIS galaxies. The field of view for each image is $8\farcm9 \times 8\farcm9$, and the typical seeing is 1\arcsec. The details are described in \citet{Ho11} and \citet{Li11}. The survey also provides a comprehensive set of photometric measurements including total fluxes, sizes, and radial surface brightness (SB) profiles.  We will use this dataset to compare with our stellar mass ($M_*$) estimate for the high-mass galaxies (Section~\ref{sec:estimate}).

\citet{Kirby12} obtained deep $H$-band (1.65$\mu$m) images with the 3.9-m Anglo Australian Telescope (AAT) and performed photometry for 57 galaxies in the Local Volume. \citet{Young14} followed up another 40 dwarf galaxies. The field of view for each image is $7\farcm7 \times 7\farcm7$, and the typical seeing is 1\farcs3. The two samples combined include 37 LVHIS galaxies.  We will use this dataset to compare with our stellar mass ($M_*$) estimate for the low-mass galaxies (Section~\ref{sec:estimate}).

The SINGG \citep[Survey for Ionization in Neutral Gas Galaxies,][]{Meurer06} obtained H$\alpha$ and $R$-band images for 468 galaxies randomly selected from the HIPASS catalogues to uniformly sample the \hi\ mass function. Its sister survey, SUNGG \citep[Survey of Ultraviolet emission in Neutral-Gas Galaxies,][]{Wong16} used GALEX to image a subset of SINGG galaxies in the FUV and NUV bands. SINGG (SUNGG) have 19 (6) galaxies in overlap with LVHIS. We will compare our SFR and $M_*$ estimates with those obtained in the SINGG studies (Section~\ref{sec:estimate}). 

\citet{Lee09} estimated SFRs for 300 local galaxies using data from the 11 Mpc H$\alpha$ and Ultraviolet Galaxy Survey (11HUGS). There are 29 galaxies in common for 11HUGS and LVHIS.
We will compare our SFR estimates with those from \citet{Lee09} for validation (Section~\ref{sec:estimate}).  

\section{Analysis for LVHIS galaxies}
\label{sec:analysis}
In this section, we describe how we compute the total and radial distribution of fluxes from the WISE and GALEX images. We convert these measurements into physical properties and compare them to literature data. 

\subsection{GALEX photometry}
\label{sec:uvphot}
We use the same pipeline described briefly in \citet{Overzier11} to obtain photometric measurements for the LVHIS galaxies. Here we describe each step of the procedure:

\begin{enumerate}
\item Global background estimate and neighbour masking. We use SExtractor \citep{Bertin96} to produce a segmentation map for each image, where the pixels for all detected sources are labelled.  We then exclude the pixels belonging to detected sources and calculate a global background value. Next we manually set this background value as input and run SExtractor again. Finally, we use the new segmentation map to mask sources around our galaxies. 

\item Foreground star removal.  For galaxies with available FUV images, we convolve these to match the slightly poorer resolution of the NUV images. Pixels associated with foreground stars are identified as having flux ratios NUV$/$FUV $> 5$ and NUV signal-to-noise (S/N) $>2$. For galaxies without FUV images, we inspect the NUV images and manually identify pixels containing bright point sources \footnote{Masking foreground stars through visual inspection is subject to uncertainties, but generally speaking pixels associated with foreground stars identified from the NUV$/$FUV values are a small fraction ($<0.02\%$) of all the pixels. Errors introduced in the NUV fluxes through this step do not affect the science analysis (which uses the FUV but not the NUV fluxes) in this paper.  }.  We replace the pixels associated with foreground stars  with pixels on the other side of the galaxy centre, or the average of pixels on the galaxy isophote, depending on whether or not the pixels on the other side are also associated with foreground stars.  This step makes a clean image that will also be used in the resolution matched photometric analysis (Section~\ref{sec:phot_conv}).

\item Radial profiles and residual background removal. We measure the radial profile of surface brightness (SB) along elliptical rings with a width of at least 1 pixel ($\sim1/3$ of the PSF size). For large galaxies, the spacing is increased to sample the radial profile with at most 50 data points. The ellipticity and position angle of the rings are fixed at the optical values of the 25 mag arcsec$^{-2}$ isophote. The SB is calculated as the 3-sigma clipped mean of the distribution of the pixel values in each elliptical ring. We identify the radius where the profile flattens within the noise,  and use the annulus between one and 1.4 times this radius around the galaxy centre to estimate a local residual of the background. We remove this residual background and cut the radial profile at a S$/$N of 1.25 (equivalent to an SB error of 0.87 mag).  

\item Growth curve and total flux. We measure the growth curve (GC) of  flux enclosed by elliptical apertures out to 1.25 times the last radial point of the SB radial profile.  We use the GC to obtain the asymptotic flux.  Following the method outlined in \citet{Lee11}, we extrapolate the relation between the derivative of the GC and the GC, and the intercept is taken as the asymptotic flux. The error of the asymptotic flux is a combination of the photometric error and error from the extrapolation.  We further add a  zero-point error of 0.05 mag for FUV and 0.027 mag for NUV \citep{Morrissey07}.  The asymptotic flux is taken as the total flux for GALEX bands in this paper.
\end{enumerate}

\begin{figure*} 
\includegraphics[width=13cm]{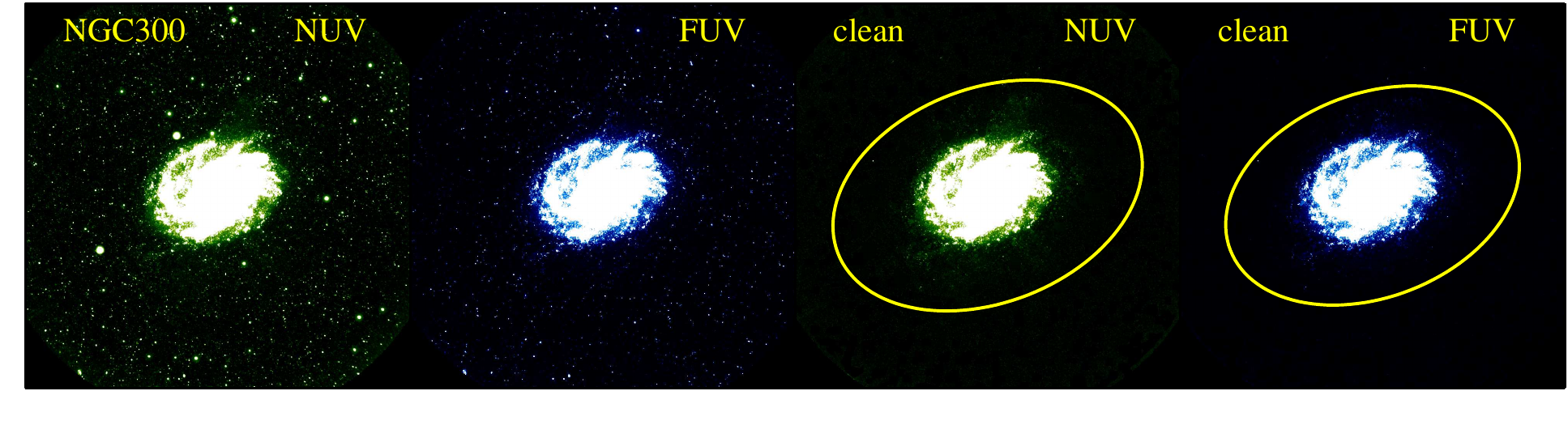}
\vspace{0.5cm}
\includegraphics[width=4.5cm]{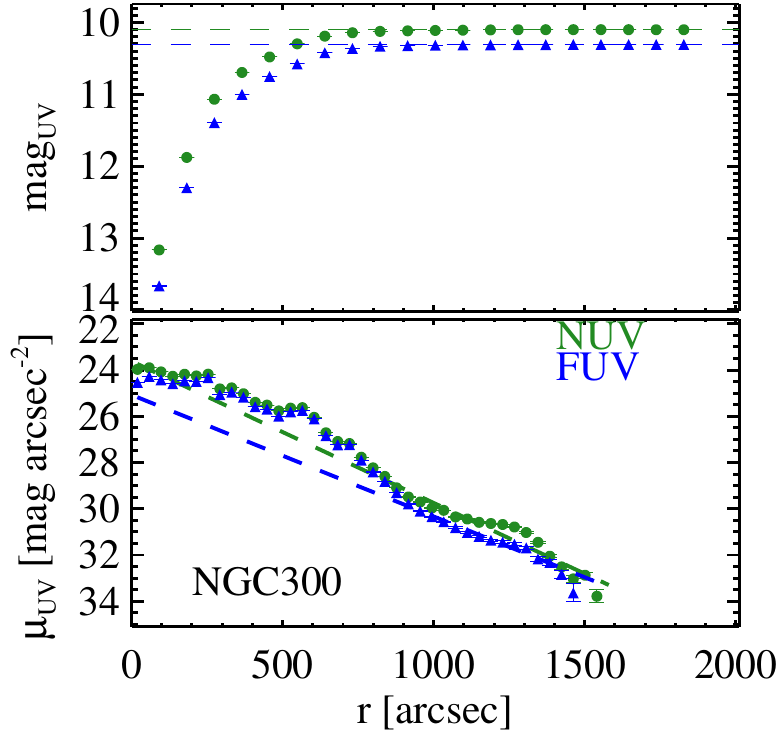}

\includegraphics[width=13.cm]{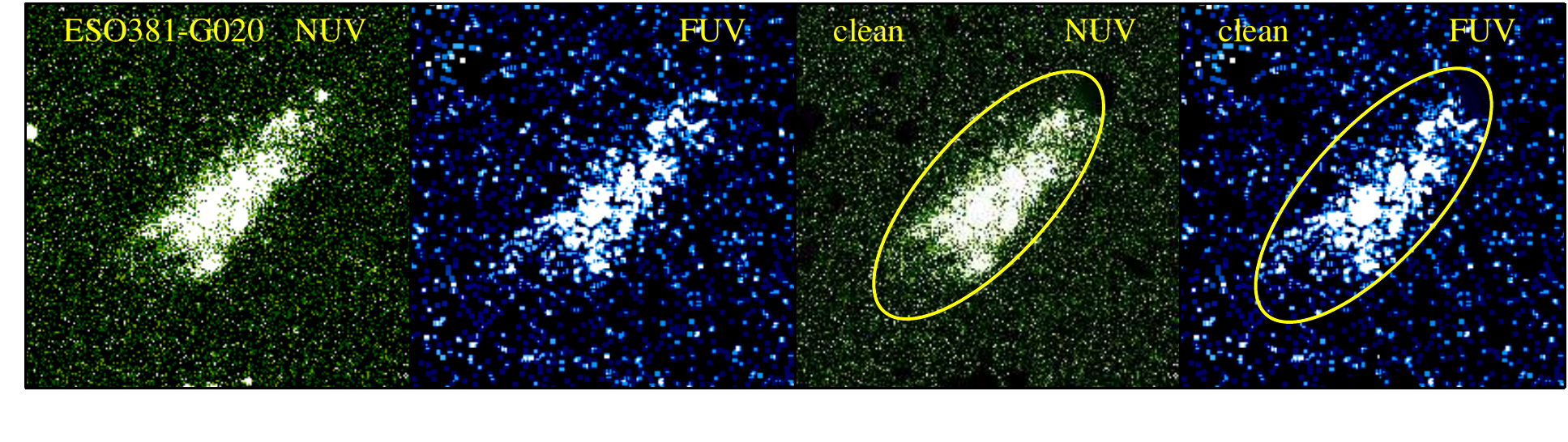}
\vspace{0.5cm}
\includegraphics[width=4.5cm]{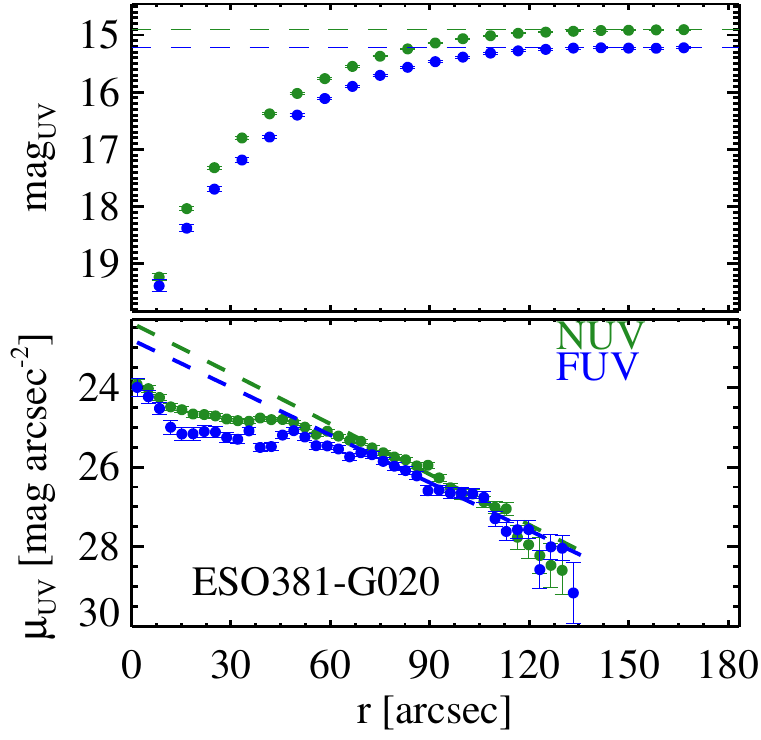}
\caption{This gallery shows examples of the intermediate products of the GALEX photometry pipeline. The top row shows NGC~300, a UV-bright galaxy with DIS images, and the bottom row shows ESO 381-G020, a relatively UV-faint galaxy with only AIS images. The left four columns as denoted in the corner of each panel display the original FUV and NUV images, and clean images with contamination sources removed. The yellow ellipses show the isophotes for the last data points in NUV radial SB profiles. The right-most panels show the GCs (top) and the SB profiles (bottom). }
\label{fig:uvphot_example}
\end{figure*}

Figure~\ref{fig:uvphot_example} shows the intermediate products of this pipeline for two galaxies, one is bright and with deep imaging and the other is faint and with shallow imaging.  We are able to reliably measure total fluxes (with an error less than 0.5 mag) for 44 and 54 galaxies in the FUV and NUV bands, respectively. We present these photometric measurements in Table~\ref{tab:uvphot}).

\begin{figure} 
\includegraphics[width=8cm]{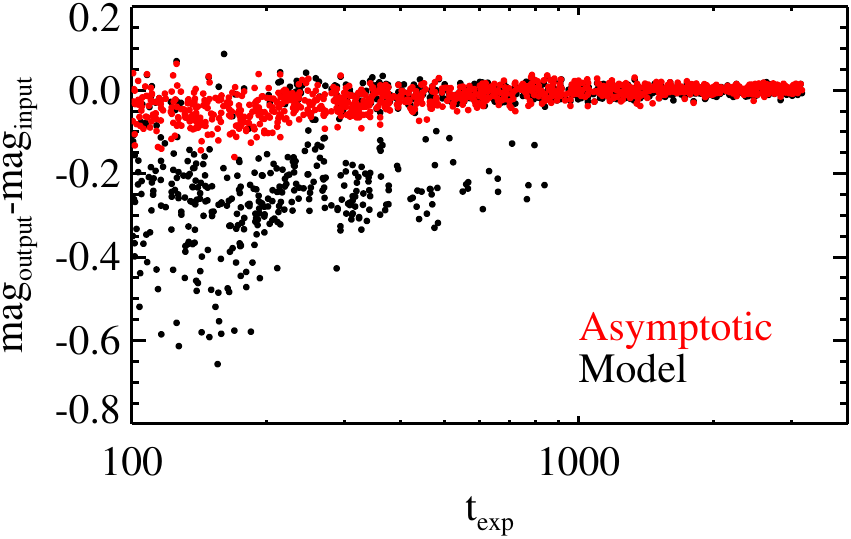}
\caption{Stabilitiy of flux measurements at different image depths.  Shallow images are simulated based on the deep GALEX FUV image of NGC~300 (see Section~\ref{sec:uvphot}). Asymptotic and model magnitudes measured from shallow simulated images (output) are compared to the magnitude measured from the original deep image (input). }
\label{fig:fluxes_asy_model}
\end{figure}

There are alternative methods to obtain the total fluxes. As adopted by \citet{Huang12}, fitting models (e.g. exponential functions) to SB radial profiles and extrapolating helps recover faint fluxes lower than the noise level, which is especially useful for low SB dwarf galaxies. Although the majority of the LVHIS sample are dwarf galaxies, we have chosen to use the asymptotic fluxes instead of the model fluxes. This is because nearly half of the LVHIS sample only have shallow AIS images (see Section~\ref{sec:data_uv}) while \citet{Huang12} selected galaxies with GALEX images of at least MIS-depth.  Model fitting could be unreliable on our shallow images. To demonstrate the reason for our choice, in Figure~\ref{fig:fluxes_asy_model}, we test the stabilities of the two methods on images of different depths. We take the DIS FUV image (with $t_{exp} = 12987.6~s$) of NGC~300 and pretend to observe it again with shorter $t_{\rm exp}$, i.e. by adding Poissonian noise to the counts.  It is clear that when $t_{\rm exp}<1000~s$ (typical for AIS),  model fluxes have large uncertainties and systematic uncertainties while asymptotic fluxes are much more stable. 

\begin{figure} 
\begin{center}
\includegraphics[width=7cm]{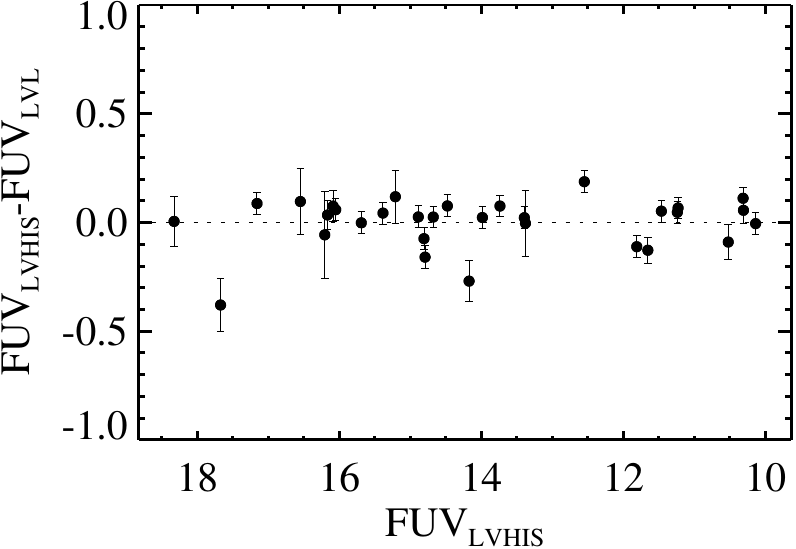}
\vspace{0.5cm}

\caption{ Comparison of GALEX FUV flux measurements between the LVHIS and LVL samples.  The dashed line mark zero.}
\end{center}
\label{fig:LVHIS_LVL_flux}
\end{figure}

As an external validation, we compare our asymptotic flux measurements with those from \citet{Lee11} for the galaxies included in both studies (53 in total). 
 GALEX magnitudes before data release 6 as presented in \citet{Lee11} are 0.045 mag brighter in NUV and 0.033 mag brighter in FUV as compared to the most recent releases, due to a less accurate zero-point calibration. We have corrected for this effect before the comparison. As can be seen in Figure~\ref{fig:LVHIS_LVL_flux}, the LVHIS and LVL measurements of FUV fluxes agree reasonably well, with a median value and scatter of 0.085 and 0.089 mag for the differences between the two. We have reviewed our photometry steps for the outliers in the comparison, and make sure that there are no obvious errors in our processing.

\subsection{WISE photometry}
\label{sec:irphot}
We use the same pipeline outlined in detail in \citet[][upgraded upon the pipeline of Jarrett et al. 2012 and Jarrett et al. 2013]{Jarrett17} to perform photometric measurements on the WISE images. Here we briefly summarise the major steps of the procedure. For each image, the pipeline first removes nearby sources and produces a clean image.  Then it obtains the aperture flux and SB radial profile within the 1 $\sigma$ (sky rms) isophote around the galaxy centre. A double S{\' e}rsic model is fitted to the SB profile, and the final total flux is calculated as the sum of the aperture flux and the extrapolation of the S{\' e}rsic  model beyond the 1 $\sigma$ isophote out to three times the scale-length of the disc (the second component of the double S{\' e}rsic model).   Calibration and spectral shape related corrections are applied to the magnitudes as described in \citet{Jarrett17}.  Structural parameters like position angle and axis ratio b$/$a (minor-to-major axis ratio)  of the outmost isophote, effective radius r$_e$ (radius that enclose half of the total flux), effective surface brightness (the averaged surface brightness within r$_e$), and concentration indices R$_{75}/$R$_{25}$ (the ratio between radii that enclose 75\% and 25\% of the total fluxes) are also derived. 

\begin{figure*} 
\includegraphics[width=12.cm]{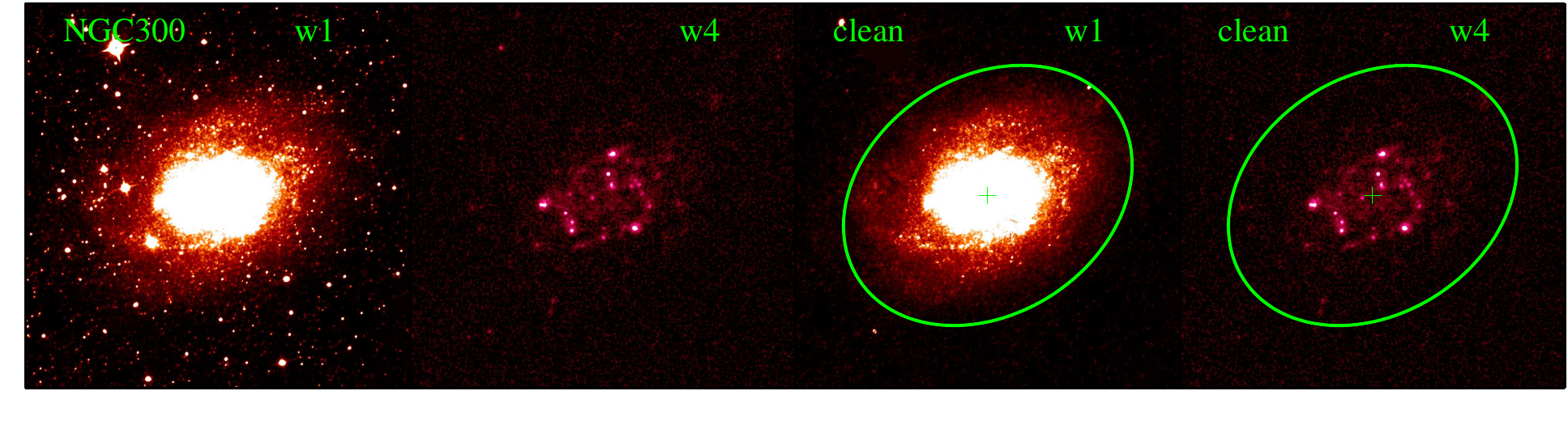}
\includegraphics[width=5.5cm]{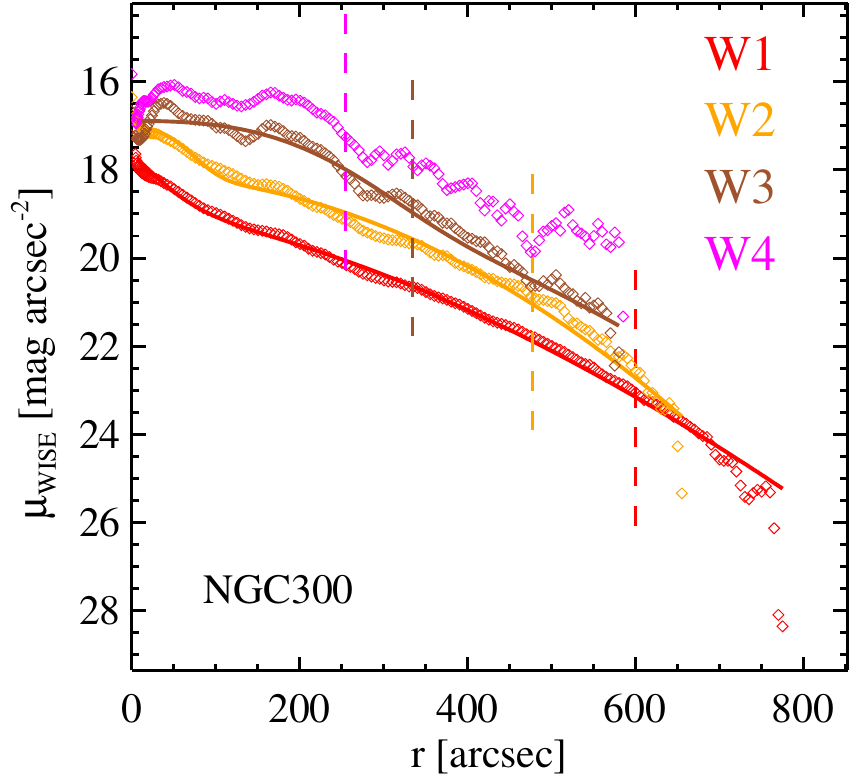}

\vspace{0.5cm}

\includegraphics[width=12.cm]{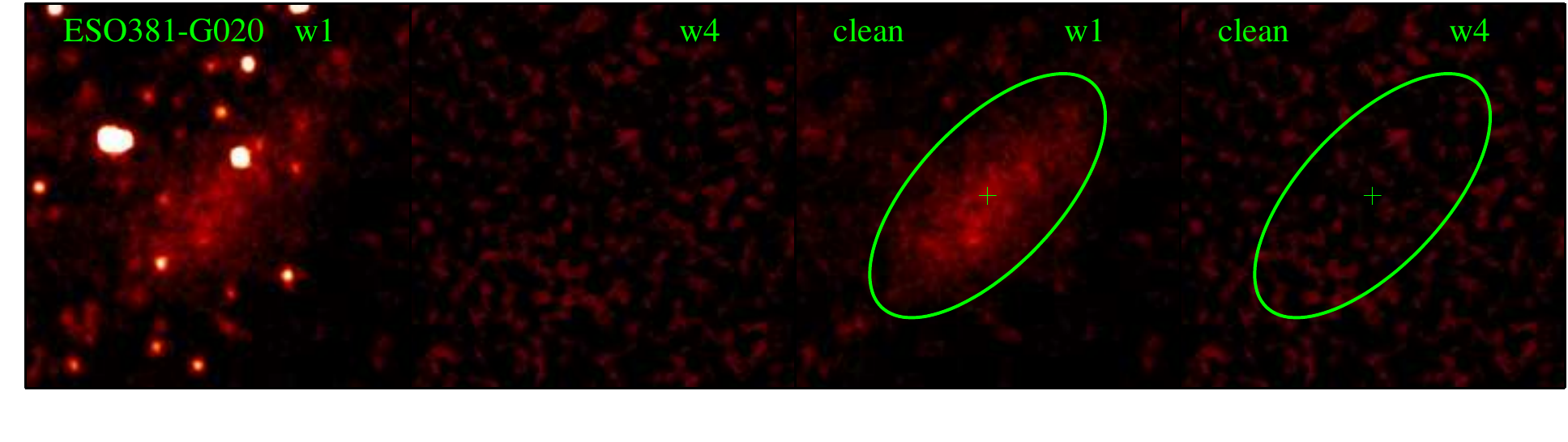}
\includegraphics[width=5.5cm]{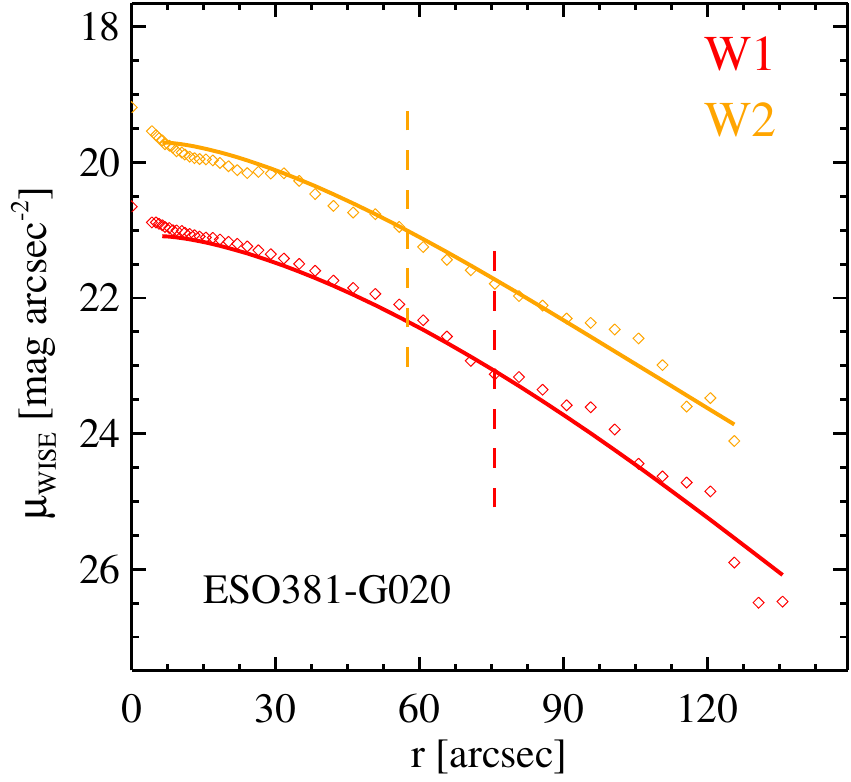}
\caption{This gallery shows examples of the intermediate products of the photometry pipeline applied to the WISE images. The top row shows the bright spiral galaxy NGC 300, and the bottom row shows the much fainter dwarf irregular galaxy ESO 381-G020. From left to right, as denoted in the corner of each panel, we show the original W1 and W4 images, and the respective clean images, where foreground and background sources are removed. The right-most panels show the SB profiles and the S{\' e}rsic fits for both galaxies. The dashed vertical lines show where the SB profile reaches the sky rms.}
\label{fig:irphot_example}
\end{figure*}

Figure~\ref{fig:irphot_example} shows the intermediate products of this pipeline for two galaxies, NGC~300 and ESO 381-G020.  We are able to reliably measure total fluxes (with errors less than 0.5 mag) for 81, 76, 65 and 57 galaxies in the W1--4 bands.  We are unable to measure W1 fluxes for only one galaxy, J1131-31, because a foreground star outshines the whole galaxy. We further impose the W1 aperture on the W2 image for the five galaxies detected in W1 but not in W2, and obtain a rough measure of the W2 flux by summing all pixels within the aperture.

As described in \citet{Jarrett17}, we obtain the rest-frame W1-4 fluxes based on these photometric fluxes.  Especially, we remove the stellar continuum to obtain the W3 fluxes of the 11.3 $\mu$m PAH (polycyclic aromatic hydrocarbon)  and the W4 fluxes originated from warm and cold small dust grains. These rest-frame and stellar continuum removed fluxes are obtained through comparing the WISE spectral energy distribution (SED) of each galaxy with extragalactic population SED templates built in \citet{Brown14} based on the SWIRE$\slash$GRASIL models \citep{Polletta06, Polletta07, Silva98}. These corrected fluxes are more directly linked to physical quantities (e.g. $M_*$ and SFR) than the photometric fluxes. 

We present these WISE rest frame luminosities and structural parameters in Table~\ref{tab:irphot}.  

\subsection{Estimate of $M_*$ and SFR}

\label{sec:estimate}
\subsubsection{Estimating $M_*$}
\label{sec:estimate_mst}
Using the method of \citet{Jarrett12}, we calculate the W1 band stellar mass-to-light ratio based on the W1-W2 colour to account for the stellar population dependence. Similar to the treatment in \citet{Jarrett17}, we put a floor/ceiling limits on the W1-W2 colour, -0.05 to 0.2 mag,  the colour range where the \citet{Jarrett12} formula was calibrated. It also minimises AGN contamination and the systematics caused by the less sensitive W2 band. Then we calculate $M_*$ for 81 LVHIS galaxies based on W1 luminosities. 

\begin{figure} 
\includegraphics[width=8cm]{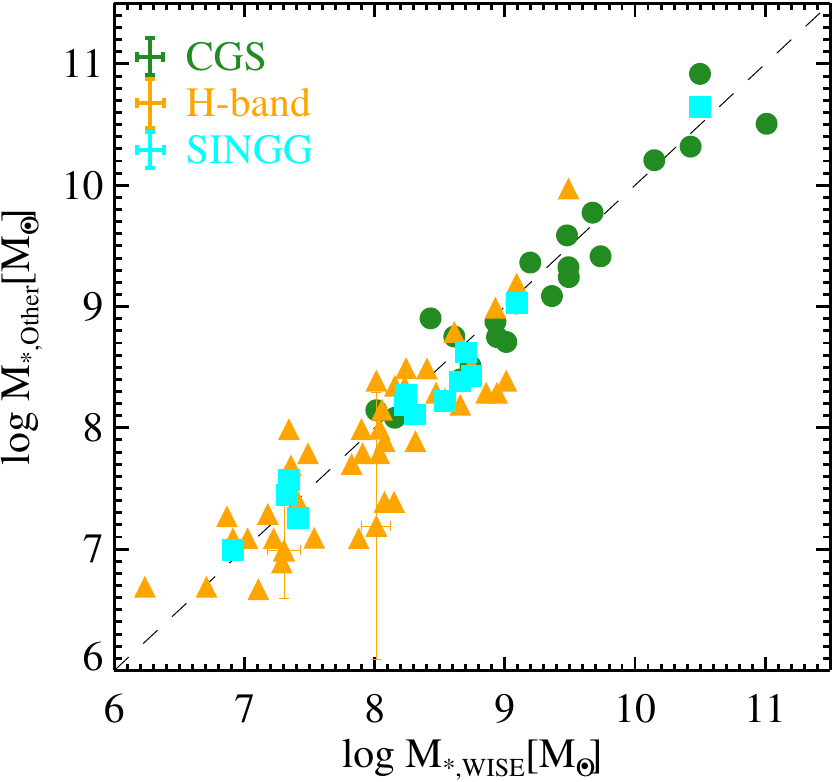}
\caption{Comparison of WISE-based $M_*$ (this work) with the estimates from other published data (see Section~\ref{sec:estimate_mst}). The dashed line marks the one-to-one relation. Typical error bars are shown in the left-top corner. Measurements from different datasets have been adjusted to have the same luminosity distance and \citet{Kroupa01} IMF before comparison.   }
\label{fig:comp_lgm}
\end{figure}

The \citet{Jarrett12} method was calibrated using nearby bright spiral galaxies, so there might be a worry that it fails for low-mass  irregular galaxies.   Hence we compare our WISE-based $M_*$ with the $M_*$ derived from optical and $H$-band data in Figure~\ref{fig:comp_lgm}. The CGS $M_*$ is estimated with the $B-R$ colour dependent $R$-band mass-to-light ratios based on the formula in \citet{Bell03}. The $H$-band based $M_*$ is taken from \citet{Kirby12} and \citet{Young14} and was estimated with a fixed mass-to-light ratio of 0.9. Finally, the SINGG  $M_*$ is estimated with a $R$-band luminosity dependent R-band mass-to-light ratio as discussed in \citet{Wong16}. The median and scatter of the differences in $M_*$ are --0.04 and 0.23 dex respectively for the comparison between WISE and CGS (19 galaxies),  --0.12 and 0.20 dex respectively for the comparison between WISE and $H$-band (22 galaxies), and 0.06 and 0.17 dex for the comparison between WISE and SINGG (13 galaxies). The scatter is consistent with the typical errors of $M_*$-to-light ratios with and without accounting for dependence on stellar populations \citep{Bell03}. As can be seen from Figure~\ref{fig:comp_lgm}, CGS contains the massive LVHIS galaxies, the $H$-band survey focusses on dwarf galaxies, while SINGG connects the two populations. The WISE-based $M_*$ estimates in this paper are in excellent agreement with the $M_*$ estimates from optical and near-infrared luminosities for both dwarf and massive galaxies.


\subsubsection{Estimating SFR}
\label{sec:estimate_sfr}
We estimate the dust un-attenuated part of the total SFR based on FUV luminosities (SFR$_{\rm FUV}$), and recover the dust attenuated part based on W4 luminosities  (SFR$_{\rm W4}$). This method is widely adopted in the literature \citep{Calzetti13, Hao11} and based on the fact that FUV fluxes are primarily contributed by the massive B stars which have a typical age of 100 Myr, and part of the FUV fluxes are absorbed and re-emitted by cool and warm dust in the far and mid-infrared,  while W4-band traces the warm dust with a temperature $\sim$150 K. 

We calculate SFR$_{\rm FUV}$ with the equation from \citet[][their table 1.2]{Calzetti13}, which is derived from stellar population models (dust-free). We calculate SFR$_{\rm W4}$ with equation 2 from \citet{Jarrett13}, which is derived from the infrared SED models of starburst galaxies (SFR fully attenuated). Then total \mbox{SFR$=$ SFR$_{\rm FUV}$+ SFR$_{\rm W4}$}.  Because FUV data are not always available  while W4 fluxes are not always detectable for all the galaxies, we use the W4 estimated SFR as a lower limit when only W4 data is available, and assume the dust attenuation to be zero when we get no detection from the W4 data.

\begin{figure} 
\includegraphics[width=8cm]{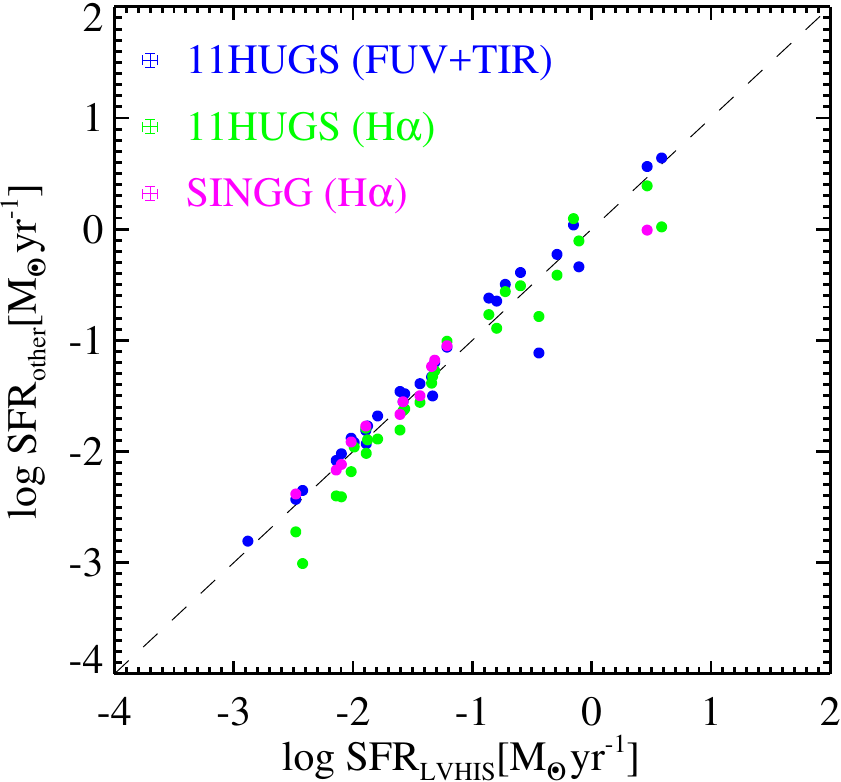}
\caption{Comparison of FUV+W4 based SFR (this work) with the estimates from other published data (see Section~\ref{sec:estimate_sfr}). The dashed line marks the one-to-one relation. Typical error bars are shown in the left-top corner. Measurements from different datasets have been adjusted to have the same luminosity distance and \citet{Kroupa01} IMF before comparison.    }
\label{fig:comp_sfr}
\end{figure}

We also caution that while we have tried our best to remove the contribution from stellar continuum (Section~\ref{sec:irphot}), W4 band as an SFR indicator has uncertainties for being a far extrapolation of the dominant bolometric emission that arises from cool dust in the far-infrared, and it has risks of being contaminated by the warm and dusty TP-AGB populations. Hence for validation we compare in Figure~\ref{fig:comp_sfr} our SFR with SFR derived in other ways (see Section~\ref{sec:otherdata}). The 11HUGS (FUV+TIR) SFR \citep[from][]{Lee09} was estimated in a similar way as in this paper, but the more accurate attenuation tracer, total infrared luminosities (TIR) instead of 22 $\mu$m were used to indicate the dust attenuated SFR. The 11HUGS (H$\alpha$) SFR \citep[also from][]{Lee09} was derived with the H$\alpha$ luminosities with a dust attenuation correction based on the Balmer decrement H$\alpha/$H$\beta$.  The SINGG (H$\alpha$) SFR is estimated from the H$\alpha$ luminosities with a dust attenuation correction determined by the R-band absolute magnitude \citep{Meurer06}. 

The LVHIS SFRs agree well with these three datasets, with a median difference of -0.08$\pm$0.09, 0.07$\pm$0.36 and -0.02$\pm$0.17 dex from the 11HUGS (FUV+TIR), 11HUGS (H$\alpha$) and SINGG (H$\alpha$) SFRs. When SFR$<$0.01 $\ms$ yr$^{-1}$, 11HUGS (H$\alpha$) SFRs tend to be systematically lower than our SFRs, which might be caused by insufficient samplings of the massive end of the initial mass function (IMF), or  by a different IMF shape at the massive end in the low gas surface density regions \citep{Meurer06, Lee09, LopezSanchez15}. On the other hand, there is no systematic offset between LVHIS and SINGG (H$\alpha$) SFRs at the low SFR end. 

\begin{table*}\small
\label{tab:galprop}
\caption{ Derived galaxy properties (to be continued).}
\centering
\begin{tabular}{lll | c c  c c c  c c c }
\hline
{\scriptsize ID} &  {\scriptsize HIPASS ID}  &{\scriptsize Galaxy Name} & SFR  & {\scriptsize log $M_*$} & {\scriptsize $\log \fHI$} &  {\scriptsize ${\rm R_{HI}/R_{25}}^r$}  & {\scriptsize $\log \Sigma_{{\rm *,eff}}$} & {\scriptsize $\log \Sigma_{\rm SFR,r<R25}$}  & {\scriptsize $\Sigma_{\rm HI,r<R25}^r$} & {\scriptsize $\Sigma_{\rm HI,r<3.2rs}^r$} \\
 
 & & & {\tiny [$\ms$ yr$^{-1}$]}  &{\tiny [$\ms$]} &  &    &  {\tiny [$\mspc$]} &  {\tiny [$M_{\odot}~{\rm yr^{-1}~kpc^{-2}}$]}   &  {\tiny [$\mspc$]} & {\tiny [$\mspc$]} \\

(1) & (2) & (3) & (4) & (5) & (6) & (7) & (8) & (9) & (10) & (11) \\

\hline
1 & J0008-34  &  ESO349-G031  &  0.0006$\pm$0.0001  &  7.3  &  -0.22  &  -  &  6.81  &  -2.92 &  -  &  0.35 \\
2 & J0026-41  &  ESO294-G010  &  0.0001$\pm$0.0001  &  6.7  &  -  &  -  &  7.36  &  -3.16 &  -  &  - \\
3 & J0015-32  &  ESO410-G005  &  0.0001$\pm$0.0000  &  6.2  &  -0.32  &  -  &  6.72  &  -3.67 &  -  &  0.35 \\
4$^l$ & J0015-39  &  NGC55  &  0.1896$\pm$0.0220  &  9.4  &  -0.03  &  1.26  &  8.01  &  -3.14 &  6.53  &  3.39 \\
5$^l$ & J0054-37  &  NGC300  &  0.1598$\pm$0.0210  &  9.5  &  -0.21  &  2.30  &  7.91  &  -2.86 &  5.44  &  4.67 \\
 & & & & & & & & & &\\
6$^l$ & J0047-25  &  NGC253  &  3.8612$\pm$0.5582  &  10.4  &  -0.99  &  0.82  &  8.84  &  -2.38 &  2.94  &  7.13 \\
7 & J0047-20  &  NGC247  &  0.1374$\pm$0.0206  &  9.7  &  -0.42  &  1.33  &  7.92  &  -3.37 &  4.74  &  1.70 \\
8 & J0135-41  &  NGC625  &  0.0455$\pm$0.0051  &  8.7  &  -0.75  &  0.77  &  8.00  &  -2.97 &  1.83  &  2.29 \\
9 & J0145-43  &  ESO245-G005  &  0.0249$\pm$0.0037  &  8.3  &  0.27  &  2.60  &  7.15  &  -2.74 &  7.73  &  2.65 \\
10 & J0150-44  &  ESO245-G007  &  0.0000$\pm$0.0000  &  6.8  &  -1.85  &  -  &  6.73  &  -3.45 &  -  &  0.03 \\
 & & & & & & & & & &\\
11$^l$ & J0237-61  &  ESO115-G021  &  0.0133$\pm$0.0020  &  8.1  &  0.76  &  2.25  &  7.28  &  -3.48 &  5.09  &  3.72 \\
12 & J0256-54  &  ESO154-G023  &  0.0366$\pm$0.0055  &  8.7  &  0.27  &  -  &  7.32  &  - &  -  &  - \\
13 & J0258-49  &  ESO199-G007  &  0.0015$\pm$0.0004  &  7.0$^h$  &  0.23  &  -  &  -  &  -3.02 &  -  &  0.62 \\
14$^l$ & J0317-66  &  NGC1313  &  0.5164$\pm$0.0642  &  9.5  &  -0.20  &  1.65  &  7.83  &  -2.40 &  8.45  &  8.31 \\
15 & J0320-52  &  NGC1311  &  0.0128$\pm$0.0019  &  8.2  &  -0.29  &  1.18  &  7.69  &  -3.15 &  3.80  &  2.22 \\
 & & & & & & & & & &\\
16 & J0321-66  &  AM0319-662  &  0.0001$\pm$0.0001  &  7.1  &  -1.14  &  -  &  -  &  -4.35 &  -  &  - \\
17 & J0333-50  &  IC1959  &  0.0261$\pm$0.0036  &  8.5  &  -0.19  &  1.55  &  8.02  &  -3.00 &  5.08  &  4.37 \\
18 & J0454-53  &  NGC1705  &  0.0481$\pm$0.0069  &  8.2  &  -0.37  &  -  &  8.15  &  -2.06 &  -  &  2.33 \\
19 & J0457-42  &  ESO252-IG001  &  -  &  7.3  &  0.76  &  2.11  &  6.77  &  - &  -  &  2.95 \\
20 & J0605-33  &  ESO364-GQ029  &  -  &  7.6  &  0.90  &  2.14  &  6.46  &  - &  -  &  2.57 \\
 & & & & & & & & & &\\
21 & J0607-34  &  AM0605-341  &  -  &  7.9  &  0.18  &  -  &  7.80  &  - &  -  &  - \\
22 & J0610-34  &  NGC2188  &  $>$0.0211$\pm$0.0036  &  8.9  &  -0.28  &  0.91  &  7.92  &  $>$-3.71 &  3.91  &  1.66 \\
23 & J0615-57  &  ESO121-G020  &  0.0033$\pm$0.0006  &  7.2  &  0.58  &  2.92  &  6.96  &  -2.66 &  -  &  1.38 \\
24 & J0639-40  &  ESO308-G022  &  -  &  7.8  &  -0.04  &  4.54  &  7.01  &  - &  -  &  - \\
25 & J0705-58  &  AM0704-582  &  -  &  6.2  &  2.09  &  -  &  6.76  &  - &  -  &  - \\
 & & & & & & & & & &\\
26 & J0731-68  &  ESO059-G001  &  -  &  8.1  &  -0.16  &  2.49  &  7.25  &  - &  4.41  &  1.03 \\
27 & J0926-76  &  NGC2915  &  0.0265$\pm$0.0038  &  8.4  &  0.10  &  4.26  &  8.22  &  -2.32 &  4.64  &  2.27 \\
28 & J1043-37  &  ESO376-G016  &  -  &  8.0  &  0.07  &  4.00  &  7.26  &  - &  -  &  - \\
29 & J1047-38  &  ESO318-G013  &  0.0057$\pm$0.0017  &  7.4  &  0.60  &  -  &  6.85  &  -3.35 &  -  &  0.62 \\
30 & J1057-48  &  ESO215-GQ009  &  0.0442$\pm$0.0067  &  8.0  &  0.89  &  17.38  &  6.92  &  -1.19 &  -  &  4.92 \\
 & & & & & & & & & &\\
31$^l$ & J1118-32  &  NGC3621  &  0.7127$\pm$0.0775  &  9.7  &  0.28  &  3.44  &  8.28  &  -2.60 &  4.15  &  3.18 \\
32 & J1131-31  &  new  &  -  &  -  &  -  &  -  &  -  &  - &  -  &  - \\
33 & J1132-32  &  new  &  -  &  7.7  &  -0.70  &  -  &  7.04  &  - &  -  &  - \\
34 & J1137-39  &  ESO320-G014  &  -  &  8.1  &  -0.84  &  -  &  7.26  &  - &  -  &  0.47 \\
35 & J1154-33  &  ESO379-G007  &  -  &  7.3  &  0.18  &  -  &  -  &  - &  -  &  - \\
 & & & & & & & & & &\\
36 & J1204-35  &  ESO379-G024  &  -  &  7.5  &  -0.37  &  -  &  7.28  &  - &  -  &  - \\
37 & J1214-38  &  ESO321-G014  &  0.0010$\pm$0.0001  &  7.0  &  0.06  &  1.79  &  7.06  &  -3.12 &  -  &  0.97 \\
38 & J1219-79  &  IC3104  &  -  &  8.2  &  -1.19  &  1.18  &  7.71  &  - &  4.08  &  0.37 \\
39 & J1244-35  &  ESO381-G018  &  -  &  7.2  &  0.04  &  -  &  7.31  &  - &  -  &  1.51 \\
40 & J1246-33  &  ESO381-G020  &  0.0097$\pm$0.0015  &  7.4  &  0.95  &  2.37  &  6.67  &  -3.06 &  -  &  2.33 \\
 & & & & & & & & & &\\
41 & J1247-77  &  new  &  -  &  7.1  &  -0.07  &  -  &  6.94  &  - &  -  &  1.39 \\
42 & J1305-40  &  CEN06  &  0.0016$\pm$0.0005  &  7.8  &  -0.20  &  -  &  6.91  &  -3.57 &  -  &  0.53 \\
43 & J1305-49  &  NGC4945  &  $>$0.9566$\pm$0.1455  &  10.1  &  -1.01  &  0.90  &  8.58  &  $>$-2.74 &  2.56  &  5.39 \\
44 & J1310-46  &  ESO269-G058  &  0.0018$\pm$0.0004  &  8.5  &  -1.21  &  0.66  &  7.78  &  -3.81 &  1.60  &  0.62 \\
45 & J1321-31  &  new  &  0.0005$\pm$0.0001  &  -  &  -  &  -  &  -  &  -3.97 &  -  &  - \\

 \hline
\end{tabular}
\begin{flushleft}
$^h$: $H$-band photometry based.\\
$^l$: Mark the sub-sample of large galaxies for detailed analysis (see Section~\ref{sec:sample} and \ref{sec:analysis2}).\\
\end{flushleft}
\label{tab:galprop}
\end{table*} 
      
\addtocounter{table}{-1}      
\begin{table*}\small
\caption{ Derived galaxy properties  (continued).}
\centering
\begin{tabular}{lll | c c  c c c  c c c }
\hline
 {\scriptsize ID} &  {\scriptsize HIPASS ID}  &{\scriptsize Galaxy Name} & SFR  & {\scriptsize log $M_*^e$} & {\scriptsize $\log \fHI$} &  {\scriptsize ${\rm R_{HI}/R_{25}}$}  & {\scriptsize $\log \Sigma_{{\rm *,eff}}$} & {\scriptsize $\log \Sigma_{\rm SFR,r<R25}$}  & {\scriptsize $\Sigma_{\rm HI,r<R25}$} & {\scriptsize $\Sigma_{\rm HI,r<3.2rs}$} \\
 
& & & {\tiny [$\ms$ yr$^{-1}$]}  &{\tiny [$\ms$]} &  &    &  {\tiny [$\mspc$]} &  {\tiny [$M_{\odot}~{\rm yr^{-1}~kpc^{-2}}$]}   &  {\tiny [$\mspc$]} & {\tiny [$\mspc$]} \\
(
1) & (2) & (3) & (4) & (5) & (6) & (7) & (8) & (9) & (10) & (11) \\

\hline
46 & J1321-36  &  NGC5102  &  -  &  9.5  &  -1.11  &  1.14  &  8.68  &  - &  1.77  &  1.39 \\
47 & J1324-30  &  AM1321-304  &  0.0004$\pm$0.0006  &  7.9  &  -0.95  &  -  &  7.15  &  -3.99 &  -  &  0.65 \\
48 & J1324-42  &  NGC5128  &  0.7782$\pm$0.1038  &  11.0  &  -  &  -  &  9.00  &  -3.05 &  -  &  - \\
49 & J1326-30  &  IC4247  &  0.0037$\pm$0.0006  &  7.5  &  -0.21  &  -  &  7.42  &  -2.97 &  -  &  1.12 \\
50 & J1327-41  &  ESO324-G024  &  0.0118$\pm$0.0019  &  8.0  &  0.20  &  2.36  &  7.23  &  -2.85 &  6.97  &  3.69 \\
 & & & & & & & & & &\\
51 & J1334-45  &  ESO270-G017  &  $>$0.0053$\pm$0.0012  &  8.7  &  0.10  &  1.36  &  7.41  &  $>$-4.33 &  4.02  &  3.34 \\
52 & J1336-29  &  UGCA365  &  0.0010$\pm$0.0003  &  7.2  &  0.25  &  -  &  7.01  &  -3.39 &  -  &  - \\
53$^l$ & J1337-29  &  NGC5236  &  2.9211$\pm$0.3408  &  10.5  &  -0.59  &  2.62  &  8.50  &  -2.11 &  4.25  &  3.92 \\
54 & J1337-39  &  new  &  0.0012$\pm$0.0018  &  -  &  -  &  -  &  -  &  -3.55 &  -  &  - \\
55 & J1337-42  &  NGC5237  &  0.0031$\pm$0.0007  &  8.3  &  -0.81  &  0.89  &  8.11  &  -3.07 &  1.99  &  1.03 \\
 & & & & & & & & & &\\
56 & J1337-28  &  ESO444-G084  &  0.0031$\pm$0.0005  &  6.9  &  1.00  &  2.89  &  6.75  &  -2.89 &  -  &  3.12 \\
57 & J1339-31  &  NGC5253  &  0.3606$\pm$0.0471  &  8.6  &  -0.66  &  0.95  &  8.38  &  -1.79 &  3.74  &  10.06 \\
58 & J1340-28  &  IC4316  &  0.0026$\pm$0.0004  &  7.9  &  -0.83  &  -  &  7.38  &  -2.26 &  -  &  0.33 \\
59 & J1341-29  &  NGC5264  &  0.0099$\pm$0.0015  &  8.7  &  -0.96  &  1.01  &  7.71  &  -3.20 &  2.65  &  0.43 \\
60 & J1345-41  &  ESO325-GQ011  &  -  &  7.1  &  0.72  &  1.79  &  6.71  &  - &  6.77  &  2.79 \\
 & & & & & & & & & &\\
61 & J1348-37  &  new  &  -  &  -  &  -  &  -  &  -  &  - &  -  &  0.29 \\
62 & J1348-53  &  ESO174-GQ001  &  -  &  7.6  &  0.57  &  -  &  7.65  &  - &  -  &  2.36 \\
63 & J1349-36  &  ESO383-G087  &  $>$0.0022$\pm$0.0005  &  8.9  &  -1.09  &  0.92  &  7.85  &  $>$-4.07 &  2.47  &  0.63 \\
64 & J1351-47  &  new  &  -  &  -  &  -  &  -  &  -  &  - &  -  &  - \\
65 & J1403-41  &  NGC5408  &  $>$0.0197$\pm$0.0031  &  8.2  &  0.19  &  2.70  &  7.73  &  $>$-2.79 &  -  &  3.88 \\
 & & & & & & & & & &\\
66 & J1413-65  &  Circinus  &  $>$3.0452$\pm$0.4645  &  10.1  &  -0.26  &  5.63  &  -  &  $>$-1.46 &  3.72  &  3.88 \\
67 & J1428-46  &  UKS1424-460  &  -  &  7.5  &  0.18  &  -  &  7.04  &  - &  -  &  3.71 \\
68 & J1434-49  &  ESO222-G010  &  -  &  7.8$^h$  &  -0.09  &  -  &  -  &  - &  -  &  0.96 \\
69 & J1441-62  &  new  &  -  &  7.8  &  -0.47  &  -  &  8.16  &  - &  -  &  - \\
70 & J1443-44  &  ESO272-G025  &  -  &  8.0  &  -0.93  &  -  &  7.72  &  - &  -  &  - \\
 & & & & & & & & & &\\
71 & J1501-48  &  ESO223-G009  &  -  &  8.9  &  0.13  &  -  &  7.67  &  - &  -  &  5.32 \\
72 & J1514-46  &  ESO274-G001  &  $>$0.0101$\pm$0.0016  &  9.0  &  -0.52  &  1.26  &  7.96  &  $>$-3.78 &  3.67  &  3.94 \\
73 & J1526-51  &  new  &  -  &  6.9  &  0.66  &  -  &  -  &  - &  -  &  - \\
74 & J1620-60  &  ESO137-G018  &  -  &  8.4  &  0.19  &  1.36  &  7.55  &  - &  8.19  &  0.84 \\
75 & J1747-64  &  IC4662  &  0.0485$\pm$0.0058  &  8.0  &  0.15  &  4.68  &  7.92  &  -1.59 &  -  &  8.48 \\
 & & & & & & & & & &\\
76 & J2003-31  &  ESO461-G036  &  -  &  8.1  &  -0.12  &  5.12  &  7.09  &  - &  -  &  1.73 \\
77$^l$ & J2052-69  &  IC5052  &  0.0613$\pm$0.0068  &  9.1  &  -0.20  &  1.33  &  8.02  &  -3.30 &  3.92  &  2.42 \\
78$^l$ & J2202-51  &  IC5152  &  0.0161$\pm$0.0022  &  8.2  &  -0.20  &  1.85  &  7.88  &  -2.59 &  8.03  &  4.73 \\
79 & J2326-32  &  UGCA438  &  0.0013$\pm$0.0002  &  6.9  &  -0.25  &  -  &  6.86  &  -2.83 &  -  &  0.37 \\
80 & J2343-31  &  UGCA442  &  0.0080$\pm$0.0012  &  7.3  &  1.01  &  1.81  &  6.95  &  -3.33 &  7.02  &  3.47 \\
 & & & & & & & & & &\\
81 & J2352-52  &  ESO149-G003  &  0.0073$\pm$0.0011  &  7.4  &  0.36  &  -  &  7.52  &  -3.14 &  2.12  &  - \\
82$^l$ & J2357-32  &  NGC7793  &  0.2526$\pm$0.0307  &  9.5  &  -0.53  &  1.30  &  8.03  &  -2.64 &  6.22  &  4.22 \\

\hline
\end{tabular}
\label{tab:galprop}
\end{table*}

\section{Analysis for the large LVHIS galaxies}
\label{sec:analysis2}
\subsection{Resolution-matched multi-wavelength photometry}
\label{sec:phot_conv}
We have selected ten well-resolved LVHIS galaxies for an analysis of radial distributions of multi-wavelength light. We use the pipeline of \citet{Wang10} to perform resolution-matched photometric measurements from the GALEX, WISE and \hi\ images. Instead of using the original GALEX and WISE images, we use the cleaned images (see Section~\ref{sec:uvphot} and \ref{sec:irphot}). We remove the background and set all pixels beyond the ellipse defined by the last measured point in the SB profile to zero. By doing so, we minimise possible contamination from nearby sources in the following step of resolution matching.

Because the PSFs of WISE images are known to have significant side-lobes, we convolve the WISE cleaned images with the kernels from \citet{Aniano11}, which convert the WISE PSFs to PSF shapes described by Gaussian functions. After this convolution process, the W1, 2 and 4 images have Gaussian-shaped PSFs with FWHM of 7.5, 8.0 and 15.0 arcsec, respectively. 

 For each galaxy the \hi\ image, which has the lowest resolution, is used as the reference image. All other images are registered and convolved to have the same geometry and PSF resolution as the reference image. 
 The resolution matched images for the whole sub-sample have a physical spatial resolution between 0.3 to 2.5 kpc, and a median value of 1.5 kpc. We show the stellar mass and SFR images derived based on these images, as well as the relation matched \hi\ images in Figure~\ref{fig:atlas_conv}.
 
 We measure radial SB profiles from these images. SB is azimuthally averaged in elliptical rings with a position angle and axis ratio  determined from the WISE W1 images (Section~\ref{sec:irphot}). We apply a projection correction to SB by multiplying them with the axis ratio b$/$a (assuming very thin discs). The major axis of each ring is taken as the profile radius.  We limit the outermost radius of each profile to be the same as the profile measured from the original image before convolution, because we have removed all pixels outside that radius before the image resolution matching step. We use the background noise from the original images to estimate photometric errors, as the noise has become correlated in the resolution-matched images. 
 
 Finally, $M_*$ and SFR profiles are calculated based on these SB profiles. When deriving $M_*$, we use the W1 SB profile and a fixed mass-to-light ratio determined by the total W1$-$W2 colour, because the W2 SB profiles do not extend as far as the W1 SB profiles in most of the galaxies while our study focuses on the outskirts of galaxies. When deriving SFR profiles, the W4 SB profiles also end at much smaller radius than the FUV SB profiles, so we exponentially extrapolating the SFR$_{\rm W4}$ profiles in order to obtain total SFR profiles. In most galaxies, the extrapolated SFR$_{\rm W4}$ profile in the outer region is lower than the SFR$_{\rm FUV}$ profile, which alleviates the uncertainties caused by extrapolation.
  
 We obtained the surface densities ($\Sigma$) at the radius where $\SHI=3~ \ms {\rm pc^{-2}}$,  where $\Sigma_*=3~ \ms {\rm pc^{-2}}$ and at $r$ = 1.25 R$_{25}$. This allows us to study the  $\Sigma_{\rm SFR}$-$\SHI$ relation when fixing $\Sigma_*$ at $3~ \ms {\rm pc^{-2}}$, the $\Sigma_{\rm SFR}$-$\Sigma_*$ relation when fixing $\SHI$ at $3~ \ms {\rm pc^{-2}}$, and both relations at 1.25 R$_{25}$\footnote[1]{We caution that there is faint Galactic cirrus in the foreground of NGC 1313 which may affect the $\Sigma_{\rm SFR}$ measurements in the outskirts \citep{Contursi02}. On the other hand, removing NGC 1313 from our sample does not change our conclusion.}. These investigations will be presented in Section~\ref{sec:loc_SFE}. When the radial positions are outside the radial range of a profile (in all cases, $\Sigma_*$ profiles), we obtain the surface densities through linearly extrapolating in logarithmic space the exponential outer part of the radial profiles. Most of the extrapolated radii are within 1.2 times the last data point of the profiles. The error of an extrapolated surface density is a combination of the extrapolated measurement error and the f-test based model error. 
 
 These resolution-matched $\mHI$, $M_*$ and SFR radial profiles and measurements are presented in the last column of Figure~\ref{fig:atlas_conv}.
 
 Regarding extrapolating the $\Sigma_*$ profiles, one may worry about the fact that optical and mid-infrared radial profiles frequently bend up or downward in the galactic outskirts \citep{Pohlen06}. However, there is evidence that $\Sigma_*$ radial profiles are much less bent than the surface brightness profiles \citep{Bakos08}, and we are not extrapolating far away from the last measurable data points. These two facts to some extent alleviate the problem of bending profiles. Nevertheless, optical or MIR images that are deep, of high-resolution (to suppress confusion noise), and of large field-of-view (to homogeneously cover the outskirts of the galaxies) are needed for a future confirmation of our results.  Such kind of data will be available with the completion of  wide-field optical imaging surveys, like the Skymapper Southern Sky Survey \citep{Keller07}.

\subsection{HI tilted ring models}
\label{sec:tiltedring}
\subsubsection{Building the HI tilted ring models}
We make use of Fully Automated TiRiFiC \citep[FAT]{Kamphuis15}, which builds on the 3D Tilted Ring Fitting Code \citep[TiRiFiC][J07 hereafter]{Jozsa07}, to derive tilted ring models and rotation curves for the large LVHIS galaxies.  The approaching and receding velocity sides are modelled separately. Part of the sample has already been processed by \citet{Kamphuis15} using FAT, including NGC~253, NGC~1313 and IC\,5152.  We  present the model rotation curves and radial profiles of position angles, inclinations and surface densities in Figure~\ref{fig:atlas_highrs} in the appendix. 

In Figure~\ref{fig:atlas_highrs}, we can see that at the optical radius R$_{25}$, the \hi\ and optical inclinations usually agree within 20$\degree$. At R$_{25}$, PA (position angle) from the two bands also agree within 20$\degree$ for most of the galaxies, except for NGC 5236 (PA from the two bands differ by $\sim60\degree$). Tilted ring models mostly provide a fit considering kinematics, while in the optical they are not taken into account. Below an inclination of 40$\degree$ it becomes very difficult to accurately measure PAs without kinematical information, which might explain the discrepancy between the optical and kinematical PAs for NGC 5236.


 If we assume the extent of fluctuations to reflect fitting uncertainties, the 3D de-projected radial \hi\ profiles (as an output of FAT) generally agree well with the radial \hi\ profiles which are measured and corrected for projection effects from the 2D moment-0 images (see Section~\ref{sec:sample} and \citet{Wang16}).  

We use the model results on both the approaching and receding velocity sides of galaxies to analyse \hi\ warp related properties (see next section), and use the average of the two sides to study the relation between  SFR and mass radial surface densities (Section~\ref{sec:best_res} and \ref{sec:loc_SFE}).

\subsubsection{Tiltograms and identification of warps}
As summarised in the introduction of \citet{Jozsa07b}, an \hi\ warp is the inner plane of an \hi\ disc kinematically bending in the outer region, and very often transiting into another plane.  \citet{Jozsa07b} demonstrated that the ``tiltograms'' are very useful for identifying \hi\ warps.

As outlined in \citet{Jozsa07b}, a tiltogram is a pixel map showing the mutual inclination angle ($\theta_{tilt}$) between rings at different radius.  The pixel value ($\theta_{tilt}$) at x$=$r1 and y$=$r2 in the map measures the angle between the normal vectors of the tilted rings at radius r1 and r2. The direction of the normal vector for each tilted ring is fully described by the position angle and inclination angle of the tilted ring.  A completely flattened disc plane has a tiltogram with zero value in all pixels.  

A tiltogram can be divided into four regions at a radius $r$, where the bottom-left square region with $x<r$ and $y<r$ shows the intrinsic mutual inclination of the disc within $r$ ($\theta_{tilt}^{in}$),  the top-right square region with $x>r$ and $y>r$  shows the intrinsic mutual inclination of the disc outside $r$ ($\theta_{tilt}^{out}$),  and the remaining two rectangles (which are identical to each other) showing the mutual inclination between the discs within and beyond $r$ ($\theta_{tilt}^{mut}$).  

If a galaxy has an \hi\ warp starting from radius $r_{warp}$, $r_{warp}$ will divide the tiltogram into four regions as described above, where $\theta_{tilt}^{in}$ are close to zero and nearly constant, while $\theta_{tilt}^{mut}$ are large. When an outer disc plane exists beyond $r_{warp}$, $\theta_{tilt}^{out}$ are also close to zero and nearly constant. So we calculate the average $\theta_{tilt}$ ($\overline{\theta_{tilt}}$) in the four regions divided by each radius $r$, and set the criteria that an \hi\ disc is classified as warped if a radius $r_{warp}$ can be found where
\begin{enumerate}
\item $|\overline{\theta_{tilt}^{in}}-\overline{\theta_{tilt}^{mut}}|$ is a local maximum,
\item $\overline{\theta_{tilt}^{in}} < 5$ deg, and
\item $\overline{\theta_{tilt}^{mut}} > 10$ deg.
\end{enumerate}
$\overline{\theta_{tilt}^{in}}$ indicates the flatness of the inner disc, $\overline{\theta_{tilt}^{mut}}$ indicates the warp angle, and $\overline{\theta_{tilt}^{out}}$ indicates the flatness of the outer disc.

We set the criteria in a relatively arbitrary way, which can be biased toward selecting the strongly warped \hi\ discs. But it is sufficient for our scientific aim of investigating the influence of warps on SFE, as strong warps should have stronger effects than weak warps.

Because a galaxy can be warped on only one side of the galaxy, we analyse the approaching and receding side of the velocity field seperately.
We display the tiltograms for the approaching velocity side of the LVHIS large galaxies in the fifth column of  Figure~\ref{fig:atlas_highrs}. The tiltogram of NGC~300 is a perfect example showing warp features, where $r_{warp}$ is marked by purple lines. We notice from the tiltograms that sometimes one ring deviates abruptly from adjacent rings (e.g. at the radii 7.5\arcmin\ in IC5052). This kind of feature occurs when the tilted rings warping dramatically from one side to the other side of the inner plane.

We list the $\overline{\theta_{tilt}}$ and $r_{warp}$ measurements in Table~\ref{tab:warp_app} for the approaching and receding velocity side respectively. We find \hi\ warps on the approaching velocity side in seven out of the ten large LVHIS galaxies except for NGC~253, ESO115-G021 and NGC~1313 for which $\overline{\theta_{tilt}^{mut}}$ are too small. The same result is found on the receding velocity side.

\begin{table*}
\caption{Measurements from the tiltograms of the ten large LVHIS galaxies.}
\centering
\begin{tabular}{l|ccccc	| ccccc}
\hline
\multicolumn{1}{c|}{ } & \multicolumn{5}{c|}{Approaching side}&  \multicolumn{5}{c}{Receding side} \\
\hline
 & & & & & & & & & & \\	
Galaxy & $\overline{\theta_{tilt}^{in}}$  & $\overline{\theta_{tilt}^{out}}$ 
       & $\overline{\theta_{tilt}^{mut}}$ & $r_{warp}^1$  & $r_{warp}/$R$_{25}$ 
             & $\overline{\theta_{tilt}^{in}}$  & $\overline{\theta_{tilt}^{out}}$ 
       & $\overline{\theta_{tilt}^{mut}}$ & $r_{warp}^1$  & $r_{warp}/$R$_{25}$ \\
 name & [deg] & [deg] & [deg] & [arcmin] & 
 		 [deg] & [deg] & [deg] & [arcmin] &\\
\hline
NGC~55      &  2.6 &  10.3 &  21.7 &  14.7 & 0.93 			&3.2&4.9&12.6&20.0&1.36\\
NGC~300     &  0.9 &   9.4 &  20.5 &  10.2  & 1.05			&0.6&8.3&17.8&10.2&1.05\\
NGC~253$^*$ &  0.0 &   1.1 &   2.3 &   4.2  & 0.28			&0.0&1.0&2.1&4.2&0.28\\
ESO115-G021$^*$ &  0.2 &   5.7 &   8.9 &   3.7 & 1.50		&0.4&4.7&7.2&3.5&1.42\\ 
NGC~1313$^*$&  0.3 &   4.2 &   3.8 &   2.2  & 0.41			&0.2&6.8&7.5&2.7&0.50\\
NGC~3621    &  1.4 &   5.5 &  15.4 &  12.4 & 2.56			&3.6&5.5&21.0&14.7&3.02\\
NGC~5236    &  3.0 &   7.3 &  22.6 &  10.0 & 1.30			&1.8&20.5&29.5&6.9&0.89\\
IC\,5052    &  2.0 &  17.2 &  18.6 &   3.1 & 0.88				&0.9&7.4&14.5&3.4&0.95\\
IC\,5152    &  0.0 &  10.0 &  17.5 &   1.9 & 0.76				&0.2&8.8&17.3&2.0&0.82\\
NGC~7793    &  3.1 &   5.1 &  17.5 &   6.3  &1.22			&1.6&3.6&11.1&6.2&1.19\\
\hline
\end{tabular}
\begin{flushleft}
$^1$: the radius identified based on the warp identification criteria 1.\\
$^*$:  galaxies that do not meet the warp identification criteria 2 and 3 in Section \label{sec:tiltedring}. As we have adopted relatively strict criteria that are biased toward strong warps, we provide warp related parameters for these un-warped galaxies to show the full parameter space of the sample.
\end{flushleft}
\label{tab:warp_app}
\end{table*}

\subsubsection{Dark matter and stellar volume densities}
\label{sec:best_res}
In Section~\ref{sec:loc_SFE}, motivated by the mid-plane pressure models for star formation \citep{Ostriker10, Krumholz13}, we will study the dependence of SFR surface densities on the volume densities of stars and dark matter added together ($\sigma_{\rm sd}$). In order to obtain a rough estimate of the parameter $\sigma_{\rm sd}$, we make use of the tilted ring models derived in Section~\ref{sec:tiltedring}, which have been de-convolved with their PSF. So when we derive the $\Sigma_*$ and $\Sigma_{\rm SFR}$ radial profiles, we perform a resolution matched photometry as in Section~\ref{sec:phot_conv}, but at the resolution of the W4 images.

We use the radial \hi\ profiles from the tilted ring models. We use Equation 3 provided by \citet{Bigiel08} to estimate the molecular hydrogen surface density $\SH2$ from $\Sigma_{\rm SFR}$. The atomic and molecular hydrogen gas surface densities are multiplied by 1.36 to account for the helium gas.  We use the sum of gas and stellar surface densities to approximate the baryonic mass densities. We use the rotation curves to estimate the total mass at each radius ($M_{\rm tot}$). We subtracted from $M_{\rm tot}$ the contributions of stars and gas, and obtain the mass of the dark matter halo. We then assume a simple spherical model to calculate the volume density of the dark matter halo, $\sigma_{\rm dm}$. These calculations are crude, while a more detailed decomposition of the different masses is beyond the scope of this paper.

Following the method in \citet{Leroy08}, we assume a fixed ratio of 7.3 between the scale-length and the scale-height ($h$) for the stellar discs, and calculate the volume density of stars as $\sigma_*=\Sigma_*/(4h)=\Sigma_*/(0.54 {r_s})$.  Finally $\sigma_{\rm sd}=\sigma_*+\sigma_{\rm dm}$.

\section{LVHIS galaxies and the HI radial distributions}
We have built a multi-wavelength dataset for the LVHIS sample, based on which we have derived $\mHI$, $M_*$ and SFR.
In this section, we compare the distribution of these parameters with other surveys to understand the selection effect and the advantage of LVHIS. We also investigate how \hi\ is radially distributed with respect to the optical discs and how it depends on the \hi-richness of galaxies. By assessing the diversity of radial \hi\ distributions, we better understand what new information is gained with interferometry data, and why they are useful for understanding global scaling relations in the next section. 

\subsection{LVHIS compared to other HI samples}
\label{sec:scale_relation}
This section compares the the $\fHI$ and SFR distributions of the LVHIS sample with other \hi\ samples.

\begin{figure*} 
\centering
\includegraphics[width=14cm]{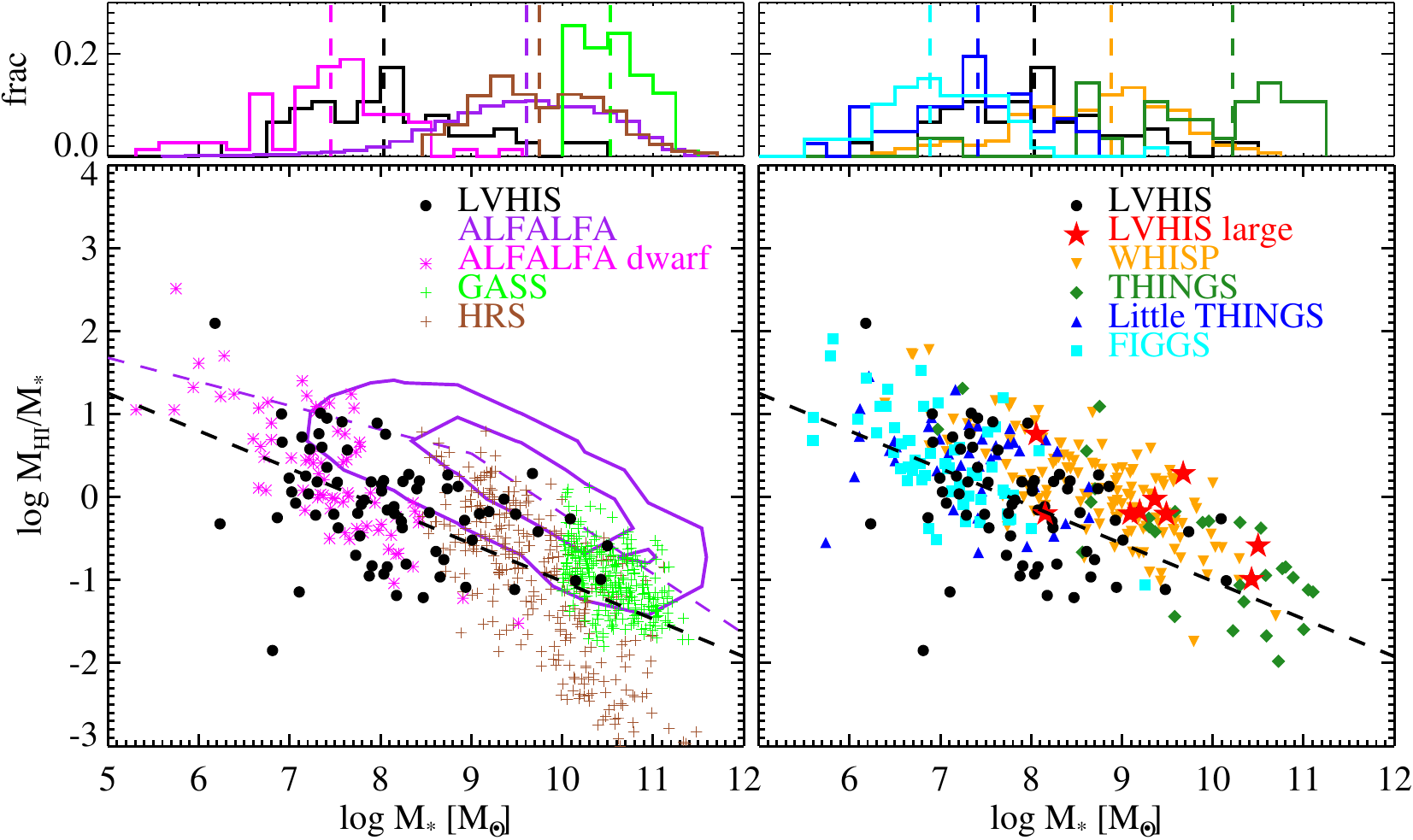}
\caption{\hi\ fraction ($\fHI=\mHI/M_*$) as a function of $M_*$ for all LVHIS galaxies compared to other \hi\ surveys obtained with single dish telescopes (left) and interferometers (right).
 The left panel compares LVHIS with surveys with single-dish \hi\ data, including LVHIS, ALFALFA \citep{Huang12b}, ALFALFA dwarf sample \citep{Huang12}, GASS \citep{Catinella13} and HRS \citep{Boselli14}.  $M_*$ values are taken from the MPA$\slash$JHU catalog for the ALFALFA and GASS samples, from \citet{Huang12} for the ALFALFA dwarf sample, and from \citet{Cortese12} for HRS.
The right panel shows samples from \hi\ interferometric surveys, including LVHIS, WHISP \citep{Swaters02}, THINGS \citep{Walter08}, Little THINGS \citep{Hunter12} and FIGGS \citep{Begum08}. $M_*$ are calculated from the $B$-band luminosities with mass-to-light ratios determined by the $B-V$ colour \citep[formula from ][]{Bell03} for the THINGS, Little THINGS and FIGGS samples. 
The THINGS $B$-band magnitudes are from SIMBAD astronomical database \citep{Wenger00}, and V-band magnitudes are from \citet{Walter08}. The Little THINGS $B$- and $V$- magnitudes are from \citet{Hunter06}. 
The FIGGS $B$- and $V$- magnitudes are from \citet{Begum08}.  The WHISP  $M_*$ are calculated from the $B$-band luminosities with mass-to-light ratio determined by the $B-R$ colour \citep[formula from ][]{Bell03}.  The WHISP $B$-band magnitudes are from \citet{Swaters02}, and R-band magnitudes are from \citet{Swaters02b}.
 Different samples are shown in different colours and symbols as denoted in the right-top corner of each panel.  The purple contours on the left panel show the distributions of 65\% and 95\% of the data points from the ALFAFLA sample \citep{Huang12b}. The dashed purple line in the bottom-left panel shows the mean relation for the ALFALFA sample \citep{Huang12b}.The dashed black lines in the bottom panels show the linear fit to the log $M_*$-log $\fHI$ relation  for LVHIS. The dashed lines in the top panels show the median values of $M_*$ distributions. 
}
\label{fig:surveys}
\end{figure*}

In Figure~\ref{fig:surveys}, we can see that $M_*$ of LVHIS ranges from 10$^6~\ms$ to 10$^{10.5}~\ms$ and has a median $\sim 10^8~\ms$. The sample covers nearly two orders of magnitude in $\fHI$ ($=\mHI/M_*$) at a fixed $M_*$. For example,  at $M_*\sim10^8~\ms$, $\fHI$ ranges from -1.2 to 1 dex. 

The left panel of Figure~\ref{fig:surveys} compares LVHIS with several single-dish \hi\ datasets. We can see that most of the LVHIS galaxies have lower $M_*$ than the median $M_*$ of ALFALFA, GASS and HRS. Their $\fHI$ are typically lower the median $\fHI$ of ALFALFA at a fixed $M_*$, and comparable to an extrapolation of the $\fHI$-$M_*$ distribution to the lower $M_*$ side for GASS and HRS.

In the right panel of Figure~\ref{fig:surveys}, we compare LVHIS to several \hi\ interferometry surveys. We can see that both LVHIS and WHISP tend to connect the existing resolved \hi\ samples of irregular galaxies (like LITTLE THINGS and FIGGS) and massive spiral galaxies (like THINGS) in the $\fHI$-$M_*$ space. WHISP provides a wider $M_*$ distribution than LVHIS, while LVHIS detects galaxies with on average lower $\fHI$ at a fixed $M_*$ than WHISP. 
On the other hand, the LVHIS large galaxies sample, in comparison to the rest LVHIS galaxies, are biased toward systems with high $\fHI$ for their $M_*$.

These comparisons suggest LVHIS as (or close to) a volume-limited \hi\ sample covers a wide range of both $M_*$ and $\fHI$. It is not significantly biased to high $\fHI$ galaxies compared to single dish samples, hence has the ability to explore the cause for global scaling relations using spatially resolved \hi\ distributions; it includes both spiral and irregular galaxies, hence has the ability to explore general physics at work in these two types of galaxies. This paper will make use of these advantages.

\begin{figure} 
\centering
\includegraphics[width=8cm]{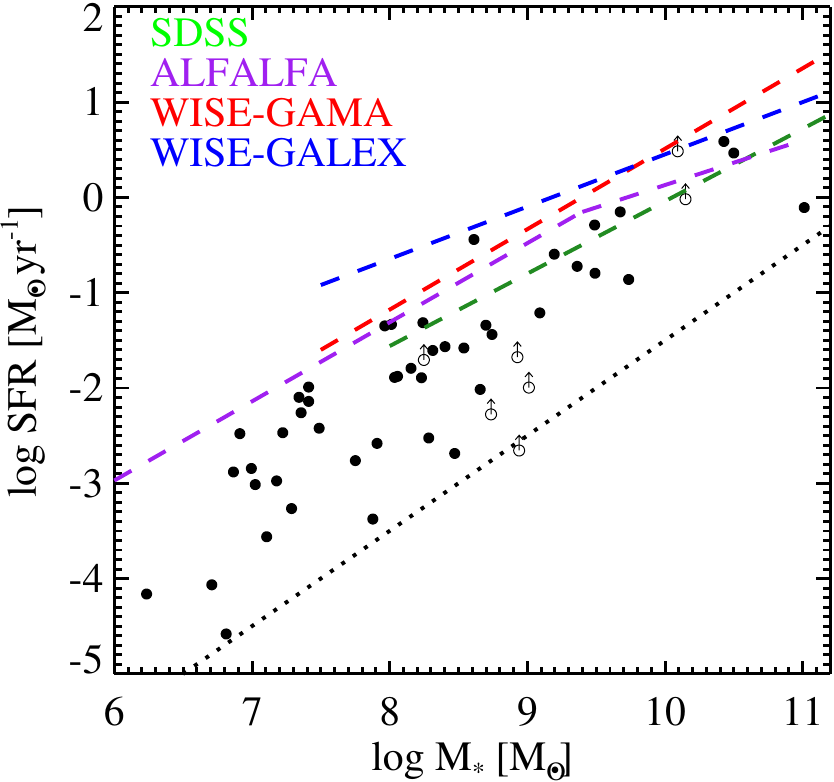}
\caption{The distribution of LVHIS galaxies around the star-forming Main Sequence (MS). The open circles and upward arrows show the lower limits of SFR estimate. The green line is the MS from the SDSS sample \citep{Renzini15}, the purple, red and blue dashed lines are the mean relations from ALFALFA  \citep{Huang12b}, the WISE-GAMA sample \citep{Jarrett17} and the GALEX-GAMA sample \citep{Grootes13}. The black dotted line shows the position for log sSFR ($={\log \rm SFR}/M_*$)$=-11.5$.}
\label{fig:SFMS}
\end{figure}

In Figure~\ref{fig:SFMS}, we show that most of the LVHIS galaxies lie below the star-forming Main Sequence (MS) from SDSS \citep{Renzini15} and also below the mean SFR-$M_*$ relations for ALFALFA \citep{Huang12b}, WISE-GAMA \citep{Jarrett17} and WISE-GALEX \citep{Grootes13}.  This is consistent with the result that most LVHIS galaxies lie below the mean $\fHI$-$M_*$ relation for ALFALFA,  which is close to the MS of low redshift galaxies \citep{Huang12b}. It confirms the previous finding that the SFR-$M_*$ and $\fHI$-$M_*$ relations mimic one another \citep{Saintonge16, Brown15}. On the other hand, most of the LVHIS galaxies have log sSFR ($={\log \rm SFR}/M_*$)$>-11.5$, indicating most of them to be star-forming galaxies.

\subsection{Radial distribution of HI gas}
\label{sec:dist_HI}
This section investigates the diversity of \hi\ radial distributions in the LVHIS galaxies.

\begin{figure*} 
\centering
\includegraphics[width=14cm]{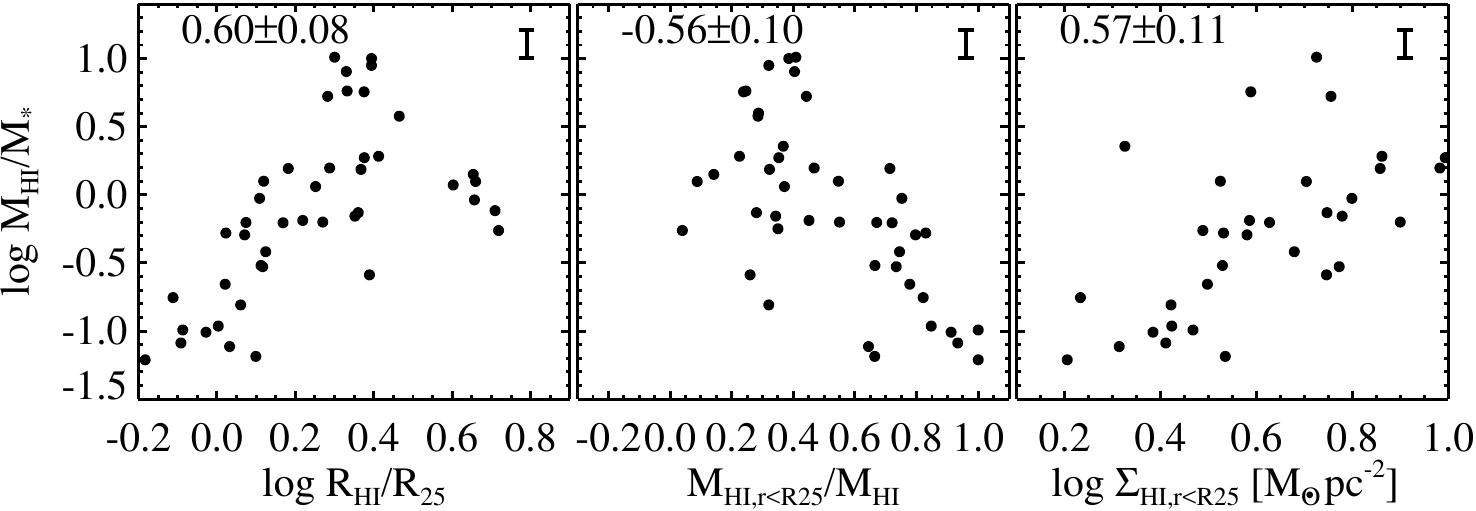}
\vspace{0.4cm}
\caption{Distribution of \hi\ gas with respect to the optical disc.  The optical disc edge is quantified as R$_{25}$. Three parameters are considered: the sizes of \hi\ disc with respect to the optical disc (${\rm R_{HI}/R_{25}}$, left panel) ,  the fraction of total \hi\ mass within the optical disc (${M_{{\rm HI,r<R25}}/M_{{\rm HI}}}$, middle panel), and the average \hi\ surface density within the optical disc (${\rm \Sigma_{HI,r<R25}}$, right panel). When the \hi\ related parameter is defined with a radius smaller than 2$\times$B$_{\rm maj}$ (the \hi\ PSF size), the \hi\ data is considered unresolved and the galaxy is excluded. In each panel, the correlation coefficient $\rho$ is denoted on the left-top corner and the measurements error bar is shown in the right-top corner.}
\centering
\label{fig:HI_distr}
\end{figure*}

In Figure~\ref{fig:HI_distr}, we find that galaxies with higher $\fHI$ on average have larger \hi\ discs with respect to the optical discs (${\rm R_{HI}/R_{25}}$, top panel),  and lower fractions of total \hi\ masses enclosed within the optical discs (${M_{{\rm HI,r<R25}}/M_{{\rm HI}}}$, middle panel), but higher average \hi\ surface densities within the optical discs (${\rm \Sigma_{HI,r<R25}}$, bottom panel). However, the three correlations have considerable scatter. For example, when $\fHI=1$, ${\rm R_{HI}/R_{25}}$ ranges from 1.25 to 6.3, ${M_{{\rm HI,r<R25}}/M_{{\rm HI}}}$ ranges from 0.1 to 0.7, and ${\rm \Sigma_{HI,r<R25}}$ ranges from 4 to 8 $\mspc$, within the LVHIS sample. 

So there are a large variety of \hi\ radial distributions at a fixed global properties. Because the inner disc within R$_{25}$ contains the bulk of star formation in galaxies, directly measuring the properties of \hi\ gas within R$_{25}$ is important for understanding the intrinsic physics underlying the star formation-\hi\ global scaling relations.

\section{SFR-HI relations on galactic scales and within R$_{25}$}
\label{sec:glob_SFE}
In this section, we investigate on galactic scales the relation between SFR and \hi\ mass and the dependence of SFE on other parameters. The reciprocal of SFE is also referred to as the \hi\ depletion time, $t_{dep}$.  Because both SFR and SFE are higher within the optical discs (within R$_{25}$) than beyond the optical discs \citep{Bigiel10}, we also analyse SFR, $\mHI$ and SFE measured within R$_{25}$. 

\subsection{The SFR-$\mHI$ relation}
\begin{figure} 
\includegraphics[width=8.5cm]{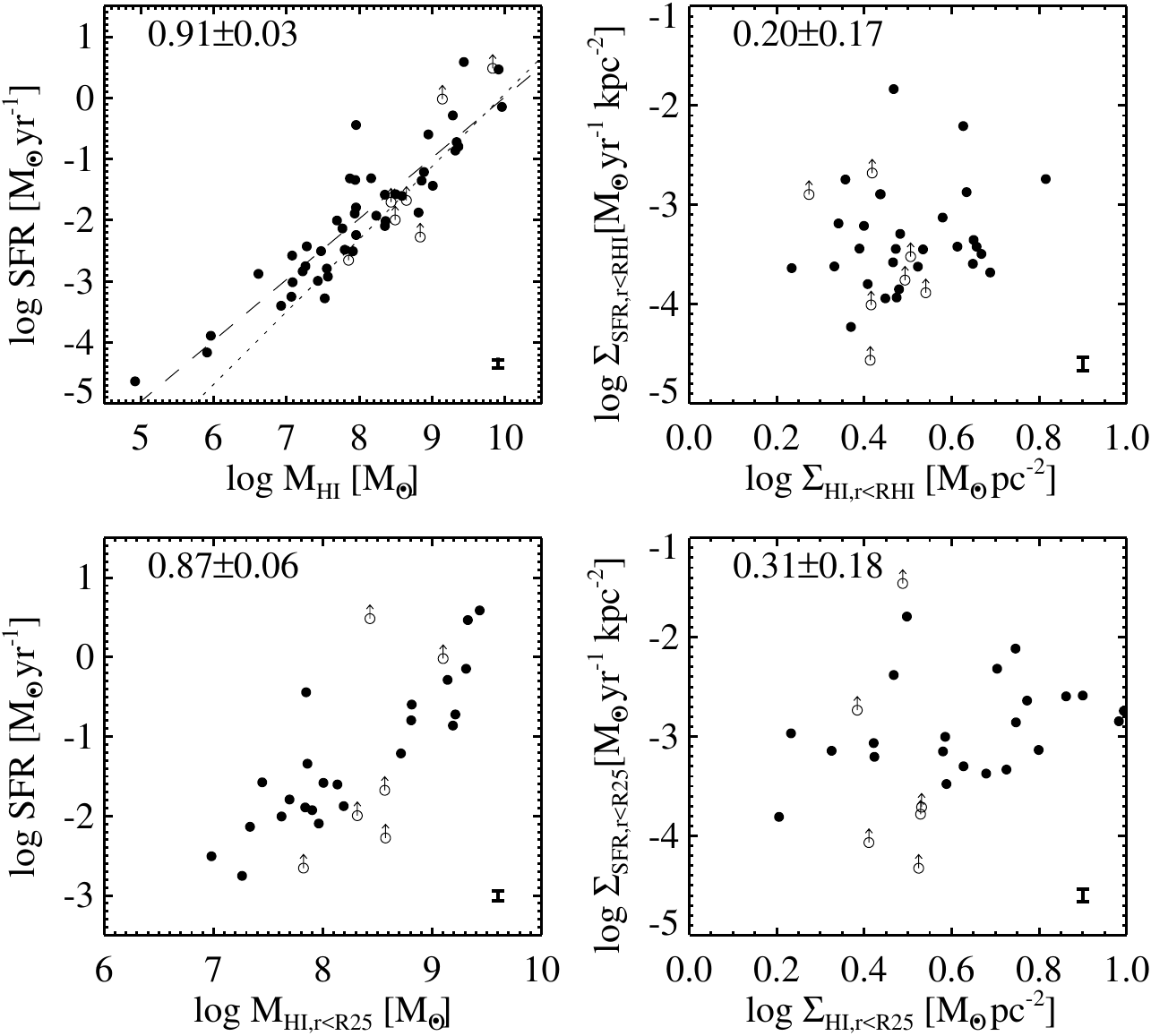}
\caption{The relations between SFR and \hi\ mass, and between their masses and surface densities within R$_{25}$. The open circles with an upward arrow mark the lower limits of SFE measurements for some galaxies; these data points are excluded when calculating the correlation coefficient $\rho$. When the \hi\ related parameter is defined with a diameter smaller than 2B$_{\rm maj}$, the \hi\ data is considered unresolved and the galaxy is excluded. 
The correlation coefficient is denoted on the top-left corner and the typical error bar is shown on the bottom-right corner of each panel. In the left-top panel, the dashed line is a liner fit to the data points (upper limits excluded), and the dotted line is the relation from \citet{Huang12b}. }
\label{fig:SFL}
\end{figure}

In the left-top panel of Figure~\ref{fig:SFL}, we show tight scaling relations between SFR and $\mHI$ and between SFR and $M_{{\rm HI,r<R25}}$ (left panels). This is consistent with the finding of ALFALFA \citep{Huang12b}. On the other hand, \citet{Bigiel08} found that $\Sigma_{\rm SFR}$ and $\SHI$ are not correlated on kpc-scales within R$_{25}$. A question is whether there is an underlying intrinsic correlation between $\Sigma_{\rm SFR}$ and $\SHI$ on galactic scales that differs from the relation on kpc-scales, because after all, \hi\ is the reservoir of material for forming molecular gas and subsequently stars.
 

We investigate the correlations between $\Sigma_{\rm SFR}$ and $\SHI$ averaged within R$_{\rm HI}$ and R$_{25}$ (right panels), and find they are weak, in agreement with findings by \citet{Bigiel08}. Hence we conclude that on galactic scales and within R$_{25}$, SFR and $\mHI$ are correlated mostly because both they both are correlated with the size of the galaxies, but not because of a physical link between star formation and \hi\ gas.

 We obtain a partial correlation coefficient of 0.84 for the relation between SFR and $\mHI$ with the effect of $M_*$ removed. So the SFR-$\mHI$ relation is not caused by a correlation of both parameters with M$_*$.

\subsection{Star Formation Efficiency}
We investigate the dependence of SFE on other galactic properties, and compare the trends measured globally and within R$_{25}$. If the global trend reflects a mechanism that directly regulates the efficiency for \hi\ to fuel star formation, then we should observe a correlation within R$_{25}$ that is stronger than or similar to the one measured globally; but if the global trend is caused by the fact that the \hi\ discs are much more extended than the stellar discs (so the gas cannot directly fuel the bulk of the star formation), then the trend should disappear when the relevant properties are measured within R$_{25}$.

\begin{figure*} 
\centering
\includegraphics[width=14cm]{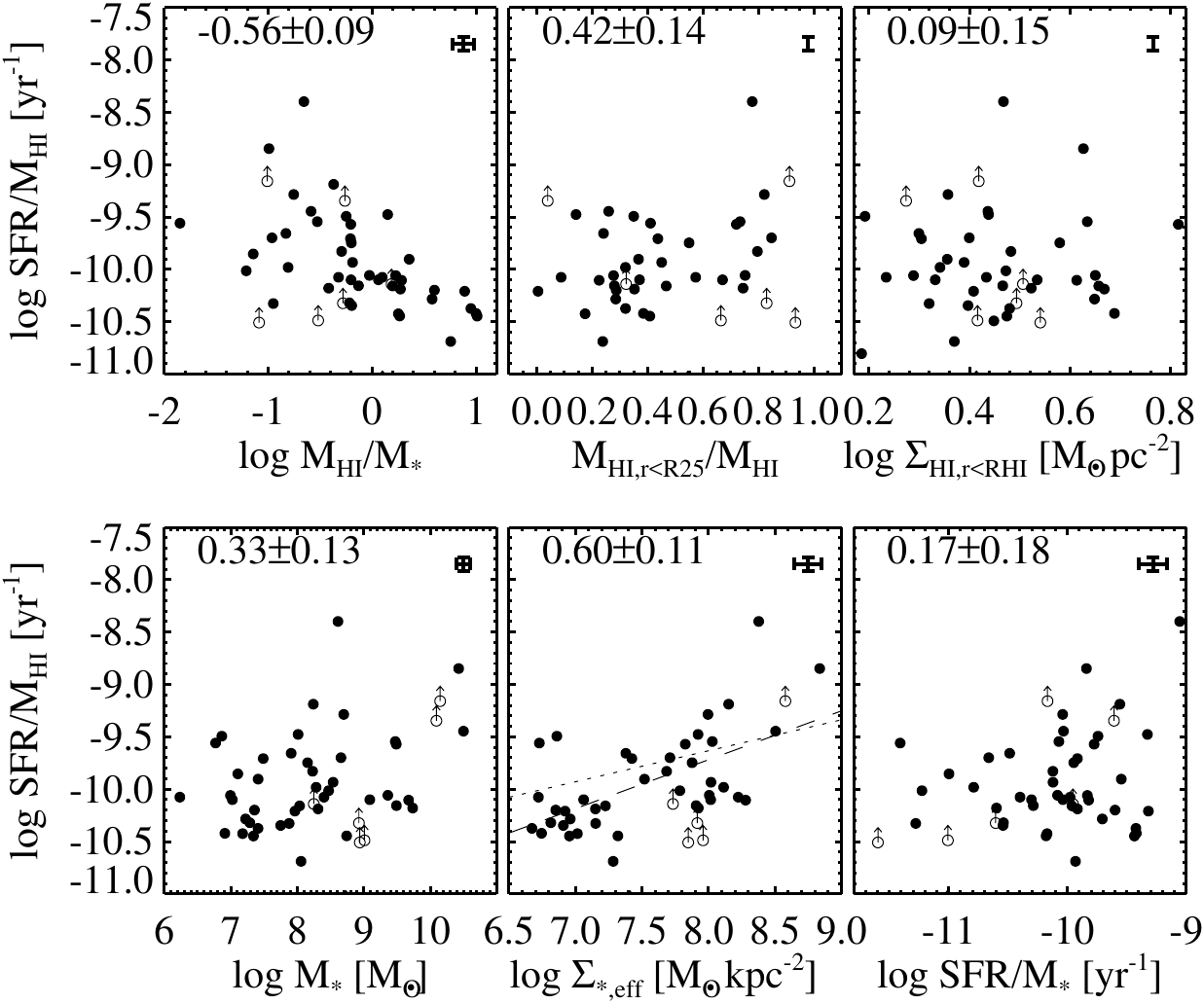}
\caption{The dependence of global SFE on other galaxy properties. The open circles with an upper-ward arrow mark the lower limits of SFE measurements for some galaxies; these data points are excluded when calculating the correlation coefficient $\rho$. The correlation coefficient  and typical error bar are shown in the corners of each panel. The  dashed line in the bottom-middle panel is the robust linear fit to the data, and the dotted line is taken from \citet{Wong16}.}
\label{fig:SFeff_global}
\end{figure*}

\begin{figure*} 
\centering
\includegraphics[width=14cm]{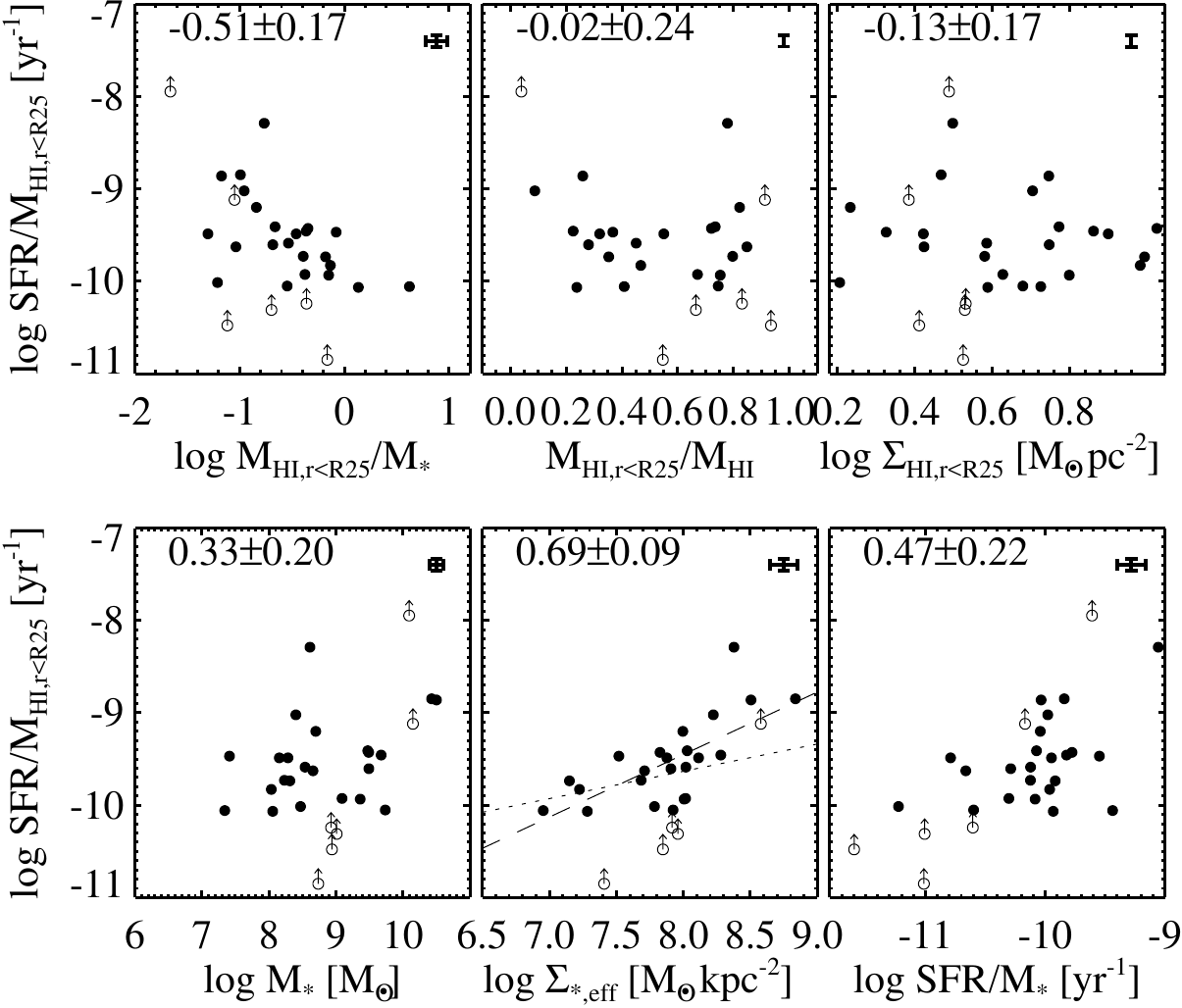}
\caption{Similar as Figure~\ref{fig:SFeff_global}, but all \hi\ masses are measured within R$_{25}$. When the \hi\ related parameter is defined with a diameter smaller than 2B$_{\rm maj}$, the \hi\ data is considered unresolved and the galaxy is excluded. }
\label{fig:SFeff_r25}
\end{figure*}

In Figure~\ref{fig:SFeff_global}, we can see that the SFE is significantly correlated with $\Sigma_{\rm *,eff}$, moderately anti-correlated with $\fHI$, weakly anti-correlated with $M_{{\rm HI,r<r25}}/M_{{\rm HI}}$, hardly correlated with $M_*$, and not correlated with $\Sigma_{\rm HI,r<RHI}$ or sSFR. In Figure~\ref{fig:SFeff_r25}, we can see that when we use $M_{{\rm HI,r<r25}}$ instead of $\mHI$ in these relations, {\rm SFE$_{\rm r<R25}$ ($=$SFR$/M_{{\rm HI,r<R25}}$) are still significantly correlated with $\Sigma_{\rm *,eff}$ and moderately (weakly) anti-correlated with ${\rm f_{HI,r<R25}=M_{HI,r<R25}/M_*}$, while the weak anti-correlation with $M_{{\rm HI,r<r25}}/M_{{\rm HI}}$ disappears. All other correlations remains weak considering the error bars.} The direct implications seem to be that $\Sigma_{\rm *,eff}$ and $\fHI$ are related to mechanisms that directly regulate star formation fuelling in the stellar disc, while the \hi\ radial distributions (indicated by $M_{{\rm HI,r<r25}}/M_{{\rm HI}}$) may only have a rather minor effect on the global SFE values. 

Because $\fHI$ and $\Sigma_{\rm *,eff}$ are strongly correlated, we calculate the partial correlation coefficient between ${\rm SFE_{r<R25}}$ and ${\rm f_{HI,r<R25}}$ with the dependence on $\Sigma_{\rm *,eff}$ removed, and the value is $-0.03\pm0.23$. It suggests that SFE depends on ${\rm f_{HI,r<R25}}$ only through their dependence on $\Sigma_{\rm *,eff}$. 

The correlation between SFE and $\Sigma_{\rm *,eff}$ is consistent with the best-fit linear relation from \citet{Wong16}. The slope of the relation was explained in \citet{Wong16} with a model where the galaxy discs are in a marginally stable status described by a constant two-fluid (gas and stars) Toomre $Q$ parameter and the mid-plane pressure determines the molecular-to-atomic ratio. In such a model, a galaxy with high stellar surface density will have low gas surface density as adjusted by the constant $Q$;  the combined effects of high stellar surface density and low gas surface density on the mid-plane pressure model result in very little change in the molecular gas mass (which SFR directly scales with); hence finally SFE as the ratio between SFR and \hi\ surface densities increases. 


To summarize, on galactic scales and within R$_{25}$, among the parameters investigated here SFE depends most strongly on the averaged stellar surface density $\Sigma_{\rm *,eff}$.  

\section{Star formation beyond R$_{25}$}
\label{sec:loc_SFE}
In this section we study the star formation on local scales in the outskirts of galaxies (beyond R$_{25}$), where both gas and stars are of low surface density and \hi\ dominates the cold gas. 

\subsection{The dependence of $\Sigma_{\rm SFR}$ on $\Sigma_*$ and $\SHI$}
\begin{figure} 
\centering
\includegraphics[width=8cm]{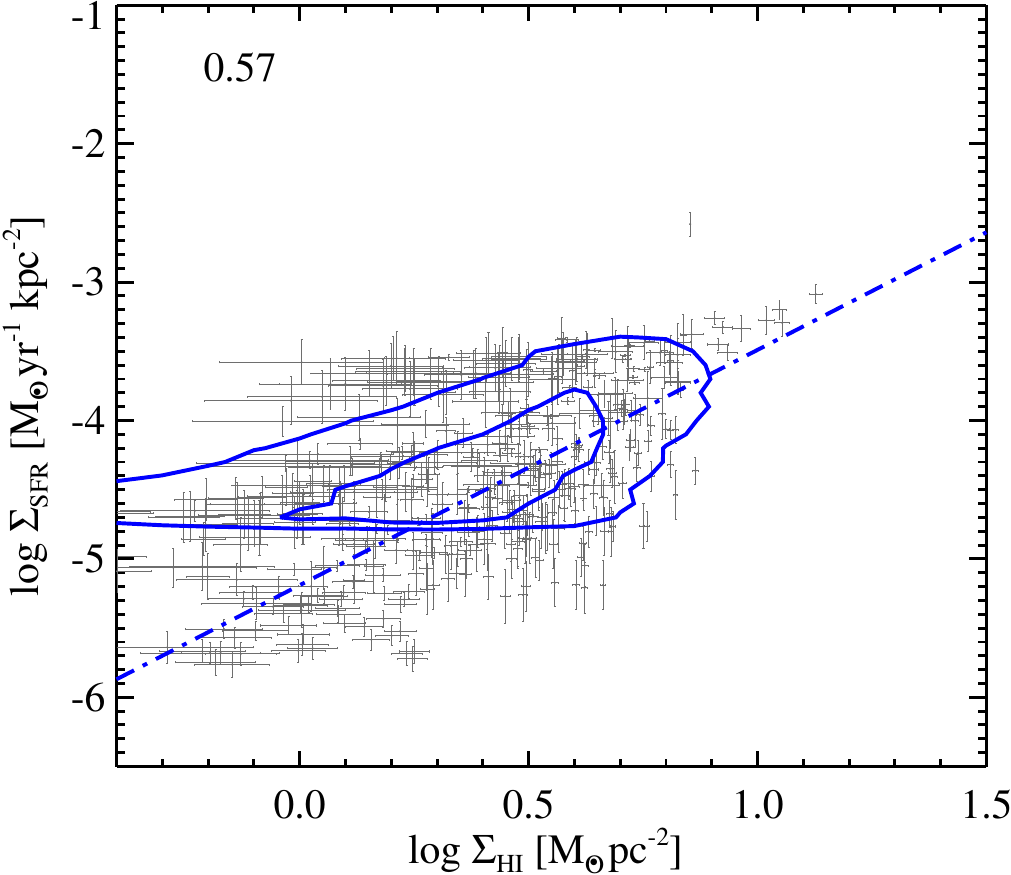}
\vspace{0.6cm}
\caption{The dependence of $\Sigma_{\rm SFR}$ on $\SHI$.   Measurements are derived from resolution-matched images. $\SHI$ and $\Sigma_{\rm SFR}$ are measured in pixels with sizes equivalent to the PSF FWHM beyond R$_{25}$. We exclude those elements in which SFR$<$0.0003 $M_*~{\rm yr^{-1}}$, to avoid the problem of insufficient IMF sampling \citep{Lee09, Lee11}.
The blue contours show the distribution of 50\% and 90\% of the data points, and the  blue dot-dashed line show the median relation from \citet{Bigiel10}. 
Correlation coefficient $\rho$ is denoted in the left-top corner. }
\label{fig:SFeff_inpix}
\end{figure}

The investigation is based on the sub-sample of ten large galaxies (Section~\ref{sec:sample}).
First of all, we show in Figure~\ref{fig:SFeff_inpix} the relation between $\SHI$ and $\Sigma_{\rm SFR}$ measured in pixels with sizes equivalent to the PSF FWHM beyond R$_{25}$. The distribution of data points and the mean relation are consistent with those from \citet{Bigiel10}.

As discussed in Section~\ref{sec:introduction}, in dwarf galaxies the radial variation of $\Sigma_{\rm SFR}$ follows the variation of $\Sigma_*$ better than that of $\SHI$ \citep{Hunter98}. We ask whether we can observe the same phenomenon for the galactic outskirts of more massive galaxies. Because $\Sigma_*$, $\SHI$ and $\Sigma_{\rm SFR}$ all generally decrease with increasing radius, we fix (control) $\SHI$ when studying the dependence of $\Sigma_{\rm SFR}$ on $\Sigma_*$, and fix $\Sigma_*$ when studying the dependence of $\Sigma_{\rm SFR}$ on $\SHI$. Limited by the data depth for $\Sigma_*$, we rely on the azimuthally averaged measurements in order to get enough signal-to-noise (S$/$N) ratio for $\Sigma_*$ measurements in these low surface brightness galactic outer regions. Radial analysis may have the risk of smoothing out$\slash$integrating correlations, but is the best we can do for now. We note that very deep optical or MIR images (to trace stellar mass) will be needed in the future to confirm our results with a pixel-based analysis in a similar way as in Figure~\ref{fig:SFeff_inpix}.

We analyse our data in two different ways, basing on resolution-matched radial profiles and profiles at the best resolution possible, respectively.

\subsubsection{Analysis based on the resolution-matched radial measurements}
We use the resolution-matched radial profiles derived in Section~\ref{sec:phot_conv}. The result is shown in Figure~\ref{fig:SFeff_conv}. 
The top-left panel shows the data points measured at 1.2R$_{25}$ confirming the  $\Sigma_{\rm SFR}$-$\SHI$ relation from pixel-based analysis (Figure~\ref{fig:SFeff_inpix} and Bigiel et al. 2010).  The data points suggest an \hi\ depletion time longer than the Hubble time, but are systematically higher in $\Sigma_{\rm SFR}$ than the median relation from \citet{Bigiel10}. This difference comes from two effects. The first effect is the trend of SFE decreasing with galactocentric radius (the \citet{Bigiel10} relation is a median relation for the radial region 1-2 R$_{25}$); the second effect is a difference between the mean and median relations noticed before in \citet{Roychowdhury15}. The best-fit relation from the data in this figure has a power law slope of 1.24$\pm$0.36. The slope is close to unity but with a large error bar, suggesting that it is hard to find any dependence of SFE on $\SHI$ in our data.

The top-right panel shows that, as in irregular galaxies, there is also a trend of  $\Sigma_{\rm SFR}$ to increase with higher $\Sigma_*$ at 1.2R$_{25}$. Which correlation might be more intrinsic? We find that the partial correlation coefficient between $\Sigma_{\rm SFR}$ and $\SHI$ with the effect of $\Sigma_*$ removed is $-0.22\pm0.50$, while the partial correlation coefficient between $\Sigma_{\rm SFR}$ and $\Sigma_*$ with the effect of $\SHI$ removed is 0.84$\pm0.27$. The partial correlation analysis suggests that the $\Sigma_{\rm SFR}$-$\Sigma_*$ correlation is more intrinsic than the $\Sigma_{\rm SFR}$-$\SHI$ correlation and the latter is caused at least partly by the former.

To confirm this result, in the bottom panels of Figure~\ref{fig:SFeff_conv}, we further investigate the $\Sigma_{\rm SFR}$ correlations firstly at a fixed $\Sigma_*$ of $3~\mspc$ and then at a fixed $\SHI$  of $3~\mspc$ \footnote{Fixing surface densities at 3~$\mspc$ rather than lower values comes from a compromise between limiting our analysis in the \hi-dominated regions and the limited depth of the $\Sigma_*$ data. If we use 2~$\mspc$ instead of 3~$\mspc$, the general trends do not change much but we will be at the risk of relying on extrapolated $\Sigma_*$ values for more than half of the sample. If we use 4~$\mspc$ instead of 3~$\mspc$, we will be at the risk of reaching the molecular dominated high gas surface density regions.}.
When $\Sigma_*=3~\mspc$, $\SHI$ ranges from 10$^{0.49}$ to 10$^{0.88}~\mspc$, while when $\SHI=3~\mspc$, $\Sigma_*$ ranges from 10$^{-0.20}$ to 10$^{1.13}~\mspc$. 
In these regions, $\Sigma_*$ values are consistent with the value typical for \hi-dominated regions, $\Sigma_* < 81~\mspc$ \citep{Leroy08}. $\SHI$ is close to the average surface density for transition between atomic and molecular gas (5$~\mspc$, with contribution from helium not included). However as shown in \citep{Leroy08} the atomic-to-molecular conversion is much more strongly correlated with $\Sigma_*$ than with $\SHI$, hence most of these measurements are obtained from \hi-dominated regions.

 We find  that there is a strong correlation of  higher $\Sigma_{\rm SFR}$ with increasing $\Sigma_*$ at a fixed $\SHI=3~\mspc$,  but very weak correlation between $\Sigma_{\rm SFR}$ and $\SHI$ at a fixed $\Sigma_*=3~\mspc$. This result implies that $\Sigma_*$ is driving changes in $\Sigma_{\rm SFR}$ more dramatically than $\SHI$ does in the galactic outskirts. These results are consistent with the finding from \citet{Hunter98}.

In the bottom panels of Figure~\ref{fig:SFeff_conv}, the dynamic range of $\Sigma_*$ is greater than that of $\SHI$ by nearly one dex for this sample. The wide dynamic range of $\Sigma_*$ as well as the strong $\Sigma_{\rm SFR}$-$\Sigma_*$ correlation have both contributed to producing the significant $\Sigma_{\rm SFR}$-$\SHI$ correlation (without fixing $\Sigma_*$) shown in the top row of Figure~\ref{fig:SFeff_conv}. 




We notice from all panels in Figure~\ref{fig:SFeff_conv} the large scatter of $\Sigma_{\rm SFR}$ in the elliptical rings where each radial profile value is measured (the crosses in the figure). Limited by the variation in image resolutions in the sample and the small sample size, we can not evaluate how the scatter varies as a function of $\SHI$ or $\Sigma_*$, but it can be inferred that the scatter in the $\Sigma_{\rm SFR}$-$\Sigma_*$ relation might be as large as that in the $\Sigma_{\rm SFR}$-$\SHI$ relation. The large scatter indicates complex physical factors other than the gravity provided by stars and \hi\ gas that regulate star formation.

\subsubsection{Analysis based on the radial measurements at the best possible resolution}

\begin{figure} 
\centering
\includegraphics[width=8cm]{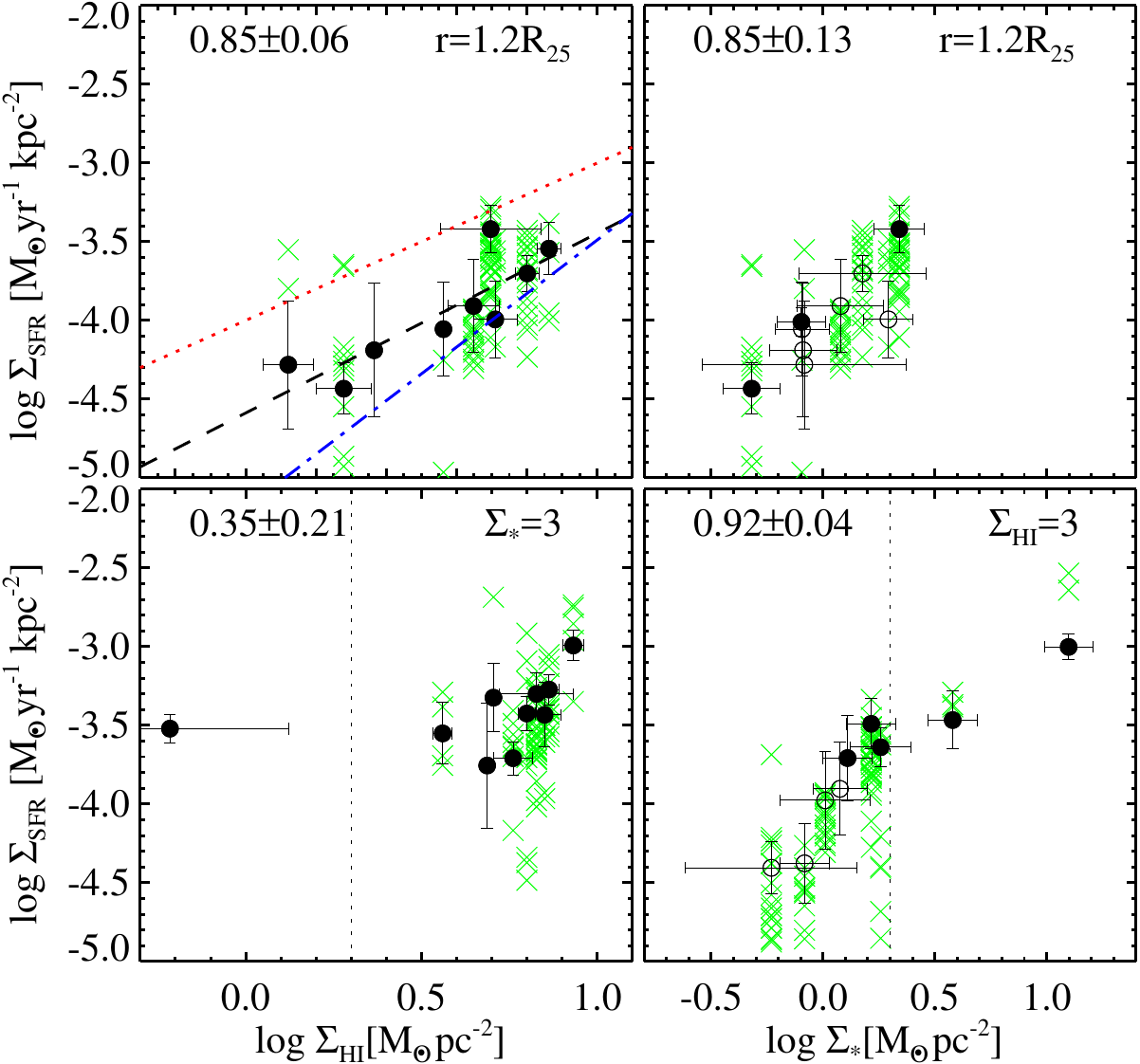}
\caption{The dependence of $\Sigma_{\rm SFR}$on $\SHI$ and $\Sigma_*$.  Parameters are derived from resolution matched images. Only measurements with an error less than 0.5 dex are displayed and considered in the analysis. Solid and open circles are measurements derived from radial profiles.  Solid circles mark direct measurements and open circles mark the $\Sigma_*$ measurements obtained through extrapolating radial profiles. Vertically from each circle, the green crosses show $\Sigma_{\rm SFR}$ measured in pixels with a size equivalent to the PSF FWHM  (see Figure~\ref{fig:SFeff_inpix}) along the elliptical ring where the radial profile value is measured. Due to the very different galactic sizes and image resolutions within the small sample, these crosses are only illustrative of the scatters in $\Sigma_{\rm SFR}$ along each elliptical ring and cannot be compared between different galaxies.   In the top panels, data points are measured at 1.2R$_{25}$.  In the bottom-left panel, data points are measured on the elliptical ring where the averaged $\Sigma_*=3~\mspc$; in the bottom-right panel, data points are measured along the elliptical ring where the average $\SHI=3~\mspc$. These characteristic positions are denoted in the top-right corner of each panel. The black dashed line in the top-left panel is a linear fit to the data points, the blue dash-dotted line is the median relation from \citet{Bigiel10}, and the red dotted line marks an SFE=10$^{10}~{\rm yr}^{-1}$ or a gas depletion time of 10 Gyr. In the bottom panels, the black dotted lines mark the density of 3 $\mspc$.  Correlation coefficient $\rho$ is denoted in the left-top corner of each panel. }
\label{fig:SFeff_conv}
\end{figure}

\begin{figure*} 
\centering
\includegraphics[width=14cm]{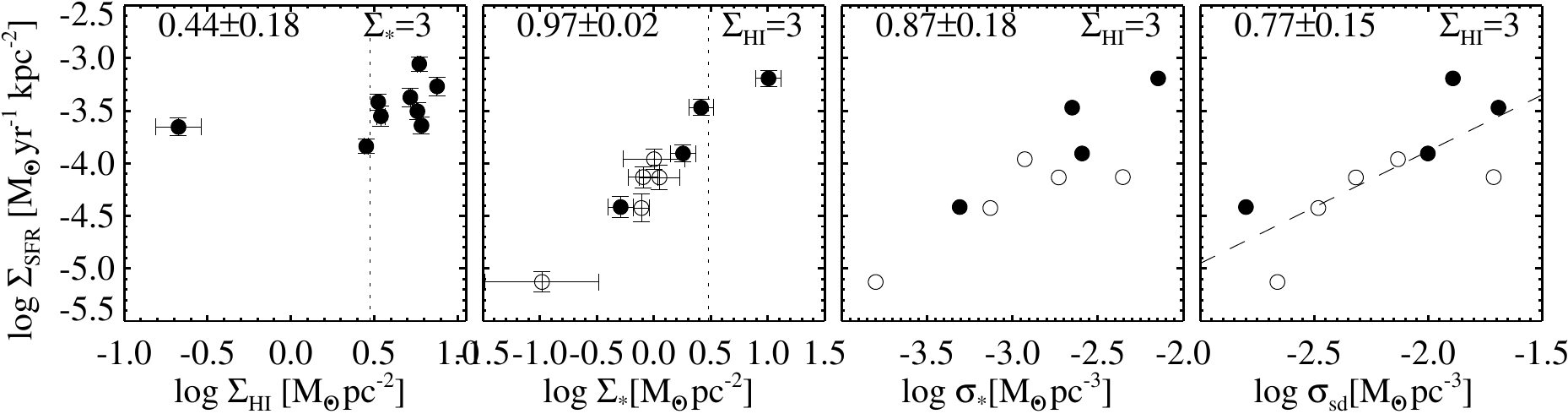}
\vspace{0.6cm}

\includegraphics[width=14cm]{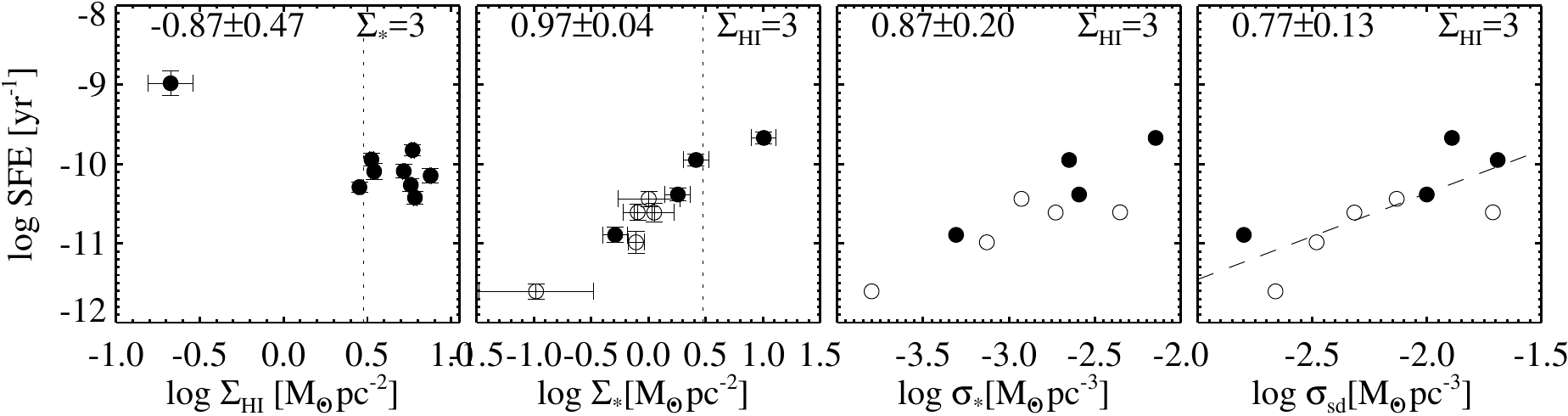}
\vspace{0.2cm}
\caption{The dependence of $\Sigma_{\rm SFR}$ (top row) and SFE (bottom row) on $\SHI$, $\Sigma_*$, $\sigma_*$ and $\sigma_{\rm sd}$.  Parameters are derived from GALEX and WISE images matched to the W4 resolution and tilted-ring fitting models. The dashed line in the top-right panel is a linear fit to the data points. Symbols, lines and denotes have the same meaning as in Figure~\ref{fig:SFeff_conv}. }
\label{fig:SFeff_highrs}
\end{figure*}

For the second way of analysing the data, we use the radial profiles derived in Section~\ref{sec:best_res}. In addition to repeating the analysis in the bottom row of Figure~\ref{fig:SFeff_conv}, we further look into the dependence of $\Sigma_{\rm SFR}$ on $\Sigma_*$ and $\sigma_{\rm sd}$. $\sigma_{\rm sd}$ has been predicted by star formation models \citep{Ostriker10, Krumholz13} to be a key parameter in setting the SFR in the outskirt of galaxies (see more details in Section~\ref{sec:SFE_model}). These results are presented in the top row of Figure~\ref{fig:SFeff_highrs}. 
We warn that we do not have a proper error estimate for $\Sigma_*$ measured from the tilted ring models, and the estimate of $\sigma_{\rm sd}$ is based on simple assumptions.

From the left two panels in the top row of Figure~\ref{fig:SFeff_highrs}, the trends are largely consistent with what we have found in the bottom panels of Figure~\ref{fig:SFeff_conv}. There is a moderate correlation between $\Sigma_{\rm SFR}$ and $\SHI$ here, but it is significantly weaker than the $\Sigma_{\rm SFR}$-$\Sigma_*$ correlation. In the last two panels of the top row, we also find significant correlations of $\Sigma_{\rm SFR}$ increasing as a function of $\sigma_*$ and $\sigma_{\rm sd}$. The slope for the $\Sigma_{\rm SFR}$-$\sigma_{\rm sd}$ relation is 1.07$\pm$0.41. 

In the bottom row of Figure~\ref{fig:SFeff_highrs}, we show the trends of SFE as function of $\SHI$, $\Sigma_*$, $\sigma_*$ and $\sigma_{\rm sd}$ (defined at Section~\ref{sec:best_res}).  SFE and $\SHI$ are not correlated. The other trends are similar as the trends of $\Sigma_{\rm SFR}$ depending on those parameters.

We note that at $\SHI=3~\mspc$, $\sigma_{\rm sd}$ is on average 3.5 times higher than $\sigma_*$ in our sample, but the two volume densities correlate with each other with $\rho=0.64\pm0.20$. $\Sigma_*$ is correlated with $\sigma_*$ ($\rho=0.85\pm0.28$) because of the way it is calculated, but is not correlated with $\sigma_{\rm sd}$ ($\rho=0.35\pm0.25$).
The partial correlation between SFE and $\sigma_*$ with the effect of $\sigma_{\rm sd}$ removed is 0.55$\pm0.45$, while the partial correlation between SFE and $\sigma_{\rm sd}$ with the effect of $\sigma_*$ removed is 0.39$\pm0.48$. 
Hence the weak statistics does not allow us to conclude whether $\sigma_{\rm sd}$ or $\sigma_*$ ($\Sigma_*$) has independent effects on setting SFE in the galactic outskirts. We will discuss it in more details in Section~\ref{sec:model_SFE}.

\subsection{Comparison with existing observations}
\label{sec:SFE_model}
As we have mentioned, \citet{Bigiel10} showed the $\Sigma_{\rm SFR}$-$\SHI$ relation for the outskirts of spiral discs. $\Sigma_{\rm SFR}$ and $\SHI$ were measured in kpc-scale pixels, and the median relation has a power law slope of 1.7.
Measuring the mean $\Sigma_{\rm SFR}$ in each $\SHI$ bin instead,  \citet{Roychowdhury15} found a uniform power law slope between 1.4 and 1.7 for the relations between log $\SHI$ and log $\Sigma_{\rm SFR}$ for irregular galaxies and the outskirts of spiral and massive disc galaxies. The non-unity slope suggests a correlation between log $\SHI$ and log SFE in the \hi-dominated regime.  On the other hand, \citet{Yildiz16} only found a close-to-unity power law slope for the mean relation in the outskirts of early-type galaxies. When we derive the $\Sigma_{\rm SFR}$-$\SHI$ relation at 1.25 R$_{25}$, we average over both star-forming and non-star-forming regions along the elliptical ring, but the power law slope of 1.24$\pm$0.36 is consistent with the literature values within the error bar. 

A large scatter in the $\Sigma_{\rm SFR}$-$\SHI$ relation has been shown in all these studies. In \citet{Bigiel10}, the distribution of $\Sigma_{\rm SFR}$ has a large scatter at a fixed $\SHI$ ($\sim$1.5 dex at $\SHI=1\mspc$). There is also large offset between the median and mean values of  $\Sigma_{\rm SFR}$ at a fixed $\SHI$ ($\sim$0.5 dex at $\SHI=1\mspc$) as revealed in \citet{Roychowdhury15}. Based on the FIGGS sample of irregular galaxies,  \citet{Roychowdhury09} found that each galaxy has its own slope and intercept for $\Sigma_{\rm SFR}$-$\SHI$ relation when such a correlation exists, while for 20\% of the irregular galaxies (most of them disturbed) in their sample, $\Sigma_{\rm SFR}$ does not correlate with $\SHI$.  The large scatter indicates complex physics regulating the relationship between SFR and \hi\ gas.
 
Despite the uncertainty in slopes and the large scatter, the $\Sigma_{\rm SFR}$-$\SHI$ relations observed from these different studies \citep[also see][]{Yim16} have one common feature: the inferred SFE is low and the gas depletion time is longer than the Hubble time. 
 
As mentioned briefly in Section~\ref{sec:introduction}, the tight $\Sigma_{\rm SFR}$-$\Sigma_*$ relation was found for the \hi-dominated irregular galaxies by \citet{Hunter98}. They tested a variety of physical quantities as possible regulators for star formation in irregular galaxies, including \hi\ gas critical densities for gravity instabilities in different disc models.  They  found  that only $\Sigma_*$ correlates well with $\Sigma_{\rm SFR}$ along radial profiles in their investigation. In this paper, we find a strong $\Sigma_{\rm SFR}$-$\Sigma_*$  correlation also exist in the outskirts of spiral galaxies.

\subsection{Comparison with theoretical models}
\label{sec:model_SFE}
In this section we try to understand our results for the $\Sigma_{\rm SFR}$-$\SHI$  and $\Sigma_{\rm SFR}$-$\Sigma_*$ relations in the context of modern star formation models. 

Simple modern star formation models consider that 1) the atomic inter-stellar medium consists of a warm and a cold phase, and only the cold mode gas phase forms molecular clouds and stars; and that the 2) SFR in the HI-dominated regime is regulated by the hydrostatic equilibrium between thermal and dynamic pressure on gas at the mid-plane (Ostriker et al. 2010, O10 hereafter, Krumholz et al. 2013, K13 hereafter). The exact recipe for determining the pressure equilibrium between gas phases differs in different models. For example, in the O10 model, the atomic gas is in two-phase equilibrium everywhere and the thermal pressure is set by the radiation field and thus, it is a consequence of star formation feedback, while in the K13 model the cold phase density, sensitively depends on an assumed maximum temperature, above which the cold phase cannot exist. Other parameters of the model, such as the gas surface density and the SFR, are regulated under this condition.  

In these models, the relation between the SFR and \hi\ surface density are of complex nature. This is because, at the densities where \hi\ dominates, other factors besides gravity contribute strongly to the equilibrium condition of the gas, such as temperature, sound speed, magnetic fields, cosmic rays and turbulence. K13 predicted a large scatter in the relation between $\Sigma_{\rm SFR}$ and $\Sigma_{\rm gas}$ (see their Figure 3 and Appendix A1). O10 predicted a complex dependence (their equation 22) of $\Sigma_{\rm SFR}$ as a function of $\SHI$, which contains several terms contributing equally. The large scatter and the complex functional forms make it hard to constrain the models, because measuring the $\SHI$-$\Sigma_{\rm SFR}$ relation from a small data set like LVHIS comes with observational uncertainties, and allow for significant ambiguity in the models. 

Because old stars and dark matter contribute to the mid-plane pressure, both O10 and K13 models predict a SFR that increases with increasing $\sigma_{\rm sd}$ at a fixed gas density, which is what we observe in our data. Especially, the O10 model predicts $\Sigma_{\rm SFR}\propto \sqrt{\sigma_{sd}}$ for the low gas surface density region \citep[also see][]{Kim11}. Our data reveal a power law index of 1.07$\pm$0.41, which does not necessarily disprove the O10 model, considering our crude calculation of the DM densities and the simplifications that go into the modelling. The $\Sigma_{\rm SFR}$-$\sigma_{\rm sd}$ relation also manifests itself as the $\Sigma_{\rm SFR}$-$\Sigma_*$ relation. Although the scatter in the $\Sigma_{\rm SFR}$-$\Sigma_*$ relation might be as large as that of the $\Sigma_{\rm SFR}$-$\SHI$ relation given the same complex physical processes involved, the much wider dynamic range of $\Sigma_*$ compared to $\SHI$ has helped us to reveal a trend between $\Sigma_*$ and $\Sigma_{\rm SFR}$.


The $\Sigma_{\rm SFR}$-$\Sigma_*$ relation might also partly reflect the SFE-metallicity relation (K13). Because metals are expected to provide the necessary shielding for star-forming clouds \citep[][and K13]{Sternberg14}, and there is a strong correlation between  $\Sigma_*$ and local metallicity \citep{Moran12,Sanchez13,LaraLopez13,Carton15,Sanchez17}. \citet{Roychowdhury15} found that the SFE in the outskirts of irregular and massive spiral galaxies are the same. Because irregular galaxies on average have much lower metallicity than the massive spiral galaxies, they concluded that the SFE in the galactic outskirts seem to be independent of metallicity. However, metallicities in the galactic outskirts might differ significantly from the global metallicities \citep{Moran12,Sanchez12, LopezSanchez15} which is what Roychowdhury et al. (2015) used for their analysis. Optical spectroscopy for the galactic outskirts is needed before we could make relevant conclusions, and we may need rely on the next generation IFU instruments like HECTOR \citep{BlandHawthorn15}. 

In the past it has been suggested that star formation can have a positive feedback effect, effectively triggering more star formation \citep{Gerola80}. In this picture, winds and the supernovae ejecta can cause gas compression at the shocks front, increasing the density of the gas and triggering new star formation. This is referred to as ``self-propagation of star formation", which seemingly could explain the $\SHI$-$\Sigma_*$ correlation.
However, modern simulations \citet{Dale11, Dale15} show that, although this form of positive feedback sometimes locally increases the gas density, enhancing star formation, the overall effect is negative. Ionising radiation effectively disperses low-to-intermediate mass molecular clouds while outflows disperse the more massive molecular clouds. In observations, there are examples of positive feedback at play and also a few of self-propagating (or subsequently triggered) star formation, such as the W5 HII region in the MW and the stellar cluster NGC 602 in the SMC \citep{Koenig08, Carlson11}. The conditions for self-propagating star formation to take place and whether this physical phenomenon could play an important role at low gas surface density regions \citep[as predicted by][]{Gerola80} remains unclear.

To summarise, the strong $\Sigma_{\rm SFR}$-$\Sigma_*$  correlation at a fixed $\SHI$ and the relatively weak $\Sigma_{\rm SFR}$-$\SHI$ correlation at a fixed $\Sigma_*$ in the galactic outskirts are broadly consistent with modern star formation models. More detailed comparisons rely on more data points from larger samples in the future.

\section{ Star Formation Efficiency near HI warps}
\label{sec:warp_SFR}
We have shown in Section~4.2 that seven out of the ten LVHIS large galaxies display warped \hi\ discs.  Theoretical studies like that of \citet{SanchezBlazquez09} have predicted a drop in SFE caused by the onset of an \hi\ warp. Because SFE generally decreases with radius because of decreasing gas and stellar surface densities \citep{Bigiel10}, we expect the additional effect from a warp would steepen the decrease. In this section, we investigate whether such a trend can be observed with our data. 

\begin{figure*} 
\centering
\includegraphics[width=16cm]{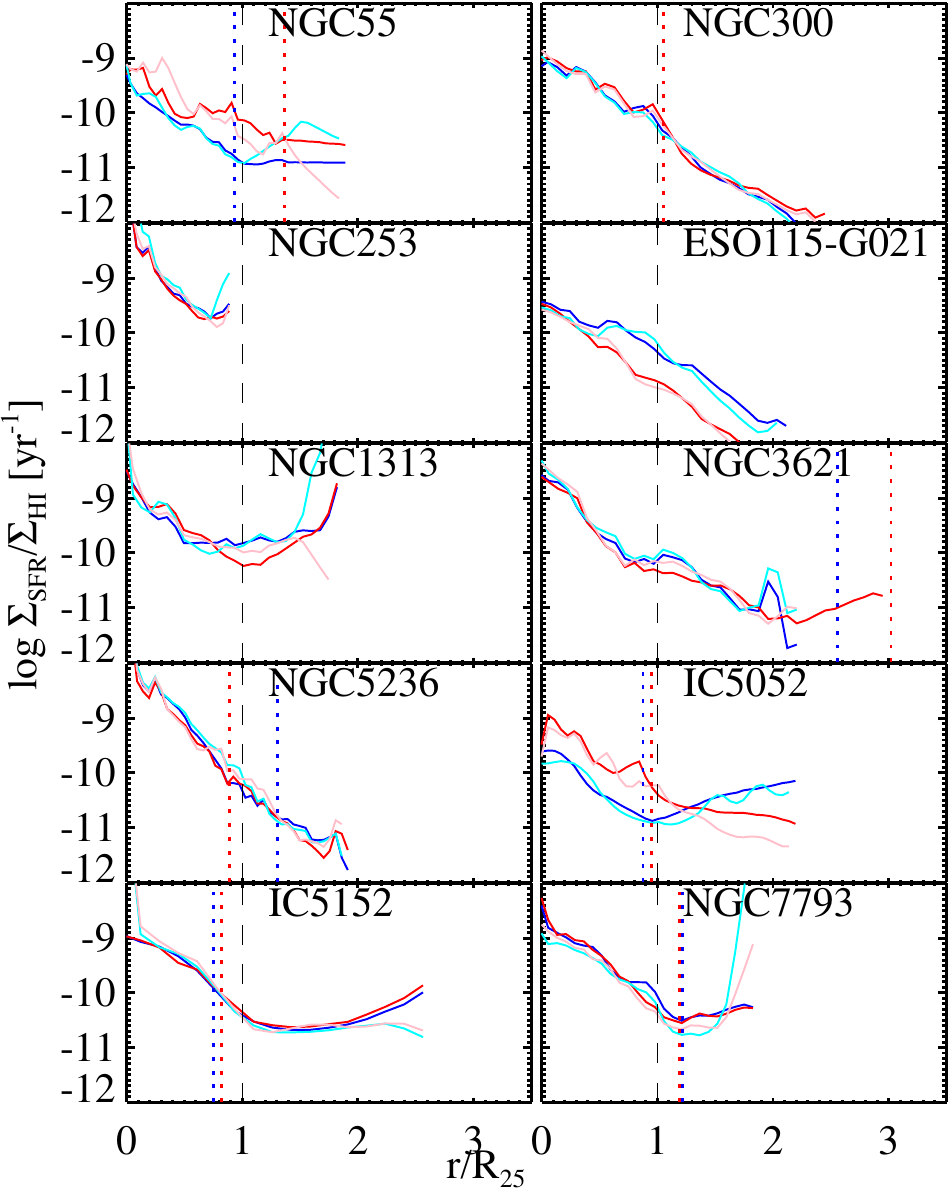}

\caption{The SFE radial profiles for the ten well-resolved LVHIS galaxies. The blue and red solid curves show SFE calculated for the approaching and receding sides of the galaxy respectively, with $\SHI$ derived from \hi\ column density images. The cyan and pink solid curves show SFE calculated for the approaching and receding velocity sides of the galaxy respectively, with $\SHI$ taken from the tilted ring models. 
The blue and red dotted lines mark $r_{warp}$ on the approaching and receding velocity sides respectively, when an \hi\ warp exists (Section~\ref{sec:tiltedring}). The radius is normalised by R$_{25}$. The black dashed lines mark R$_{25}$. }
\label{fig:SFEprofile}
\end{figure*}

We mark the warp radius $r_{warp}$ on the SFE radial profiles in Figure~\ref{fig:SFEprofile}. 
When calculating the SFE radial profiles, we derive $\Sigma_{\rm SFR}$ profiles at the resolution of the W4 images as in Section~\ref{sec:best_res}, but separate the approaching and receding velocity sides of a galaxy.  We use $\SHI$ profiles both directly derived from the mom-0 images and taken from the tilted-ring models. The former are less noisy and the latter are better resolved, but they resulted in close SFE profiles as shown in Figure~\ref{fig:SFEprofile} because $\SHI$ vary in a much narrower dynamic range than $\Sigma_{\rm SFR}$.

 We find the relations between \hi\ warps, R$_{25}$ and SFE radial profile shapes show great diversity within the very small sample. We summarise the most important results shown in Figure~\ref{fig:SFEprofile}:
\begin{enumerate}
\item $r_{warp}$ is located close to R$_{25}$ (with an offset $<$0.3 R$_{25}$) in most of the warped galaxies (6 out of 7). NGC 3621 has a warp that onsets extremely far away from the galactic centre: beyond 2.5 R$_{25}$, and beyond the detection limit of our $\Sigma_{\rm SFR}$ data. In four out of the seven warped galaxies, $r_{warp}$ on the two velocity sides differ by more than 0.4 R$_{25}$ (NGC 55, NGC 3621 and NGC5236).

\item With the limited statistics available, there is no significant difference in the level of SFE beyond R$_{25}$ between warped and un-warped galaxies.

\item Among the six \hi\ warped galaxies where SFE is measurable at $r_{warp}$, the only case for $r_{warp}$ to coincide with a steepened drop in SFE is in NGC 300. On the contrary, in NGC 55 and NGC 7793, $r_{warp}$ coincides with the onset of a shallower SFE profile or even the turning over to a rising SFE profile. In other three warped galaxies, NGC 5236, IC 5052 and IC 5152, there are no clear signs that $r_{warp}$ immediately coincides with a change in the slope of the SFE profile.

\item  R$_{25}$ shows a similar complex relation with SFE profile shapes as shown by $r_{warp}$. 
\end{enumerate}


In order to better quantify the relation between $r_{warp}$ and change in SFE profile slope, we derive the second order derivative of SFE profiles near $r_{warp}$. For a given radius $r$, we first calculate the first order derivatives of SFE for the two radial segments with distances $-0.25$ to 0 R$_{25}$ and 0 to 0.25 R$_{25}$ from r, ${\rm SFE_-'(r)}$ and ${\rm SFE_+'(r)}$ respectively. And the second order derivative of SFE at r is calculated as ${\rm SFE''(r)=SFE_+'(r)}-{\rm SFE_-'(r)}$. For comparison, we also calculate SFE$''$ at R$_{25}$ and at 100 random radii beyond R$_{25}$ for each galaxy. All these calculations use SFE profiles measured at both the approaching and the receding velocity sides of galaxies. We use the SFE profiles with $\SHI$ derived from \hi\ column density images (the red and blue curves in Figure~\ref{fig:SFEprofile}) in this analysis, but we have confirmed that, if we use the SFE profiles with $\SHI$ derived from tilted ring models instead (the pink and cyan curves in Figure~\ref{fig:SFEprofile}), the conclusion from the relevant results does not change.

\begin{figure} 
\centering
\includegraphics[width=8.5cm]{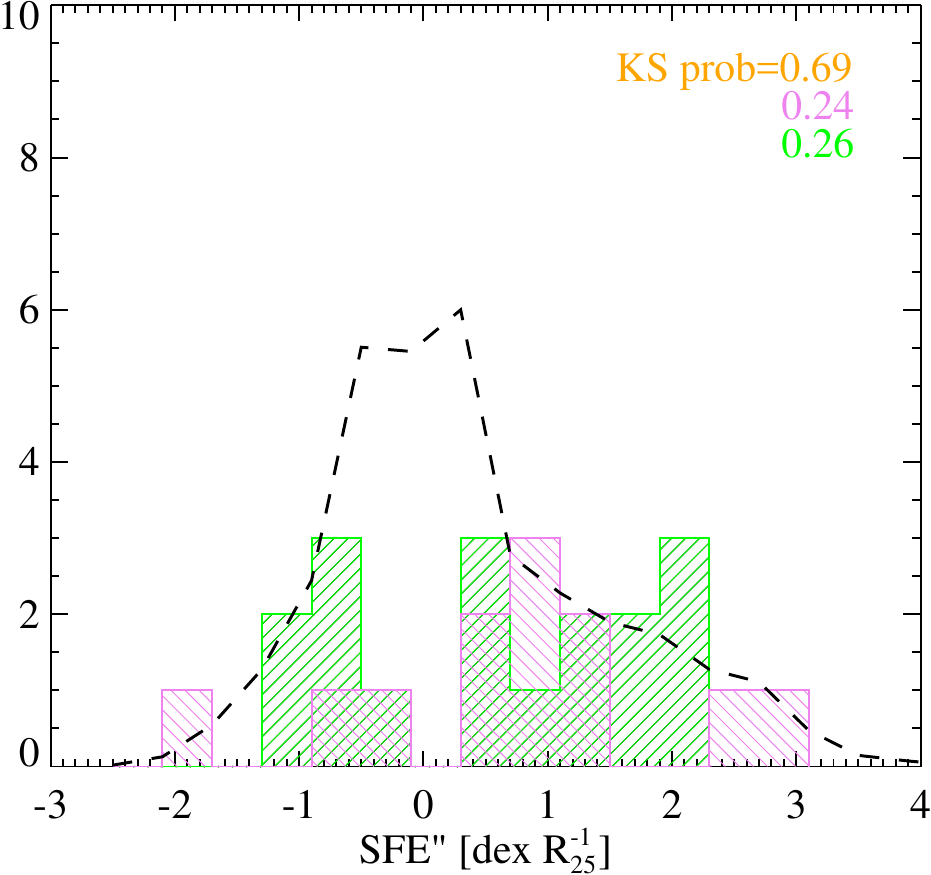}
\caption{The distributions of second order derivatives of SFE measured at $r_{warp}$ (violet), R$_{25}$ (green) and random radii ($r_{random}$) beyond R$_{25}$ (black). The black histogram has been normalised to have comparable height as the violet and green histograms. The top right corner shows the Kolmogorov-Smirnov (KS) test probabilities for the null hypothesis that the two distributions are drawn from the same parental distribution. The numbers in orange, violet and green show the KS test probabilities for the comparisons SFE$''(r_{warp})$ vs SFE$''{\rm (R_{25})}$,  SFE$''(r_{warp})$ vs SFE$''(r_{random})$ and SFE$''{\rm (R_{25})}$ vs SFE$''(r_{random})$ respectively.} 
\label{fig:SFdev_distr}
\end{figure}

Figure~\ref{fig:SFdev_distr} shows that SFE$''$ measured at $r_{warp}$ shows no significant difference with that measured at R$_{25}$ or random radii beyond R$_{25}$. It confirms what we have visually summarised from Figure~\ref{fig:SFEprofile}.

Hence what we learn from the limited statistics here is that we do not find strong evidence for \hi\ warps to be {\it always} linked to suppressed SF activity. A larger sample from future \hi\ surveys \citep[WALLABY$\slash$Apertif,][]{Koribalski12} is required to confirm this result or identify the possibly weak effect of warps.


\section{Summary and conclusion}
\label{sec:summary}
We have analysed  GALEX, WISE and \hi\ images and characterised the \hi\, $M_*$ and SFR properties for galaxies in the LVHIS sample. We use the resolved \hi\ measurements to understand previously found global relations$\slash$trends (based on the LVHIS galaxies with R$_{25}>2 {\rm B_{maj}}$) and investigate star formation properties in the galactic outskirts (beyond R$_{25}$, based on the ten large galaxies). We try to obtain a coherent picture of how SFR relates to the \hi\ gas on different scales and in different regions. The major findings are:

\begin{enumerate}
\item  The \hi\ radial distribution with respect to R$_{25}$ is correlated with $\fHI$ of galaxies (Figure~\ref{fig:HI_distr}). However the large scatter in the relations highlights the importance of using spatially resolved \hi\ data to understand global relations. 

\item The tight SFR-$\mHI$ relation is caused by each being the product of galaxy area multiplying with $\Sigma_{\rm SFR}$ and $\SHI$, respectively, and does not reflect an intrinsic link between star formation and \hi\ gas. The correlation between the globally averaged  $\Sigma_{\rm SFR}$ and $\SHI$ is weak (Figure~\ref{fig:SFL}). SFE significantly depends on the average stellar surface density $\Sigma_{\rm *,eff}$, which was explained in \citet{Wong16} with a marginally stable disc model. 

\item The $\Sigma_{\rm SFR}$-$\SHI$ correlation in the galactic outskirts becomes significantly weaker when $\Sigma_*$ is fixed, while on the other hand there is a strong  $\Sigma_{\rm SFR}$-$\Sigma_*$ correlation at a fixed $\SHI$ (Figure~\ref{fig:SFeff_conv} and \ref{fig:SFeff_highrs}). These trends are consistent with star formation models where young stars are directly formed from the molecular gas, and are indirectly linked with \hi\ gas through complex physical processes. In these models, old stars have a significant influence on the star formation process by providing gravity or through their link with metallicity.

\item  \hi\ warps hardly affect the SFE in the galactic outskirts (Figure~\ref{fig:SFdev_distr}). 
\end{enumerate}

These results suggest that the direct link between SFR and \hi\ gas is generally weak in star forming galaxies. The SFR-\hi\ relation measured globally and in the galactic outskirts are caused at least partly by side effects (both correlating to a third parameter, galaxy size for the former and $\Sigma_*$ for the latter) and are intrinsically weaker than they appear to be. The gap between them and the (lack of) SFR-\hi\ relations measured within galactic inner regions on kpc-scales \citep{Bigiel08} thus become much smaller.  On the other hand, old stars may play a significant role in regulating star formation in both inner and outer regions of galaxies. 

One major limitation of the analysis in this paper is the relatively small number of statistics, especially for the sub-sample of large galaxies. The relevant results will be strengthened by larger statistics, deeper images and follow-up optical spectroscopy. We look forward to high quality data and larger samples from future multi-wavelength surveys and instruments, such as WALLABY \citep{Koribalski12}, Skymapper  \citep{Keller07} and HECTOR \citep{BlandHawthorn15}.

\section*{Acknowledgements}
We thank the anonymous referee for constructive comments. 
We gratefully thank G. Muerer, M. Krumholz and B. Wang for useful discussions.

The \hi\  21-cm observations were obtained with the Australia Telescope Compact Array which is funded by the Commonwealth of Australia for operations as a National Facility managed by CSIRO.

GALEX (Galaxy Evolution Explorer) is a NASA Small Explorer, launched in April 2003, developed in cooperation with the Centre National d'\'{E}tudes Spatiales of France and the Korean Ministry of Science and Technology.

This publication makes use of data products from the Wide-field Infrared Survey Explorer, which is a joint project of the University of California, Los Angeles, and the Jet Propulsion Laboratory/California Institute of Technology, funded by the National Aeronautics and Space Administration.

This research has made use of the SIMBAD database\footnote[1]{http://simbad.u-strasbg.fr/simbad/}, operated at CDS, Strasbourg, France.

ZYL is supported by the National Natural Science Foundation of China under grant no. 11403072, by the China-Chile joint grant from CASSACA, and by Shanghai Yangfan Research Grant (no. 14YF1407700).

The work of LCH was supported by National Key Program for Science and Technology Research and Development grant 2016YFA0400702.

JMvdH acknowledges support from the European Research Council under the European Union's Seventh Framework Programme (FP/2007-2013)/ERC Grant Agreement no. 291531.

This project has received funding from the European Research Council (ERC) under the European Union's Horizon 2020 research and innovation programme (grant agreement no. 679627).

\bibliographystyle{mnras}
\bibliography{LVHISmw}

\appendix
\section{Measurements tables}
GALEX and WISE photometric measurements for LVHIS galaxies are presented in Tables~\ref{tab:uvphot} and \ref{tab:irphot} respectively. 

Table~\ref{tab:uvphot} gives the fitted FUV and NUV scale-length of the outer disc (r$_{\rm s}$), the apparent magnitude corrected for foreground extinction (m), the effective radius that enclose half of the light (r$_{\rm eff}$) and the effective surface brightness (the average brightness within r$_{\rm eff}$, $\mu_{\rm eff}$).  See Section~\ref{sec:uvphot}  for details.

Table~\ref{tab:irphot} the four WISE band luminosities (${\rm \log L_{W1-4}}$), the W1-W2 colour and the W1 structural parameters. The superscript $f$ marks the galaxies not detected in the W2-band, for which we have enforced the W1 aperture on the W2 images in order to measure the W2 fluxes. The expected (model) stellar continuum are removed from the luminosities ${\rm \log L_{W3}}$ and ${\rm \log L_{W4}}$. The W1 in-band luminosity (${\rm L_{W1,\odot}}$) is normalised by the equivalent solar value and used for calculating stellar mass-to-light ratio. R$_{75}$ and R$_{25}$ are the 75\% and 25\% light radius and the ratio between the two is a measure of the central concentration. See Section~\ref{sec:irphot} for details.

\begin{table*}\small
\caption{GALEX photometric measurements for the LVHIS galaxies.}
\centering
\begin{tabular}{l|llll|llll}
\hline
\multicolumn{1}{c|}{IDs} & \multicolumn{4}{c}{FUV}&  \multicolumn{4}{|c}{NUV} \\
\hline
index&r$_{\rm s}$&m&r$_{\rm eff}$&$\mu_{\rm eff}$&r$_{\rm s}$&m&r$_{\rm eff}$&$\mu_{\rm eff}$\\
&arcsec&mag&arcsec&mag arcsec$^{-2}$&arcsec&mag&arcsec&mag arcsec$^{-2}$\\
(1)&(2)&(3)&(4)&(5)&(6)&(7)&(8)&(9)\\
\hline
1&8.1$\pm$0.4&17.16$\pm$0.06&9.1&27.55&8.6$\pm$0.5&16.97$\pm$0.03&10.1&24.03\\
2&33.7$\pm$7.1&17.67$\pm$0.58&8.0&28.43&20.0$\pm$1.0&17.58$\pm$0.03&6.4&23.68\\
3&8.5$\pm$1.1&18.33$\pm$0.11&13.5&28.67&13.6$\pm$0.4&17.26$\pm$0.04&8.7&24.01\\
4&128.6$\pm$0.1&10.31$\pm$0.05&111.9&20.67&121.0$\pm$0.1&9.92$\pm$0.03&107.7&22.08\\
5&206.6$\pm$9.0&10.31$\pm$0.05&130.5&20.66&177.3$\pm$13.7&10.10$\pm$0.03&120.1&22.49\\
6&227.4$\pm$0.8&11.24$\pm$0.05&160.5&21.59&256.3$\pm$0.5&10.63$\pm$0.03&167.0&23.74\\
7&145.9$\pm$5.1&11.47$\pm$0.05&161.3&21.82&208.9$\pm$10.6&11.18$\pm$0.03&132.9&23.79\\
8&97.2$\pm$7.0&13.79$\pm$0.05&3.0&24.14&108.4$\pm$3.8&13.30$\pm$0.03&3.4&17.95\\
9&27.8$\pm$0.7&13.74$\pm$0.05&28.2&24.10&29.1$\pm$1.0&13.61$\pm$0.03&27.8&22.83\\
10&62.1$\pm$20.1&16.21$\pm$0.07&42.5&26.52&72.8$\pm$2.7&15.01$\pm$0.03&32.9&24.66\\
11&46.4$\pm$0.3&14.68$\pm$0.05&38.0&25.04&47.6$\pm$0.3&14.40$\pm$0.03&37.3&24.26\\
12&38.1$\pm$0.2&13.89$\pm$0.05&48.5&24.25&39.4$\pm$0.1&13.70$\pm$0.03&44.6&23.94\\
13&6.1$\pm$0.6&17.69$\pm$0.24&10.0&28.06&6.6$\pm$0.5&17.58$\pm$0.08&10.0&24.60\\
14&112.3$\pm$3.4&10.52$\pm$0.05&50.8&20.89&79.1$\pm$1.8&10.37$\pm$0.03&44.8&20.63\\
15&20.9$\pm$0.5&14.81$\pm$0.05&13.1&25.16&30.4$\pm$2.3&14.46$\pm$0.03&12.3&21.91\\
16&13.4$\pm$3.9&19.23$\pm$1.00&21.2&28.75&9.7$\pm$1.8&18.90$\pm$1.13&18.1&27.20\\
17&22.0$\pm$0.6&14.48$\pm$0.05&27.2&24.82&50.3$\pm$3.7&14.25$\pm$0.03&18.6&22.59\\
18&60.6$\pm$1.5&13.40$\pm$0.05&-&-&78.6$\pm$4.3&13.45$\pm$0.03&-20.4&21.99\\
19&-&-&-&-&35.9$\pm$2.9&15.26$\pm$0.05&11.6&22.59\\
23&9.1$\pm$0.3&16.62$\pm$0.13&8.6&26.93&8.4$\pm$1.9&16.43$\pm$0.08&10.4&23.49\\
27&24.3$\pm$3.0&13.38$\pm$0.05&12.2&23.71&87.7$\pm$3.5&12.97$\pm$0.03&-1.1&15.28\\
28&-&-&-&-&7.3$\pm$0.4&15.74$\pm$0.07&6.6&21.89\\
29&16.4$\pm$1.1&16.17$\pm$0.27&13.2&26.57&17.1$\pm$0.6&15.90$\pm$0.09&13.6&23.59\\
30&128.4$\pm$6.2&13.49$\pm$0.05&102.3&23.82&164.6$\pm$2.4&12.94$\pm$0.03&99.3&24.92\\
31&157.4$\pm$0.8&11.66$\pm$0.05&123.8&22.03&377.8$\pm$8.5&11.09$\pm$0.03&57.3&21.87\\
32&1.6$\pm$0.3&18.81$\pm$0.10&-&-&4.5$\pm$0.4&18.08$\pm$1.07&8.6&24.89\\
33&-&-&-&-&9.4$\pm$0.4&17.89$\pm$0.40&18.8&25.97\\
34&20.1$\pm$36.5&17.85$\pm$1.83&-&-&5.3$\pm$1.2&17.80$\pm$0.47&15.6&25.55\\
36&-&-&-&-&6.0$\pm$1.1&17.21$\pm$1.77&11.9&24.93\\
37&11.4$\pm$1.1&16.55$\pm$0.06&12.9&26.90&20.5$\pm$2.3&16.14$\pm$0.03&9.5&23.03\\
40&27.2$\pm$1.5&15.21$\pm$0.06&25.2&25.51&25.7$\pm$1.2&14.90$\pm$0.03&24.1&23.78\\
42&11.2$\pm$4.4&17.29$\pm$0.31&15.3&27.57&7.6$\pm$1.3&17.34$\pm$0.09&12.4&24.71\\
44&7.4$\pm$1.4&16.28$\pm$0.18&13.0&26.47&26.7$\pm$2.3&14.97$\pm$0.09&10.0&21.96\\
45&10.2$\pm$1.4&18.28$\pm$0.17&14.2&28.65&24.0$\pm$2.5&17.64$\pm$0.04&8.5&24.33\\
47&5.4$\pm$2.3&18.33$\pm$1.65&12.5&28.62&10.3$\pm$1.9&17.43$\pm$0.91&17.3&25.67\\
48&149.0$\pm$14.4&11.81$\pm$0.05&66.6&22.12&227.5$\pm$0.5&10.86$\pm$0.03&70.8&22.12\\
49&7.0$\pm$0.1&16.05$\pm$0.05&7.8&26.39&21.2$\pm$1.9&15.56$\pm$0.03&5.5&21.24\\
50&18.1$\pm$0.9&14.17$\pm$0.08&51.9&24.43&31.7$\pm$0.4&13.94$\pm$0.03&20.3&22.50\\
52&8.0$\pm$1.6&17.58$\pm$0.22&18.2&27.89&20.0$\pm$0.9&16.52$\pm$0.12&11.5&23.97\\
53&185.7$\pm$38.1&10.14$\pm$0.05&74.5&20.49&327.0$\pm$1.6&9.54$\pm$0.03&46.3&19.86\\
54&4.9$\pm$0.9&17.23$\pm$1.60&13.1&27.56&7.4$\pm$0.3&16.96$\pm$0.12&9.3&23.79\\
55&8.5$\pm$2.3&15.42$\pm$0.17&10.1&25.82&35.4$\pm$10.2&14.59$\pm$0.05&9.9&21.39\\
56&8.4$\pm$0.3&16.09$\pm$0.05&27.5&26.37&129.5$\pm$54.7&15.67$\pm$0.03&9.5&22.55\\
57&224.8$\pm$14.4&12.39$\pm$0.05&-&-&139.1$\pm$2.1&11.91$\pm$0.03&-17.4&20.10\\
58&6.5$\pm$0.5&16.17$\pm$0.07&4.4&26.53&9.2$\pm$0.6&15.88$\pm$0.05&4.5&21.18\\
59&136.5$\pm$30.1&14.79$\pm$0.06&9.2&25.17&66.2$\pm$7.8&14.34$\pm$0.03&11.7&21.70\\
74&-&-&-&-&29.9$\pm$2.8&13.30$\pm$0.06&25.7&22.36\\
75&200.8$\pm$1.6&12.02$\pm$0.05&22.6&22.43&305.4$\pm$12.1&12.10$\pm$0.03&-50.9&22.58\\
77&134.7$\pm$12.8&13.99$\pm$0.05&35.3&24.34&92.1$\pm$4.6&13.48$\pm$0.03&34.8&23.18\\
78&140.0$\pm$7.9&12.55$\pm$0.05&21.8&22.91&107.3$\pm$1.5&12.21$\pm$0.03&20.4&20.75\\
79&57.4$\pm$4.7&15.39$\pm$0.06&11.4&25.75&82.3$\pm$7.9&14.88$\pm$0.04&10.0&22.03\\
80&23.2$\pm$0.4&14.89$\pm$0.05&24.1&25.25&25.0$\pm$0.3&14.69$\pm$0.03&22.4&23.44\\
81&21.3$\pm$0.5&15.69$\pm$0.05&11.8&26.06&21.5$\pm$0.3&15.65$\pm$0.03&13.3&23.28\\
82&131.7$\pm$1.7&11.23$\pm$0.05&67.8&21.58&217.9$\pm$3.3&10.98$\pm$0.03&46.9&21.33\\
\hline

\end{tabular}
\label{tab:uvphot}
\end{table*}

\begin{table*}\scriptsize
\caption{WISE photometric measurements for the LVHIS galaxies (to be continued).}
\centering
\begin{tabular}{l|llllll|llll}
\hline
\multicolumn{1}{c|}{IDs} & \multicolumn{6}{c}{luminosities and colour} & \multicolumn{4}{|c}{ structural parameters } \\
\hline
index&${\rm \log L_{W1}}$&${\rm L_{W2}}$&${\rm L_{W3}}$&${\rm L_{W4}}$&${\rm L_{W1,\odot}}$&W1-W2&r$_s$&r$_{eff}$&$\mu_{eff}$&${\rm R_{75}/R_{25}}$\\

  	&   [L$_{\odot}$] & [L$_{\odot}$] & [L$_{\odot}$] & [L$_{\odot}$]&  [${\rm L_{\odot,W1}}$] & [mag] & [arcsec] & [arcsec] & [mag arcsec$^{-2}$] & \\
\hline
1$^f$&5.295$\pm$0.047&4.850$\pm$0.174&-&-&6.655$\pm$0.051&-0.123&65.1&23.0&22.281&2.18\\
2&5.472$\pm$0.015&5.068$\pm$0.049&-&-&6.832$\pm$0.025&-0.020&21.6&20.7&20.722&2.68\\
3&5.772$\pm$0.020&5.452$\pm$0.045&-&-&7.131$\pm$0.028&0.192&28.9&26.0&20.977&2.37\\
4&8.267$\pm$0.011&7.909$\pm$0.019&7.699$\pm$0.026&7.899$\pm$0.023&9.627$\pm$0.023&0.095&419.9&357.4&18.382&3.64\\
5&8.255$\pm$0.011&7.860$\pm$0.019&7.655$\pm$0.019&7.588$\pm$0.029&9.615$\pm$0.023&0.003&255.8&230.3&19.213&2.71\\
6&9.602$\pm$0.011&9.285$\pm$0.019&9.632$\pm$0.012&9.732$\pm$0.011&10.962$\pm$0.023&0.199&117.7&208.5&15.646&3.18\\
7&8.467$\pm$0.011&8.059$\pm$0.019&7.805$\pm$0.057&7.711$\pm$0.087&9.827$\pm$0.023&-0.031&348.5&282.5&19.388&2.62\\
8&7.707$\pm$0.012&7.341$\pm$0.020&7.343$\pm$0.033&7.747$\pm$0.024&9.066$\pm$0.023&0.076&51.6&64.6&18.527&3.59\\
9&6.987$\pm$0.013&6.532$\pm$0.028&5.710$\pm$0.183&-&8.346$\pm$0.024&-0.144&98.9&76.8&21.442&2.43\\
10&4.668$\pm$0.015&4.219$\pm$0.205&-&-&6.028$\pm$0.025&-0.132&139.3&107.9&22.496&2.37\\
11&7.156$\pm$0.012&6.786$\pm$0.023&5.982$\pm$0.151&5.998$\pm$0.333&8.516$\pm$0.023&0.067&78.3&66.4&20.365&2.84\\
12&7.440$\pm$0.012&6.998$\pm$0.024&6.246$\pm$0.088&6.662$\pm$0.190&8.800$\pm$0.023&-0.113&252.1&131.1&21.020&3.17\\
13&5.452$\pm$0.084&5.118$\pm$0.273&-&-&6.812$\pm$0.086&0.156&28.6&11.6&22.006&2.36\\
14&8.482$\pm$0.011&8.113$\pm$0.019&8.225$\pm$0.052&8.359$\pm$0.044&9.842$\pm$0.023&0.068&105.2&147.5&19.004&3.35\\
15&7.479$\pm$0.012&7.110$\pm$0.022&6.315$\pm$0.079&6.652$\pm$0.106&8.838$\pm$0.023&0.070&43.7&44.0&19.342&3.11\\
16&5.071$\pm$0.061&4.866$\pm$0.277&-&-&6.431$\pm$0.064&0.478&-&-&-&-\\
17&7.508$\pm$0.012&7.097$\pm$0.022&6.693$\pm$0.067&6.972$\pm$0.067&8.867$\pm$0.023&-0.035&36.2&37.3&19.175&2.47\\
18&7.298$\pm$0.012&6.941$\pm$0.021&6.513$\pm$0.047&6.726$\pm$0.060&8.658$\pm$0.023&0.097&36.6&18.2&18.005&4.53\\
19&7.898$\pm$0.022&7.534$\pm$0.053&-&-&9.258$\pm$0.030&0.080&31.5&28.4&21.569&3.16\\
20&6.849$\pm$0.021&6.519$\pm$0.050&-&-&8.209$\pm$0.029&0.166&51.3&45.7&21.793&2.30\\
21&6.944$\pm$0.014&6.580$\pm$0.029&5.894$\pm$0.110&5.928$\pm$0.163&8.303$\pm$0.024&0.081&15.3&15.4&18.984&2.60\\
22&7.981$\pm$0.012&7.625$\pm$0.021&7.395$\pm$0.033&7.622$\pm$0.037&9.341$\pm$0.023&0.100&80.3&53.6&18.575&2.40\\
23&5.679$\pm$0.029&5.269$\pm$0.075&-&-&7.039$\pm$0.035&-0.033&40.7&19.3&21.808&2.07\\
24$^f$&5.811$\pm$0.058&5.514$\pm$0.154&-&-&7.171$\pm$0.061&0.248&19.8&11.9&21.791&2.25\\
25&5.331$\pm$0.029&4.988$\pm$0.069&5.397$\pm$0.115&-&6.690$\pm$0.035&0.134&3.7&13.0&21.241&4.77\\
26&6.870$\pm$0.014&6.472$\pm$0.025&-&-&8.229$\pm$0.024&-0.005&71.5&50.1&20.894&2.71\\
27&7.322$\pm$0.012&6.942$\pm$0.021&6.159$\pm$0.035&6.404$\pm$0.043&8.681$\pm$0.023&0.043&62.7&32.0&18.168&3.33\\
28&5.710$\pm$0.044&5.290$\pm$0.115&-&-&7.069$\pm$0.048&-0.058&15.6&15.5&21.177&2.81\\
29&6.414$\pm$0.020&6.070$\pm$0.047&-&-&7.774$\pm$0.028&0.131&67.8&34.4&21.039&2.15\\
30&5.714$\pm$0.041&5.258$\pm$0.107&-&-&7.074$\pm$0.046&-0.150&27.9&31.7&22.013&3.56\\
31&8.999$\pm$0.020&8.696$\pm$0.034&8.996$\pm$0.016&8.826$\pm$0.021&10.359$\pm$0.028&0.233&184.2&64.0&17.032&2.95\\
32&-&-&-&-&-&-&-&-&-&-\\
33&5.602$\pm$0.041&5.256$\pm$0.136&-&-&6.961$\pm$0.046&0.125&16.5&12.5&21.711&2.51\\
34&5.900$\pm$0.030&5.443$\pm$0.095&5.599$\pm$0.227&-&7.260$\pm$0.036&-0.152&36.2&16.2&21.178&2.22\\
35$^f$&5.261$\pm$0.074&5.036$\pm$0.187&-&-&6.620$\pm$0.077&0.429&-&-&-&-\\
36&5.533$\pm$0.048&5.253$\pm$0.104&4.808$\pm$0.235&-&6.892$\pm$0.052&0.290&16.1&10.7&21.112&4.06\\
37&6.076$\pm$0.024&5.658$\pm$0.070&-&-&7.435$\pm$0.031&-0.052&40.6&36.5&21.664&2.31\\
38&7.010$\pm$0.012&6.616$\pm$0.021&-&-&8.369$\pm$0.023&0.007&82.8&78.0&19.671&2.72\\
39&5.878$\pm$0.027&5.458$\pm$0.073&-&-&7.238$\pm$0.034&-0.059&22.5&14.6&21.050&2.36\\
40&6.479$\pm$0.018&6.140$\pm$0.038&-&-&7.838$\pm$0.027&0.144&66.2&48.3&21.403&2.30\\
41&5.033$\pm$0.044&4.845$\pm$0.094&-&5.098$\pm$0.243&6.392$\pm$0.048&0.520&30.6&15.5&21.956&3.80\\
42&5.626$\pm$0.036&5.146$\pm$0.157&-&-&6.985$\pm$0.041&-0.208&51.7&22.5&22.045&2.35\\
43&9.323$\pm$0.011&9.008$\pm$0.019&9.242$\pm$0.011&9.146$\pm$0.011&10.682$\pm$0.023&0.203&82.4&182.2&16.285&2.95\\
44&6.907$\pm$0.012&6.540$\pm$0.021&5.983$\pm$0.060&5.819$\pm$0.083&8.266$\pm$0.023&0.075&51.8&53.8&19.062&3.30\\
45&-&-&-&-&-&-&-&-&-&-\\
46&8.252$\pm$0.011&7.855$\pm$0.019&6.322$\pm$0.073&6.288$\pm$0.106&9.611$\pm$0.023&0.000&122.6&80.4&17.290&5.32\\
47&5.840$\pm$0.032&5.377$\pm$0.121&-&-&7.199$\pm$0.038&-0.166&28.3&19.1&21.438&2.47\\
48&9.859$\pm$0.011&9.472$\pm$0.019&9.240$\pm$0.011&9.113$\pm$0.012&11.218$\pm$0.023&0.024&138.4&215.1&16.350&5.66\\
49&6.578$\pm$0.015&6.224$\pm$0.033&-&-&7.937$\pm$0.025&0.106&29.7&23.0&19.767&2.56\\
50&6.712$\pm$0.015&6.248$\pm$0.033&-&-&8.071$\pm$0.025&-0.170&59.5&57.1&21.254&2.40\\
51&8.353$\pm$0.012&7.967$\pm$0.021&7.785$\pm$0.102&7.767$\pm$0.069&9.712$\pm$0.023&0.028&131.6&231.2&20.299&3.36\\
52&6.368$\pm$0.020&6.030$\pm$0.044&-&-&7.727$\pm$0.028&0.147&29.5&21.7&20.530&2.91\\
53&9.536$\pm$0.011&9.188$\pm$0.019&9.520$\pm$0.011&9.550$\pm$0.012&10.896$\pm$0.023&0.119&160.6&144.6&16.983&3.28\\
54&4.408$\pm$0.105&4.288$\pm$0.220&4.399$\pm$0.275&-&5.767$\pm$0.107&0.693&-&-&-&-\\
55&6.632$\pm$0.012&6.243$\pm$0.021&5.002$\pm$0.037&5.078$\pm$0.095&7.992$\pm$0.023&0.019&48.3&28.1&18.595&3.07\\
56&6.018$\pm$0.027&5.668$\pm$0.073&-&-&7.377$\pm$0.034&0.117&39.8&23.6&21.394&2.17\\
57&7.952$\pm$0.011&7.889$\pm$0.019&8.374$\pm$0.019&8.814$\pm$0.011&9.312$\pm$0.023&0.832&22.1&31.9&16.786&5.67\\
58&6.690$\pm$0.015&6.288$\pm$0.031&5.640$\pm$0.113&5.649$\pm$0.198&8.049$\pm$0.025&-0.013&49.8&39.7&20.636&3.00\\
59&7.336$\pm$0.012&6.944$\pm$0.021&6.453$\pm$0.033&6.454$\pm$0.098&8.695$\pm$0.023&0.010&85.9&53.4&19.660&2.70\\
60&6.474$\pm$0.019&6.123$\pm$0.042&5.433$\pm$0.192&-&7.833$\pm$0.028&0.114&53.6&45.4&21.500&2.20\\
61&4.794$\pm$0.087&4.459$\pm$0.349&-&-&6.153$\pm$0.089&99.000&41.6&13.2&22.924&3.96\\
62&6.819$\pm$0.023&6.404$\pm$0.039&-&-&8.178$\pm$0.031&-0.046&78.0&40.3&20.169&2.50\\
63&7.003$\pm$0.012&6.591$\pm$0.021&6.044$\pm$0.035&6.002$\pm$0.063&8.362$\pm$0.023&-0.038&112.2&78.6&19.622&2.80\\
64&4.690$\pm$0.107&4.408$\pm$0.345&-&-&6.050$\pm$0.109&99.000&9.6&6.0&22.199&2.10\\
65&7.287$\pm$0.012&6.927$\pm$0.022&7.002$\pm$0.022&7.603$\pm$0.021&8.647$\pm$0.023&0.090&50.1&35.2&19.096&2.15\\
66&9.413$\pm$0.011&9.285$\pm$0.019&9.734$\pm$0.011&9.795$\pm$0.012&10.772$\pm$0.023&0.671&60.3&57.8&14.739&12.0\\
67&5.494$\pm$0.044&5.195$\pm$0.089&-&-&6.854$\pm$0.048&0.244&31.7&28.5&21.727&2.57\\
68&5.674$\pm$0.079&5.420$\pm$0.148&-&-&7.033$\pm$0.082&0.357&45.5&11.9&21.046&2.47\\
69&6.461$\pm$0.028&6.034$\pm$0.051&-&-&7.820$\pm$0.034&-0.074&8.1&10.9&18.927&4.75\\
70&6.890$\pm$0.017&6.496$\pm$0.033&5.831$\pm$0.111&6.119$\pm$0.263&8.249$\pm$0.026&0.007&31.1&23.6&19.661&2.59\\
71&7.403$\pm$0.014&6.975$\pm$0.026&-&-&8.762$\pm$0.024&-0.079&44.5&50.3&20.137&2.52\\
72&8.118$\pm$0.012&7.726$\pm$0.020&7.006$\pm$0.030&7.582$\pm$0.020&9.478$\pm$0.023&0.011&84.4&213.9&19.032&3.33\\
73&6.125$\pm$0.033&5.784$\pm$0.067&3.867$\pm$0.220&-&7.484$\pm$0.039&0.140&-&-&-&-\\
74&7.677$\pm$0.013&7.358$\pm$0.024&6.982$\pm$0.060&7.207$\pm$0.083&9.037$\pm$0.024&0.193&126.8&45.8&18.892&2.38\\
75&7.032$\pm$0.012&6.691$\pm$0.021&6.648$\pm$0.021&7.225$\pm$0.013&8.391$\pm$0.023&0.140&36.3&40.0&18.307&2.75\\

\hline
\end{tabular}

\label{tab:irphot}
\end{table*}

\addtocounter{table}{-1}
\begin{table*}\scriptsize
\caption{WISE photometric measurements for the LVHIS galaxies (continued).}
\centering
\begin{tabular}{l|llllll|llll}
\hline
\multicolumn{1}{c|}{IDs} & \multicolumn{6}{c}{luminosities and colour} & \multicolumn{4}{|c}{ structural parameters } \\
\hline
index&${\rm \log L_{W1}}$&${\rm L_{W2}}$&${\rm L_{W3}}^1$&${\rm L_{W4}}^1$&${\rm L_{W1}}^2$&W1-W2&r$_s^3$&r$_{eff}^4$&$\mu_{eff}^4$&${\rm R_{75}/R_{25}}^5$\\

  	&   [L$_{\odot}$] & [L$_{\odot}$] & [L$_{\odot}$] & [L$_{\odot}$]&  [${\rm L_{\odot,W1}}$] & [mag] & [arcsec] & [arcsec] & [mag arcsec$^{-2}$] & \\
\hline
76$^f$&6.110$\pm$0.048&5.608$\pm$0.203&-&-&7.470$\pm$0.052&-0.265&32.6&17.2&21.587&2.11\\
77&8.122$\pm$0.011&7.748$\pm$0.019&7.506$\pm$0.026&7.769$\pm$0.029&9.481$\pm$0.023&0.059&99.0&93.8&18.584&2.77\\
78&7.266$\pm$0.040&6.911$\pm$0.059&6.555$\pm$0.049&6.493$\pm$0.079&8.625$\pm$0.045&0.103&74.3&60.1&18.653&3.02\\
79&5.823$\pm$0.018&5.463$\pm$0.043&-&-&7.183$\pm$0.027&0.091&55.5&42.9&21.272&2.45\\
80&6.475$\pm$0.015&6.130$\pm$0.034&-&-&7.835$\pm$0.025&0.129&66.8&52.5&20.794&2.52\\
81&6.261$\pm$0.026&5.827$\pm$0.075&-&5.492$\pm$0.329&7.620$\pm$0.033&-0.092&17.4&15.6&20.520&2.91\\
82&8.514$\pm$0.011&8.153$\pm$0.019&8.235$\pm$0.014&8.063$\pm$0.018&9.873$\pm$0.023&0.090&131.5&119.2&18.357&2.61\\
\hline
\end{tabular}
\label{tab:irphot}
\end{table*}

\section{Atlas of the large galaxies}
\label{sec:appendix_figure}
In Figure~\ref{fig:atlas_conv} we present the resolution-matched distributions of $M_*$, SFR and $\mHI$ (left side) and radial SB profiles (right side) for the ten large LVHIS galaxies (see Section~\ref{sec:phot_conv} and ~\ref{sec:loc_SFE}). Ellipses in the images and vertical lines in the SB profile figures mark three radii, at 1.25 R$_{25}$ (dotted), $\Sigma_*=3~\mspc$ (dashed), and $\SHI = 3~\mspc$  (dash-dotted), which are also annotated with star symbols in the $\Sigma_*$ and $\SHI$ profiles.

In Figure~\ref{fig:atlas_highrs}  we present images for the same sample of 10 large LVHIS galaxies showing the distribution of $M_*$ (at the resolution of the WISE W1-band), SFR (at the resolution of the WISE W4-band), and the mean ATCA HI velocity field at the original resolution (left side) as well as our tilted ring models and derived tiltograms  (right side). Ellipses in the images highlight the onset of the warp ($r_{warp}$, solid line), if present, and the R$_{25}$ isophote (dashed line). For each tilted ring model we show Ð from top to bottom Ð the rotational velocity ($v_{\rm rot}$), position angle (PA), inclination angle (i) and HI surface brightness ($\SHI$) as a function of radius. The approaching, receding and mean values are coloured in blue, red and black, respectively. The tiltograms are for the approaching velocity side of each galaxy; the magenta dashed lines mark $r_{warp}$, if a warp is present.

\newpage

\begin{figure*} 
\includegraphics[width=12cm]{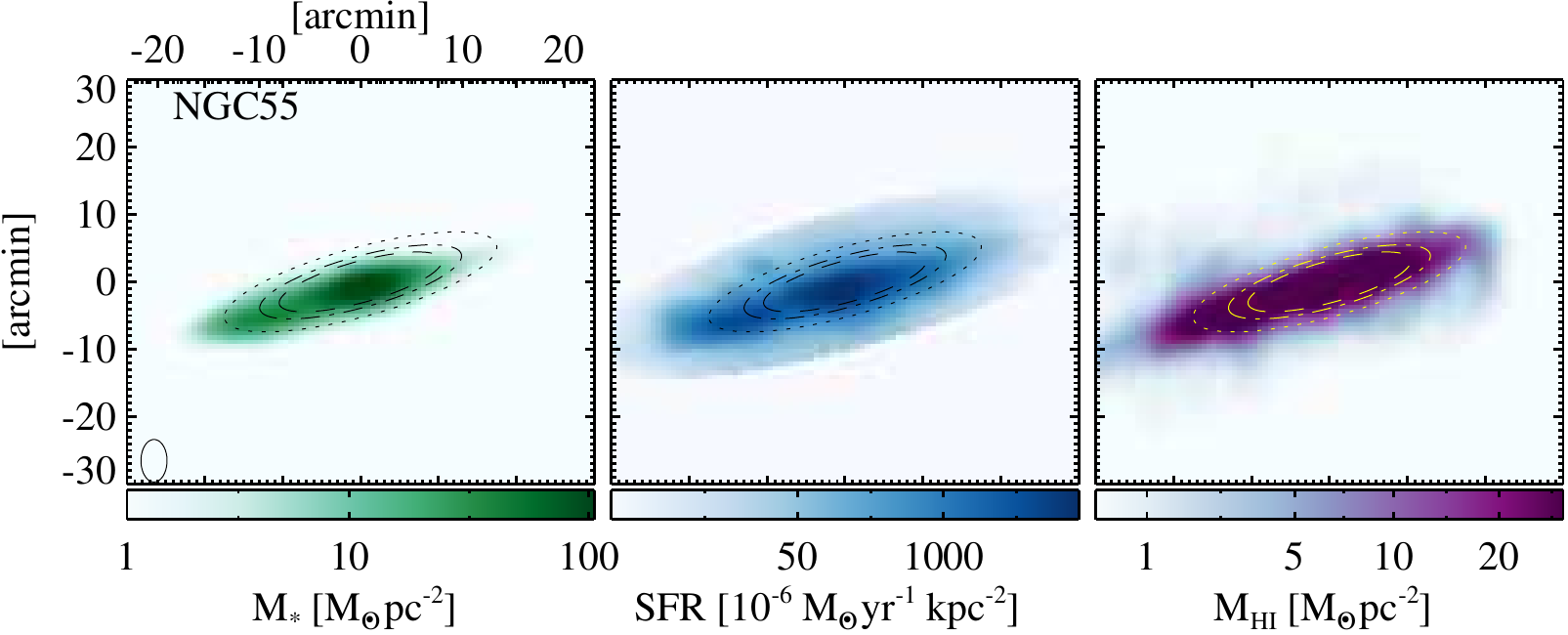}
\includegraphics[width=5cm]{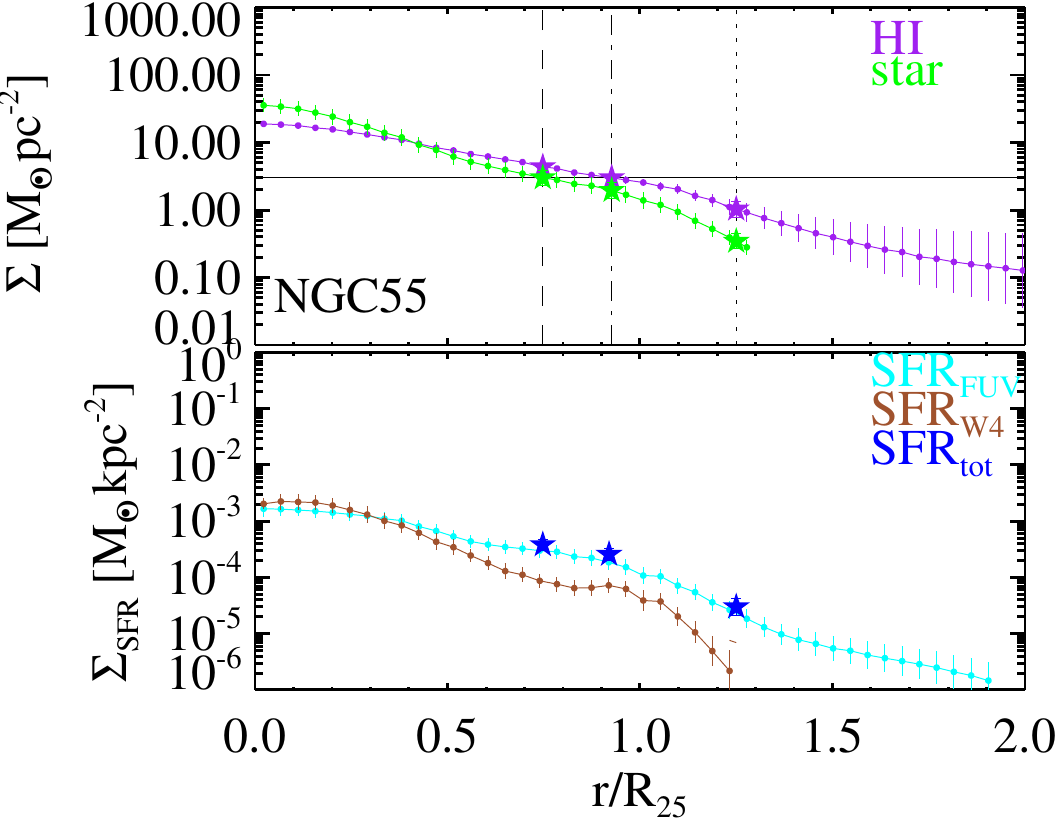}
\vspace{0.4cm}

\includegraphics[width=12cm]{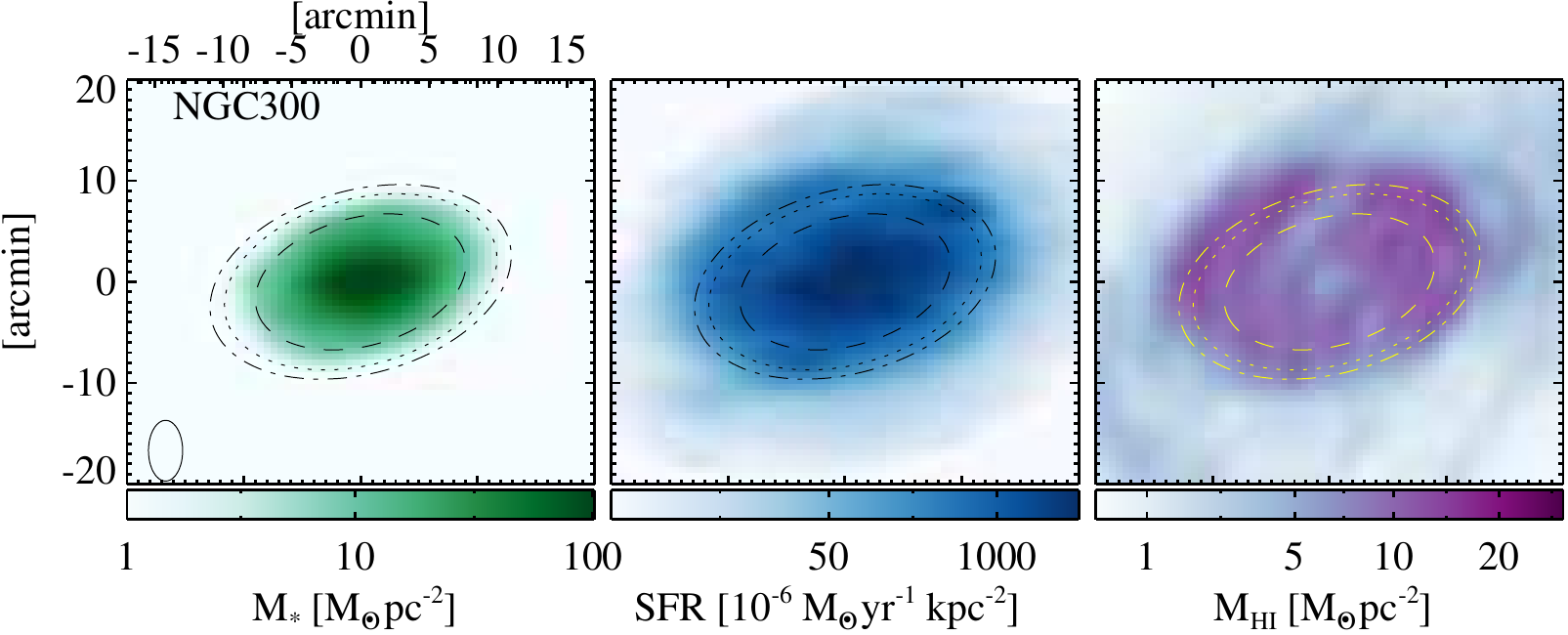}
\includegraphics[width=5cm]{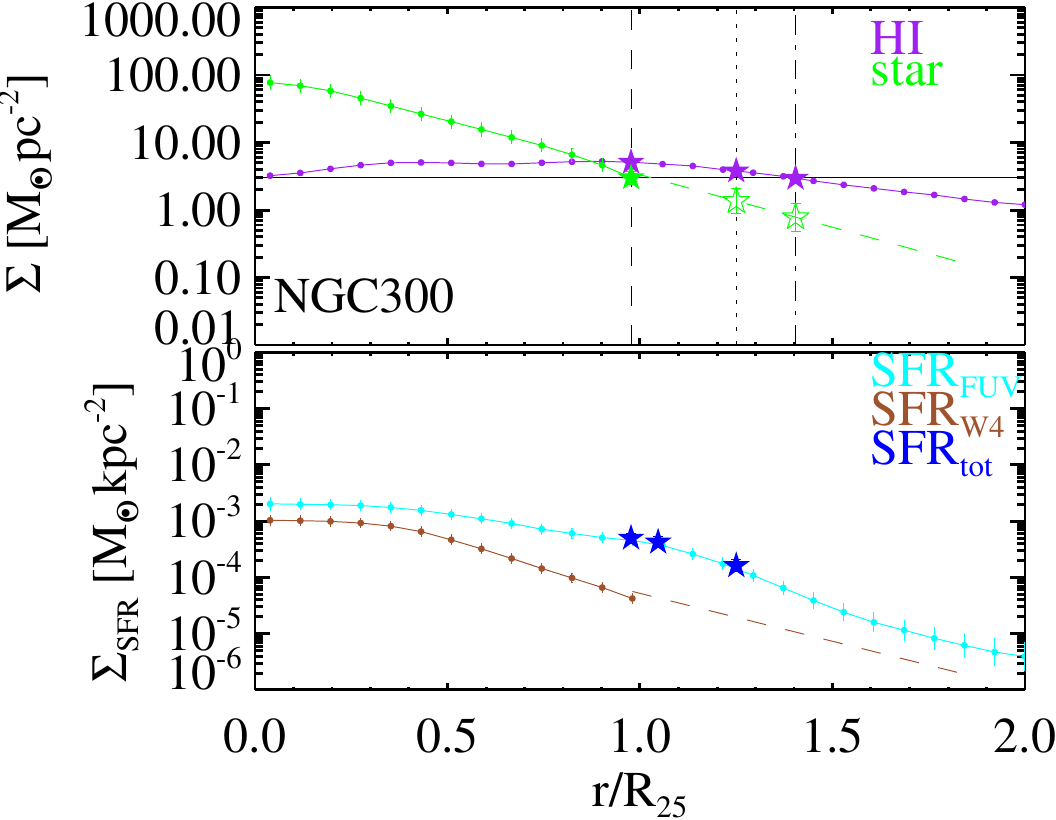}
\vspace{0.4cm}

\includegraphics[width=12cm]{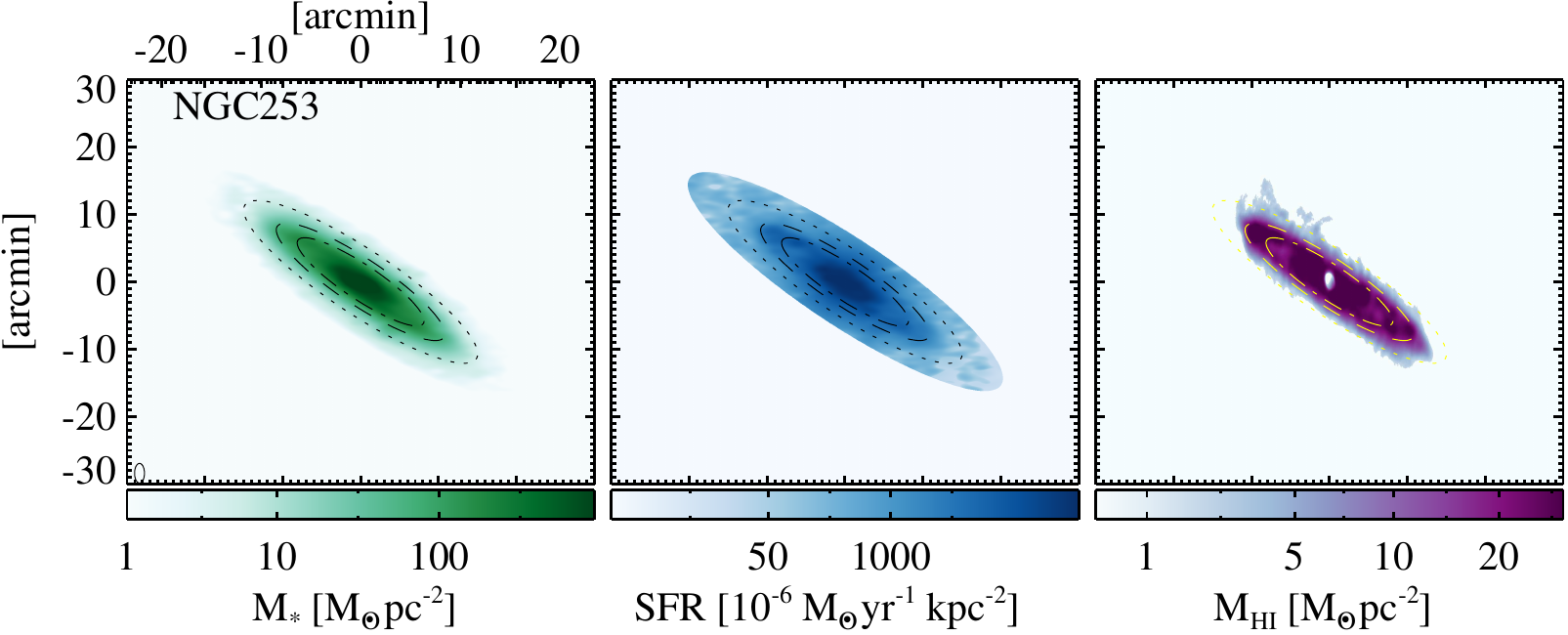}
\includegraphics[width=5cm]{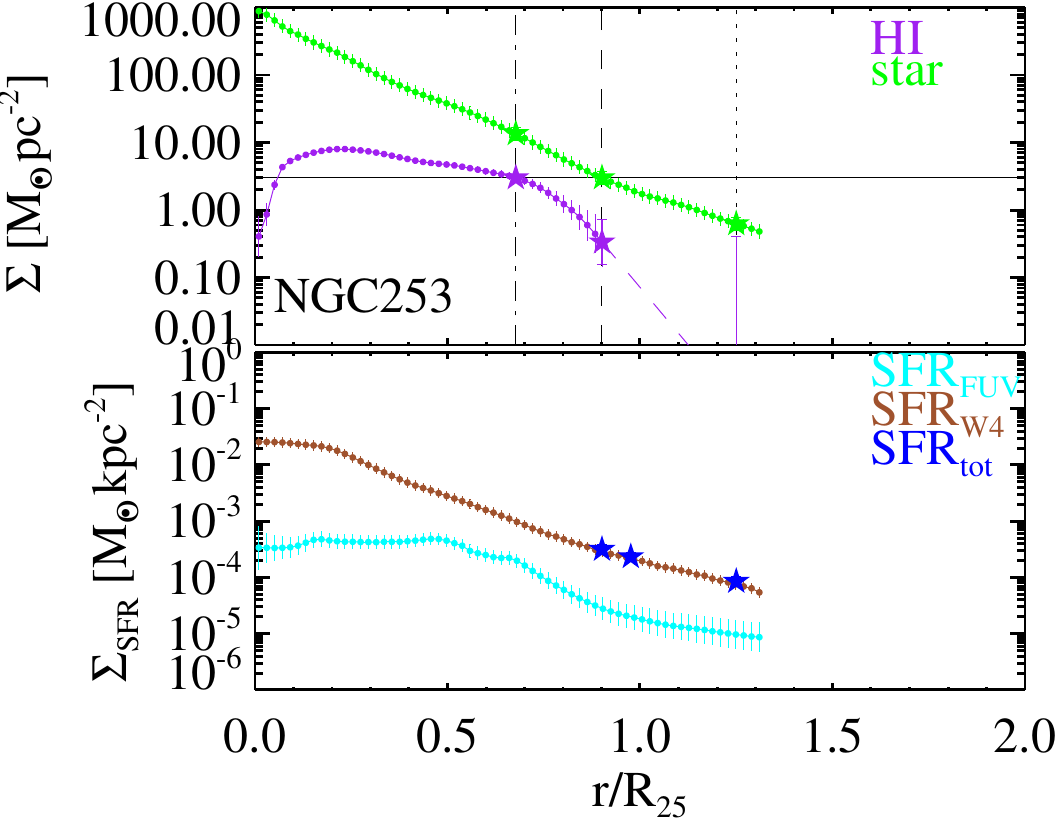}
\vspace{0.4cm}

\includegraphics[width=12cm]{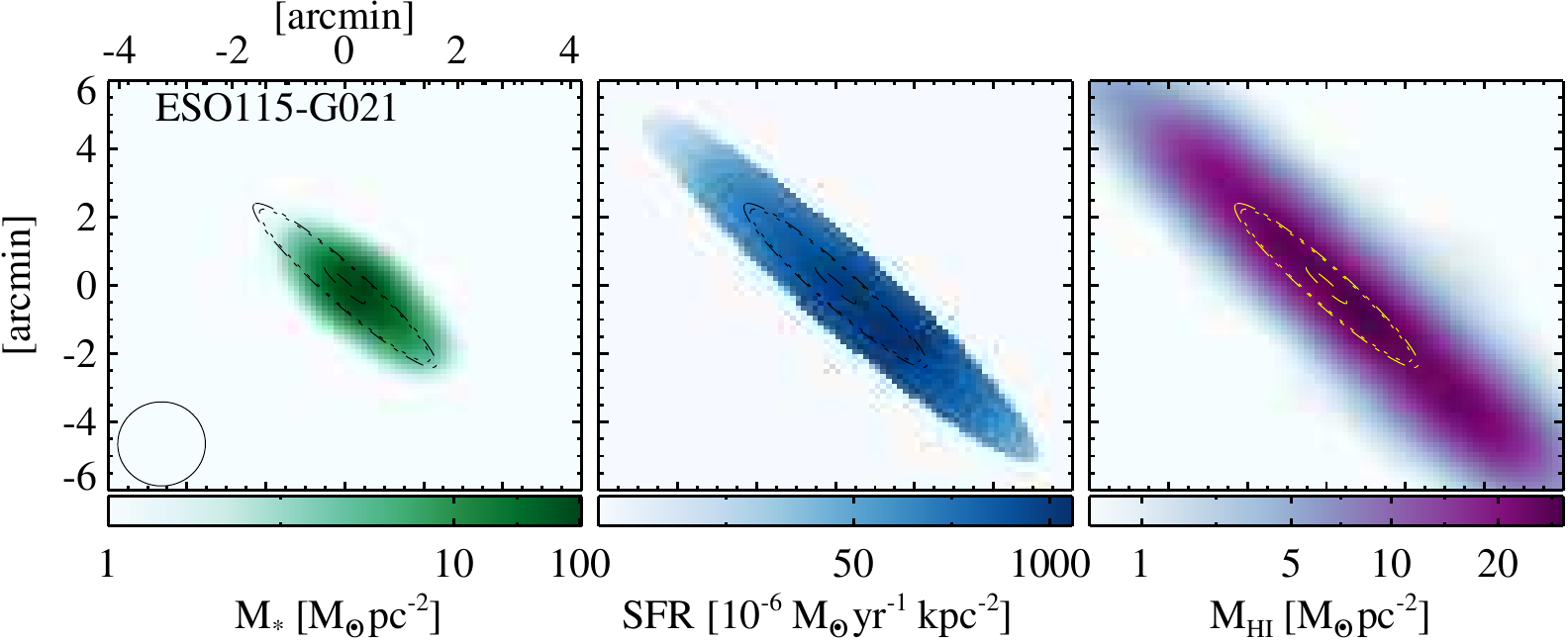}
\includegraphics[width=5cm]{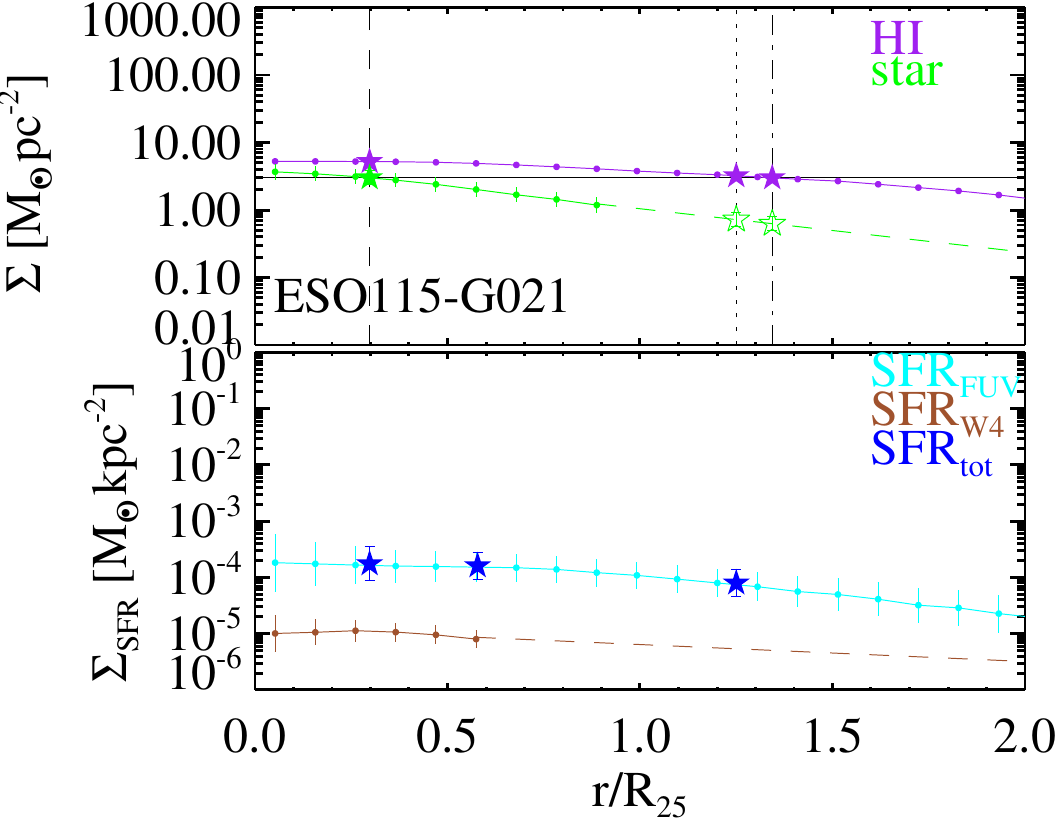}
\vspace{0.4cm}

\includegraphics[width=12cm]{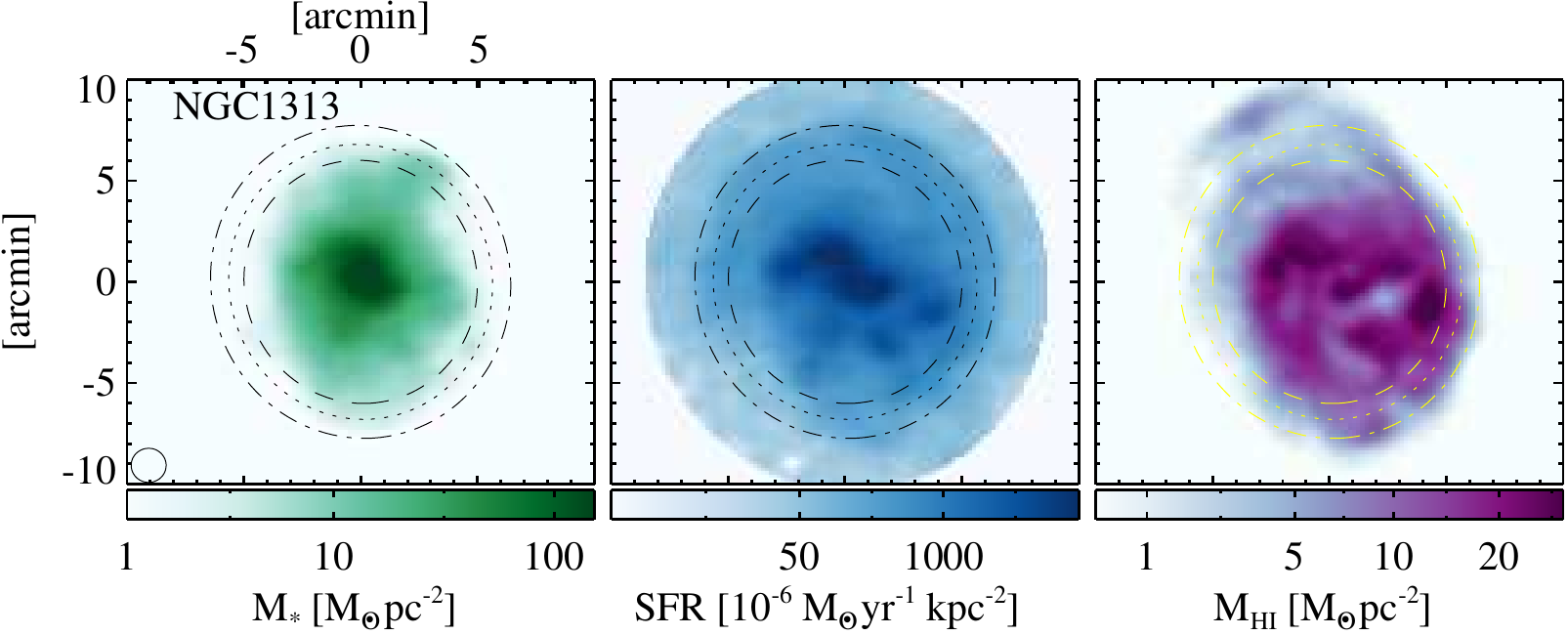}
\includegraphics[width=5cm]{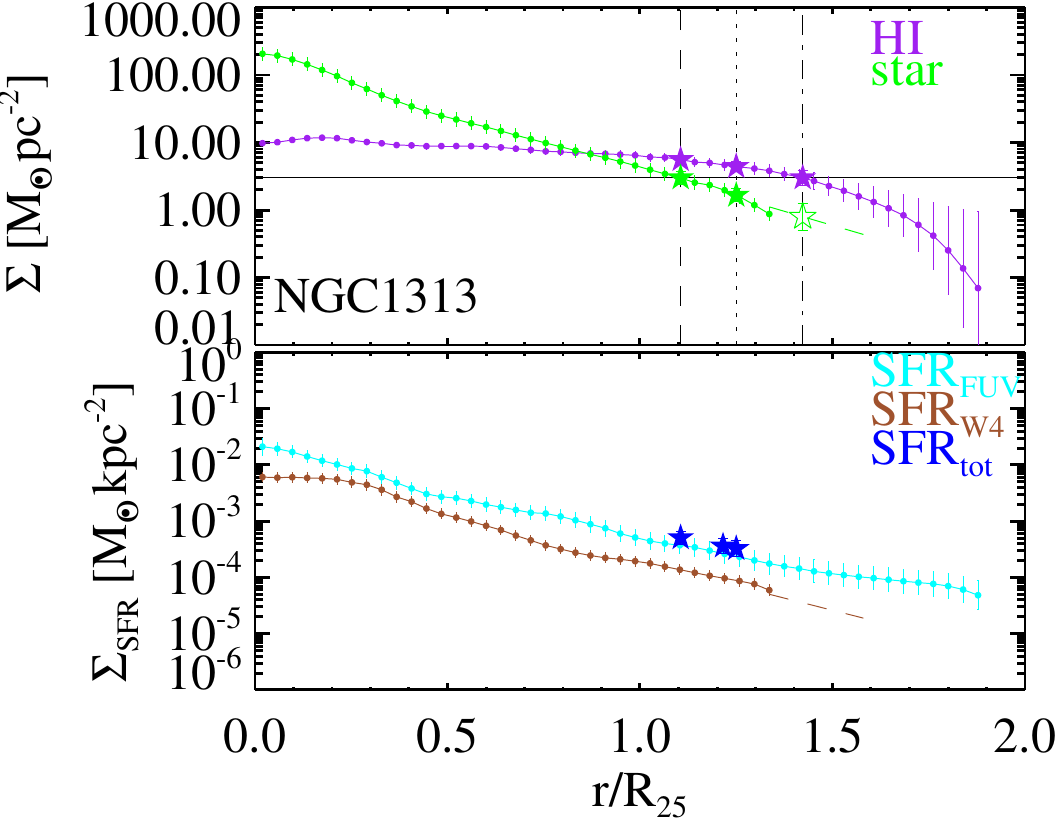}
\vspace{0.4cm}

\caption{ From left to right: resolution-matched distributions of $M_*$, SFR and $\mHI$ and radial SB profiles for the ten large LVHIS galaxies (see Appendix~\ref{sec:appendix_figure}). To be continued. }
\label{fig:atlas_conv}
\end{figure*}

\begin{figure*} 
\addtocounter{figure}{-1}

\includegraphics[width=12cm]{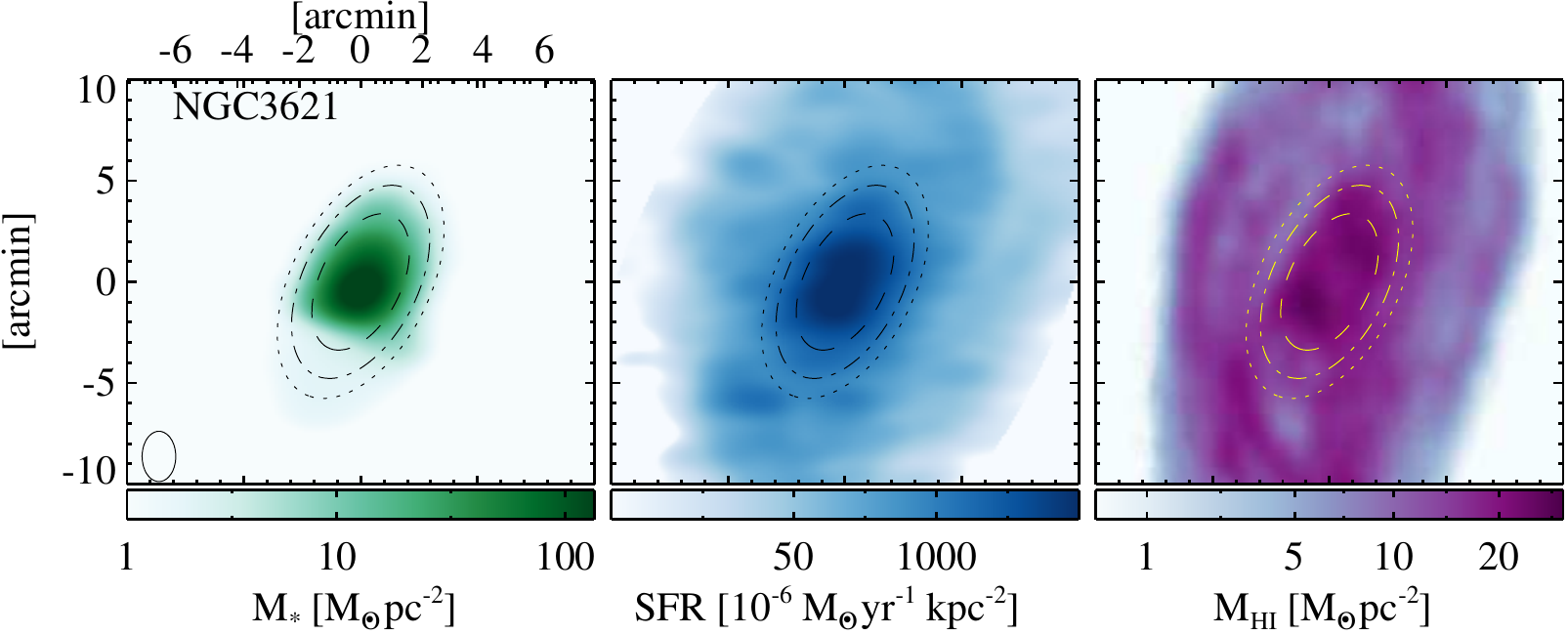}
\includegraphics[width=5cm]{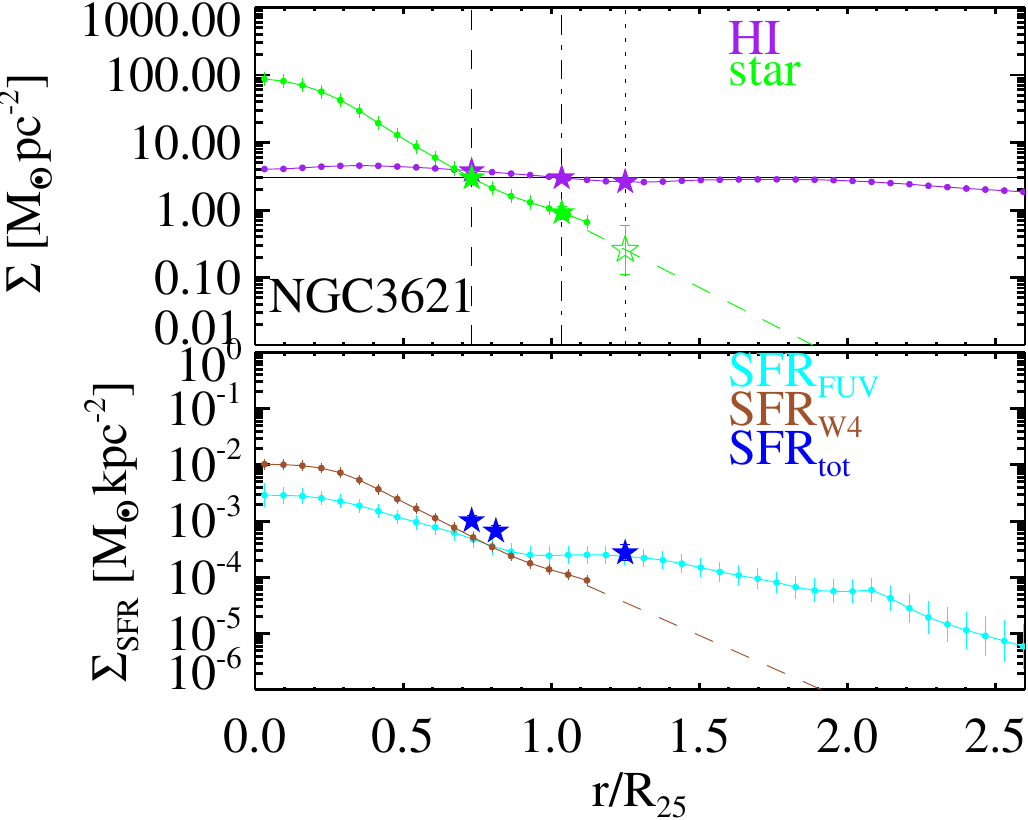}
\vspace{0.4cm}

\includegraphics[width=12cm]{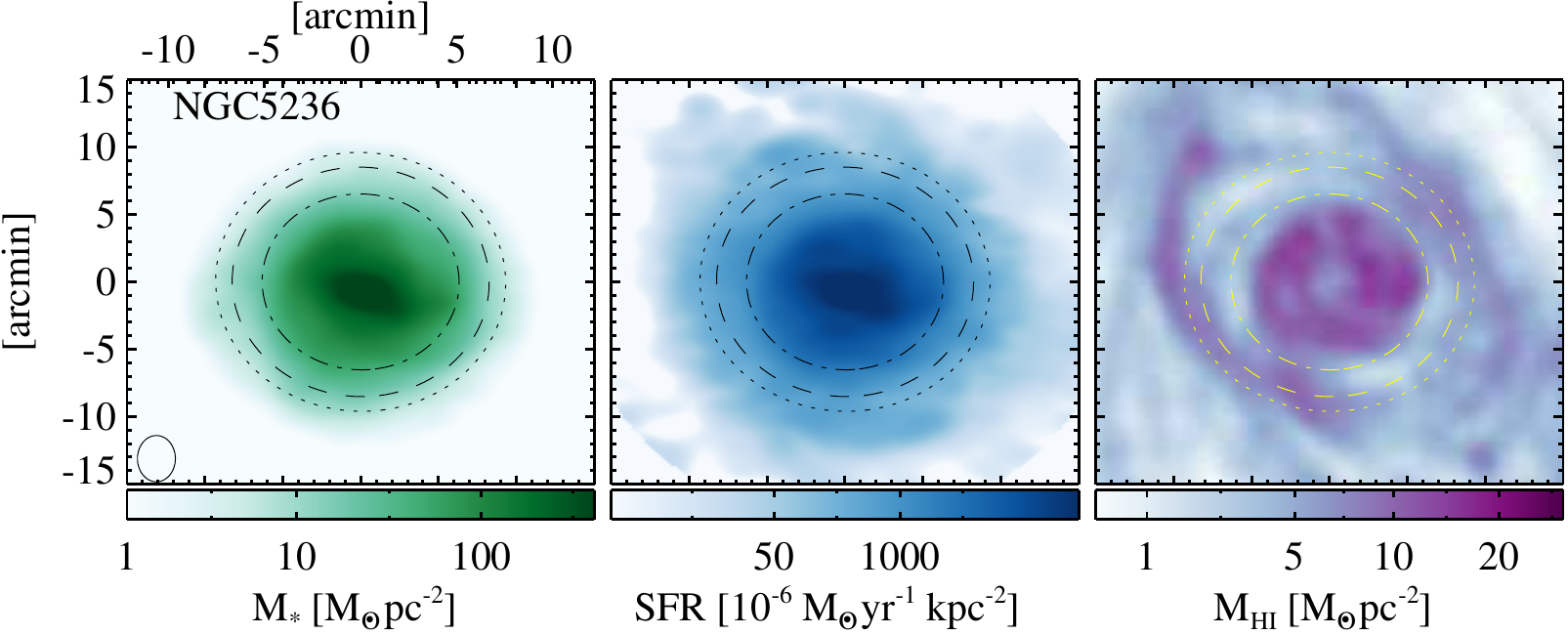}
\includegraphics[width=5cm]{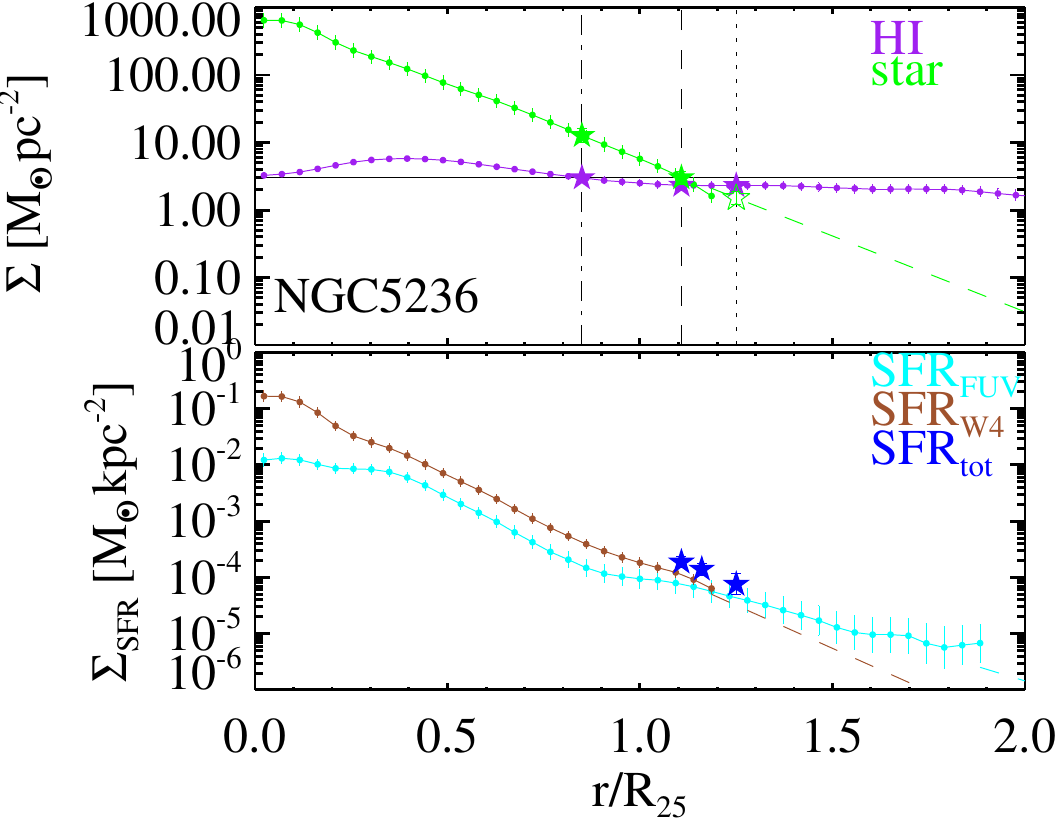}
\vspace{0.4cm}

\includegraphics[width=12cm]{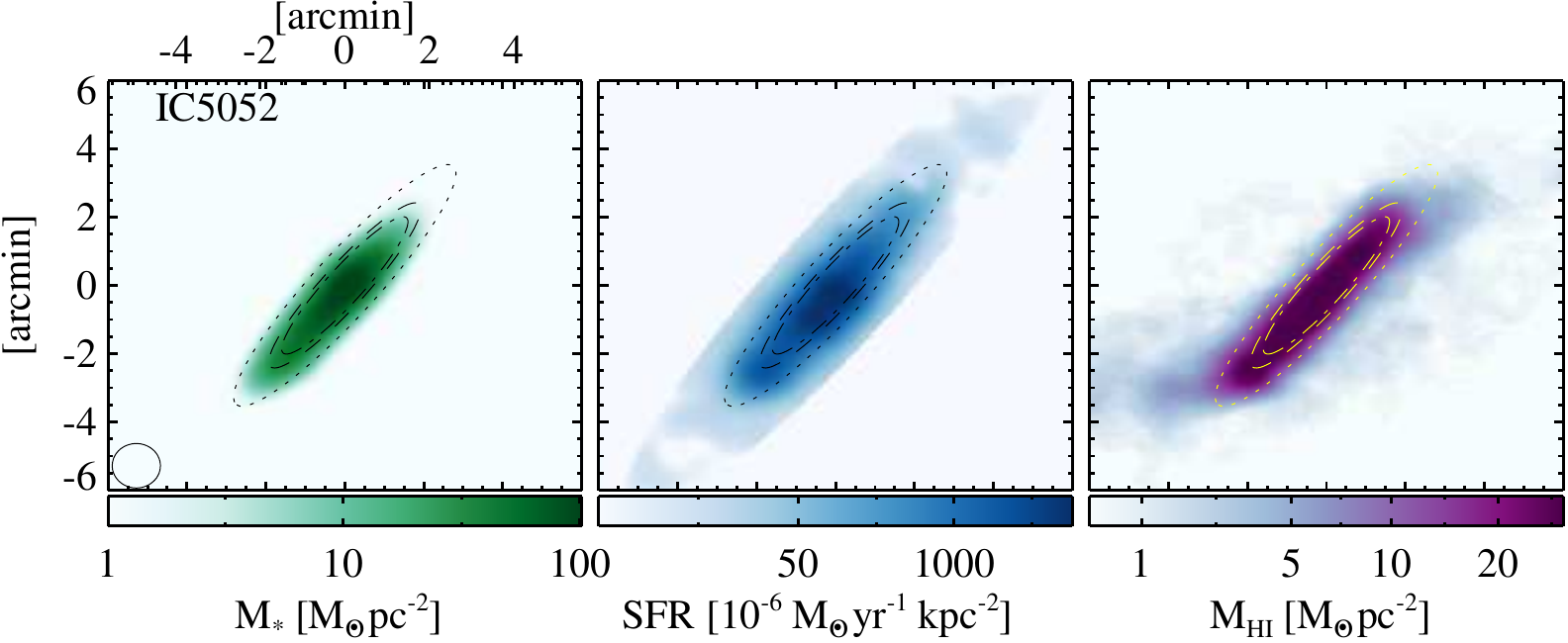}
\includegraphics[width=5cm]{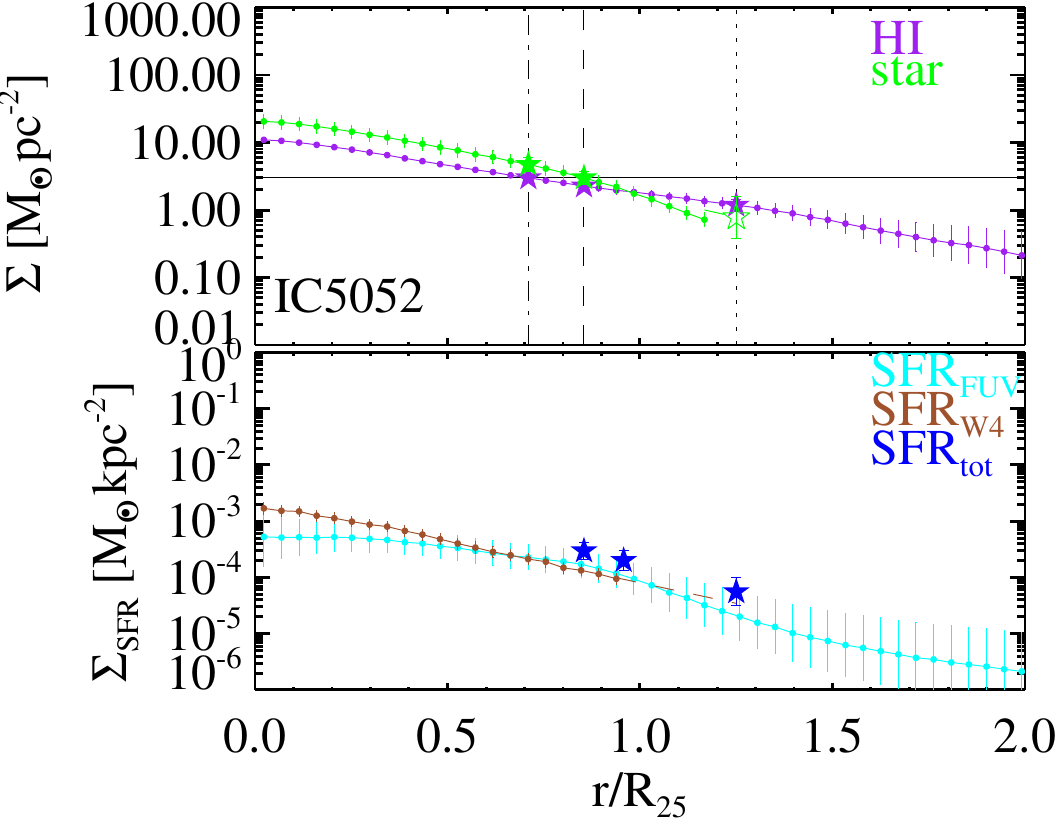}
\vspace{0.4cm}

\includegraphics[width=12cm]{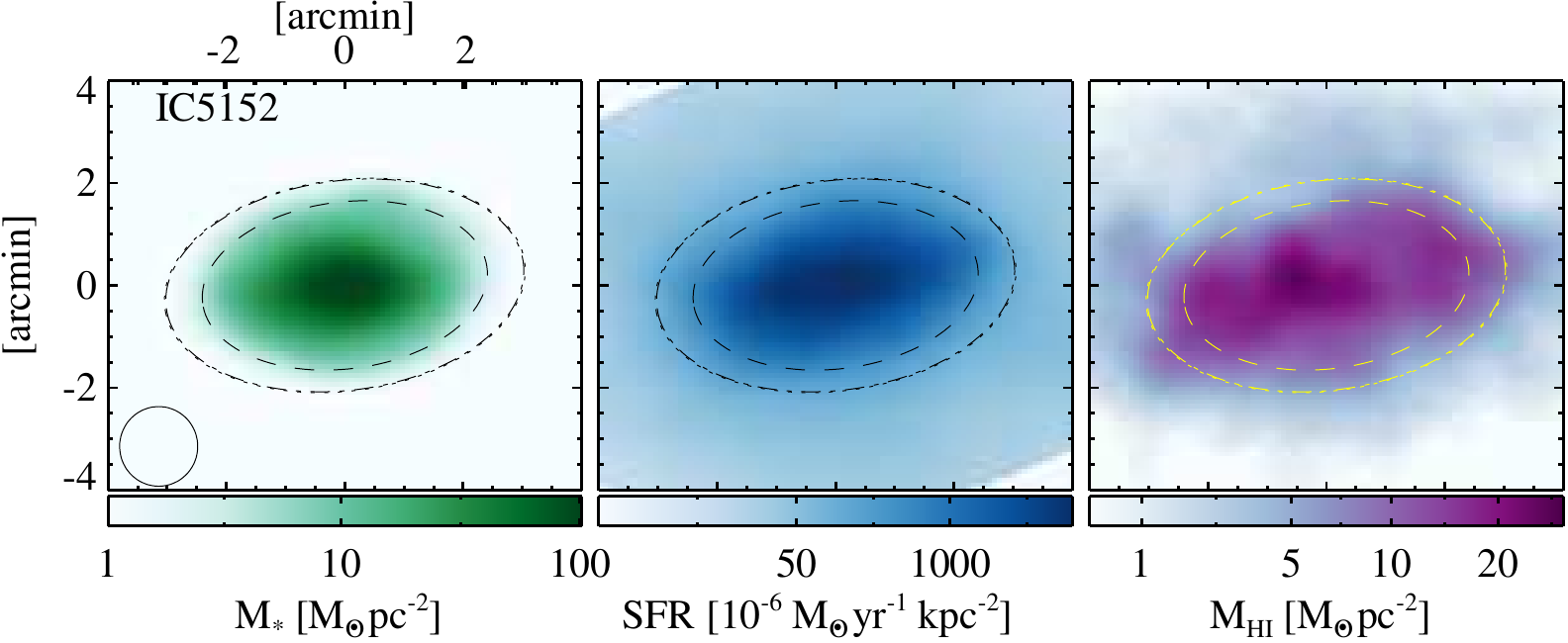}
\includegraphics[width=5cm]{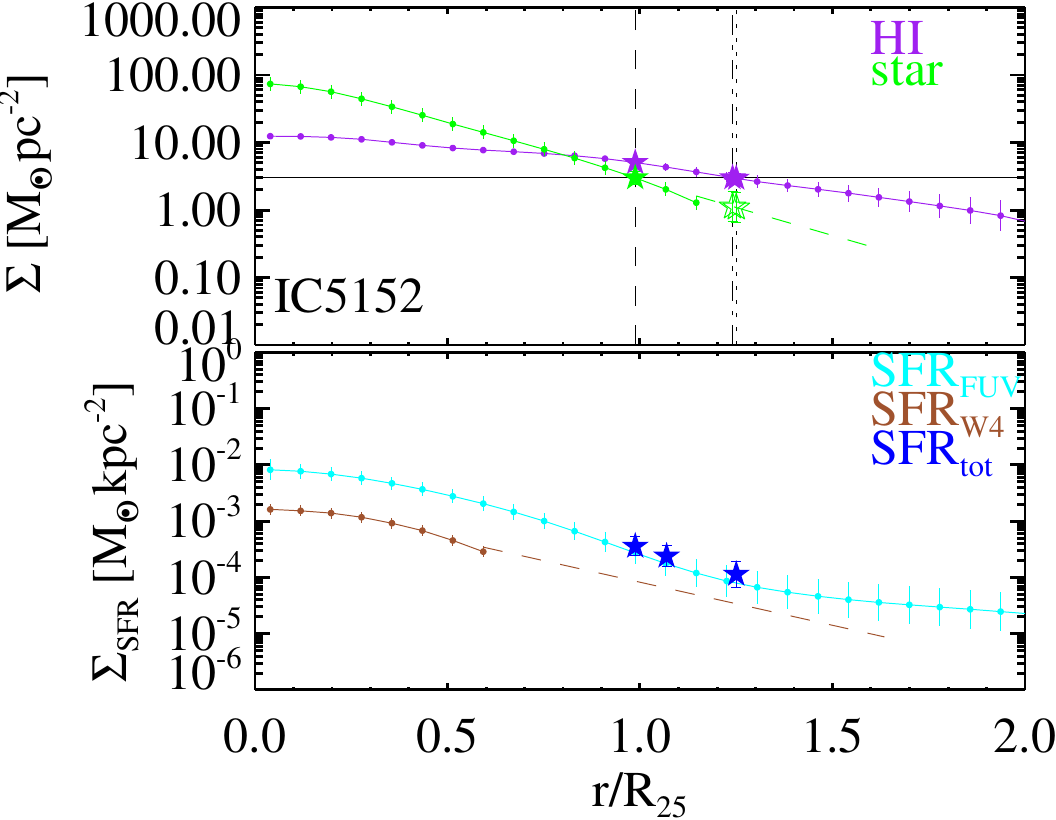}
\vspace{0.4cm}

\includegraphics[width=12cm]{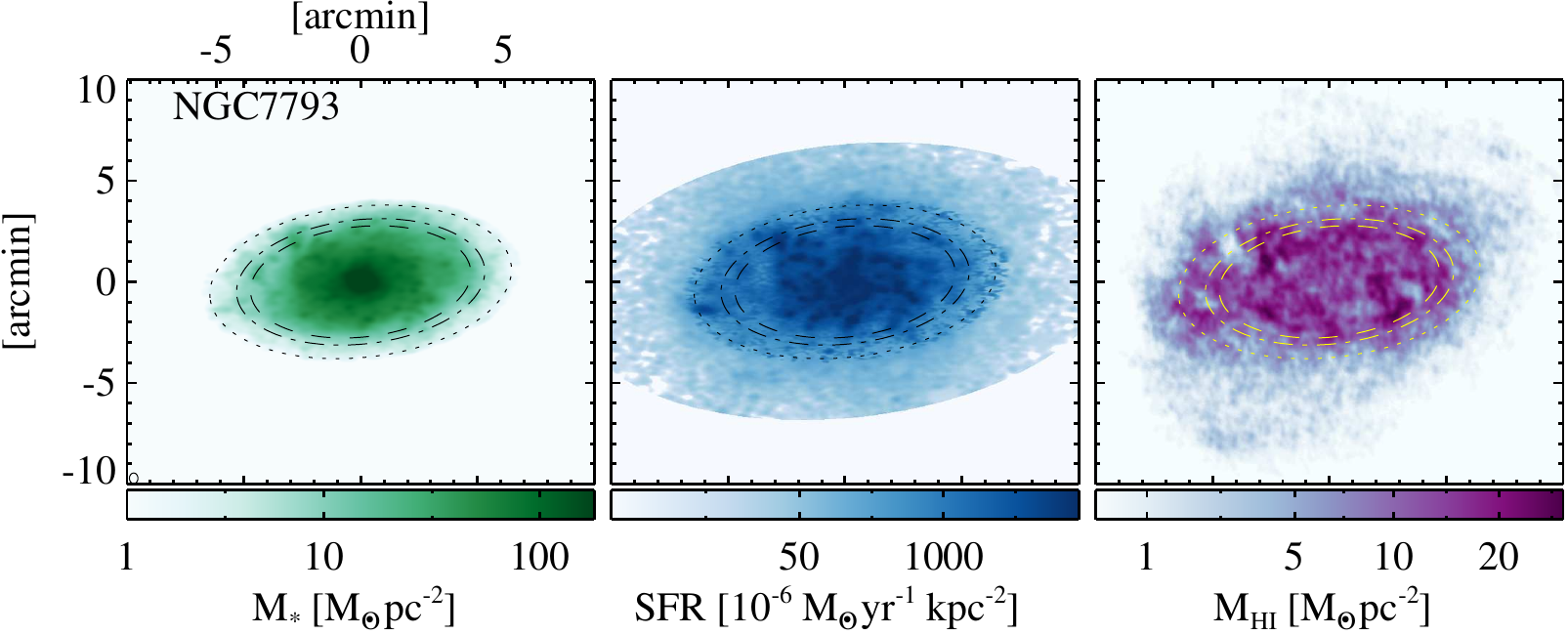}
\includegraphics[width=5cm]{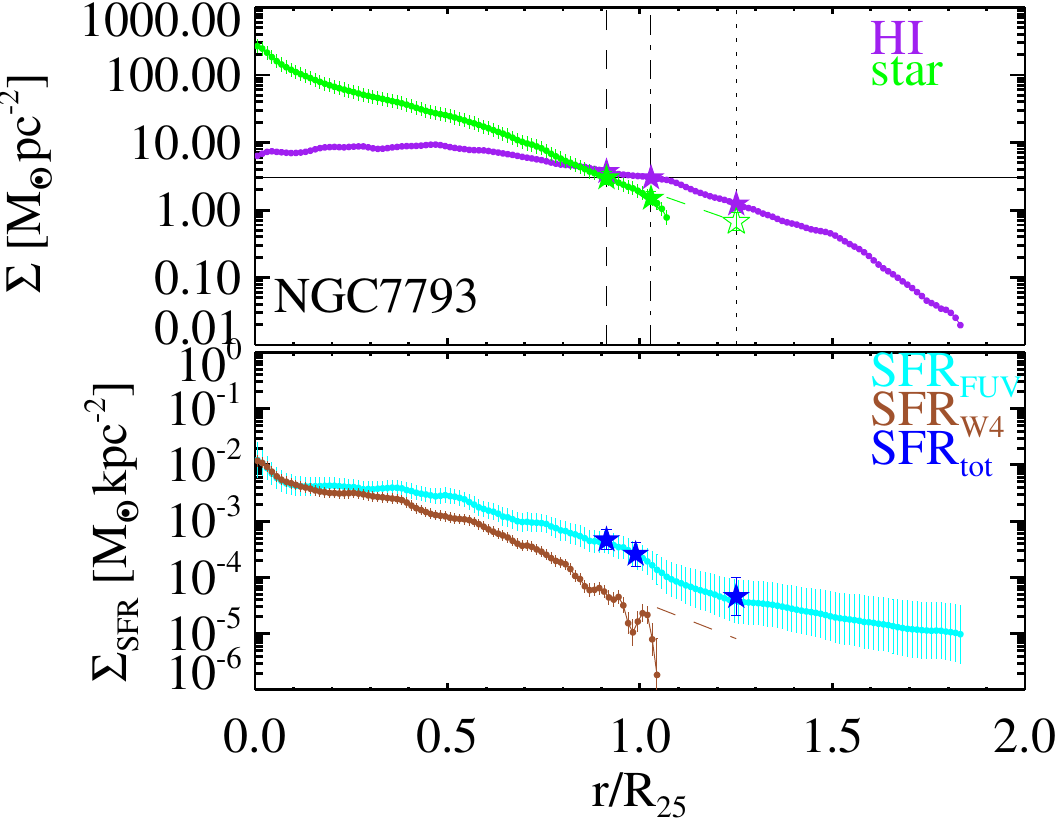}
\vspace{0.4cm}

\caption{Continued. From left to right: resolution-matched distributions of $M_*$, SFR and $\mHI$ and radial SB profiles for the ten large LVHIS galaxies.}
\label{fig:atlas_conv}
\end{figure*}

\begin{figure*} 
\includegraphics[width=10cm]{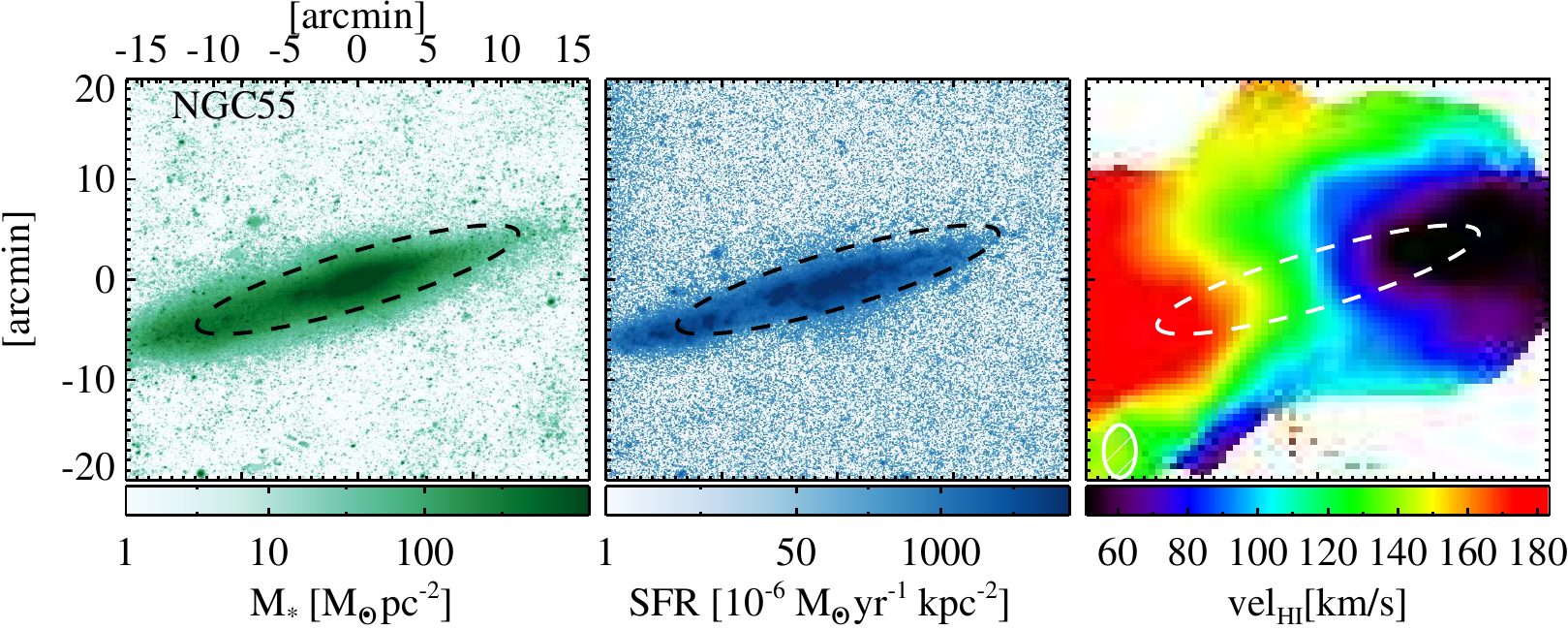}
\includegraphics[width=3cm]{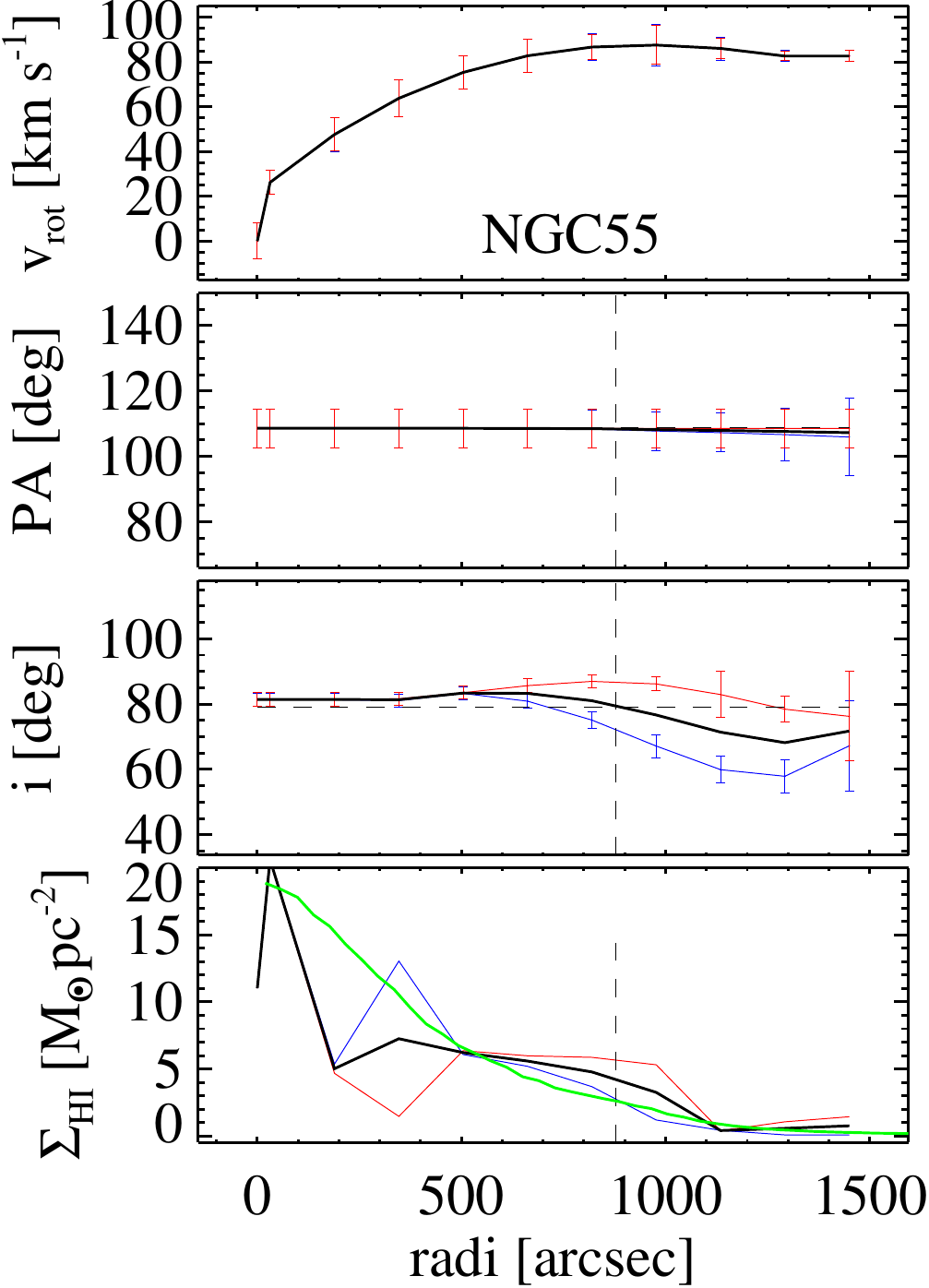}
\includegraphics[width=4cm]{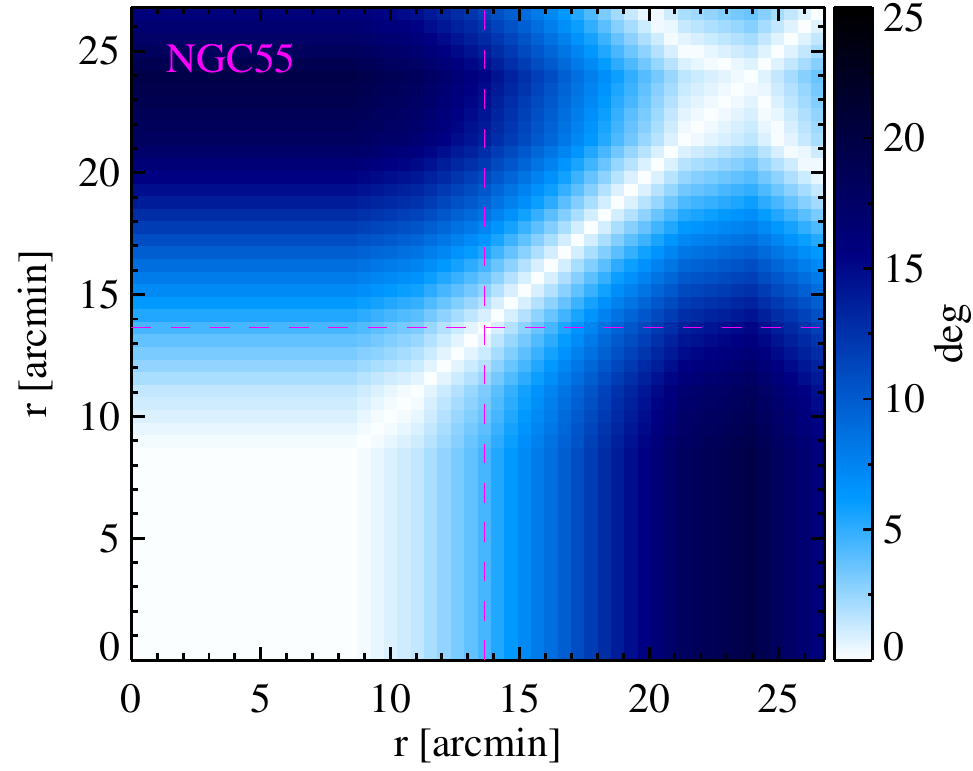}
\vspace{0.2cm}
\includegraphics[width=10cm]{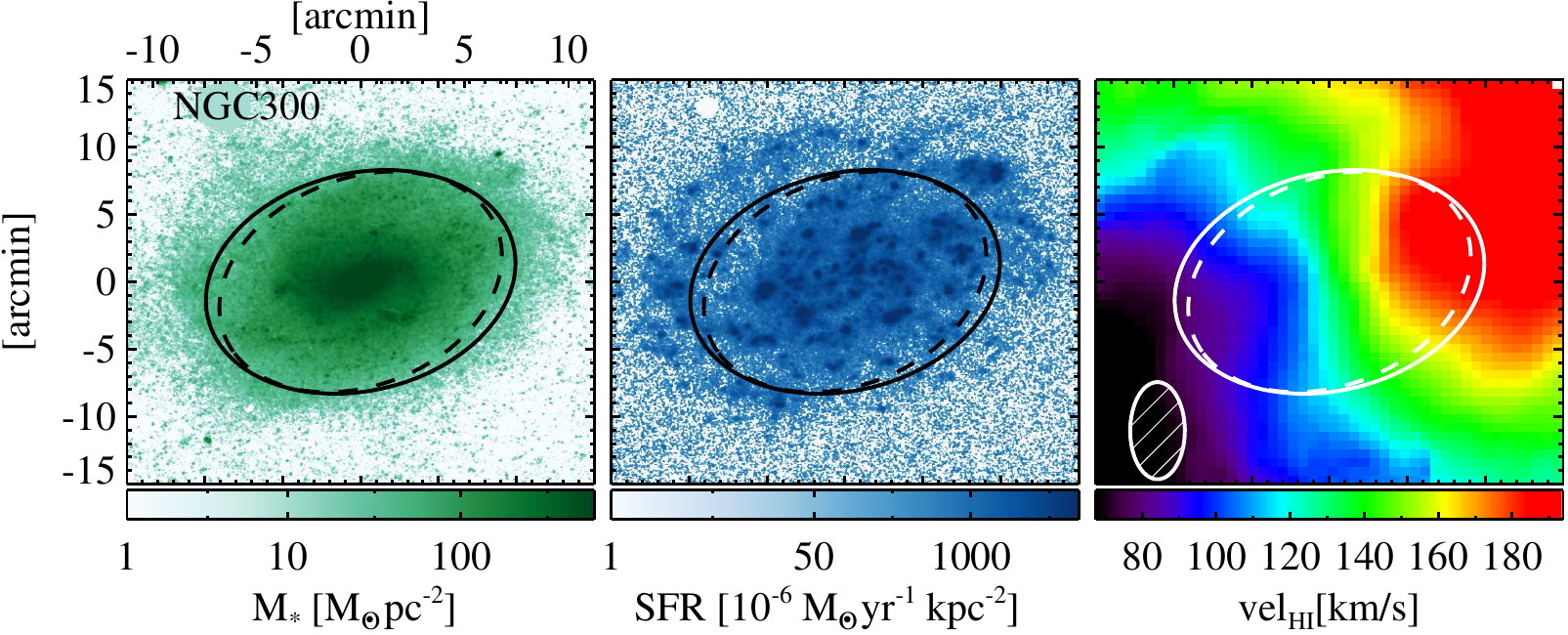}
\includegraphics[width=3cm]{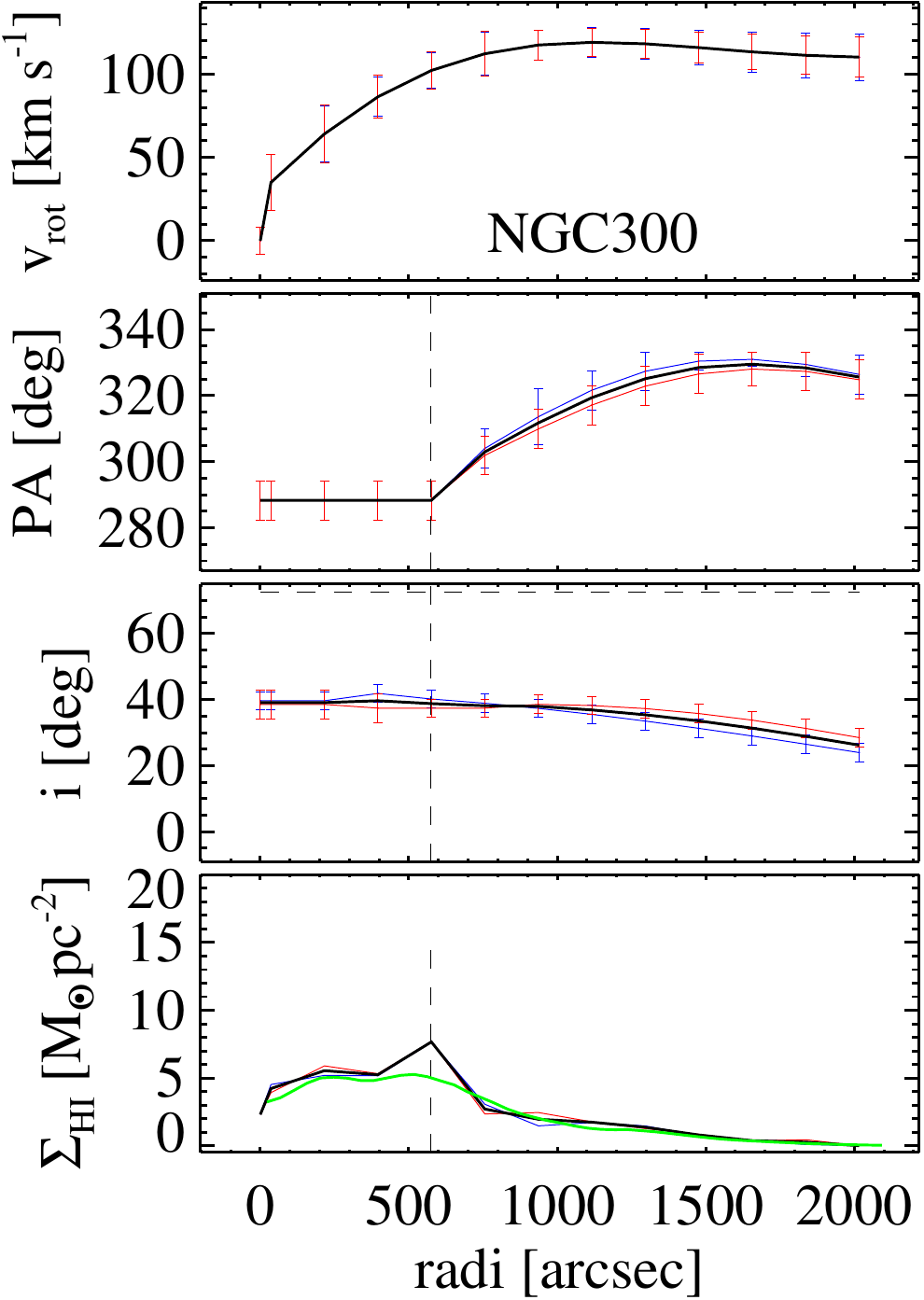}
\includegraphics[width=4cm]{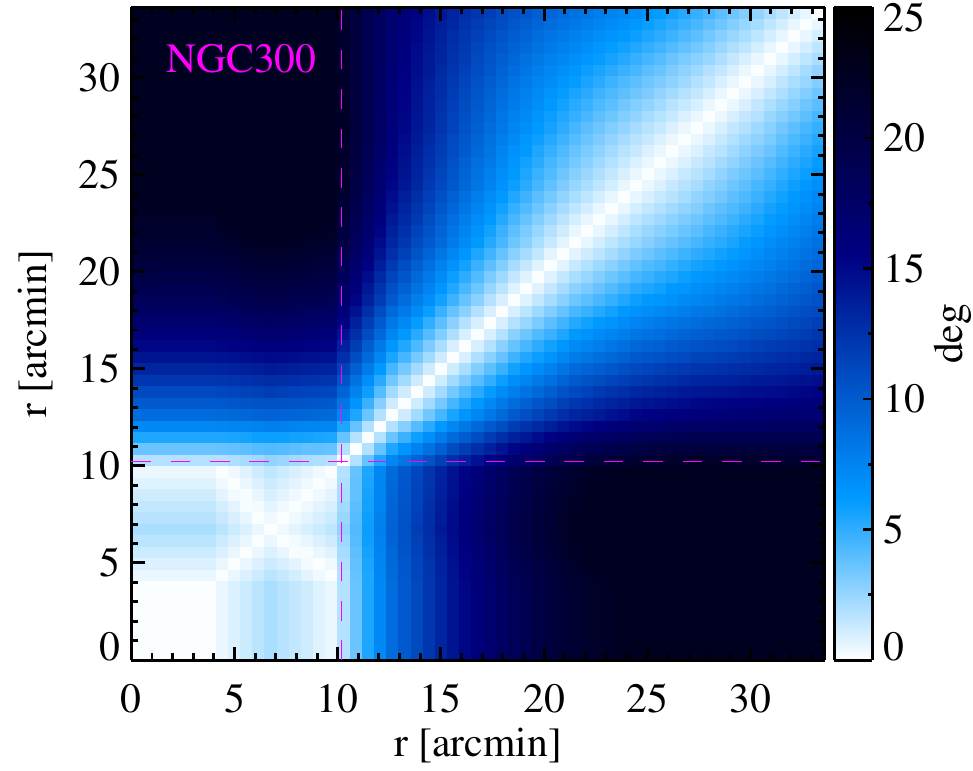}
\vspace{0.2cm}
\includegraphics[width=10cm]{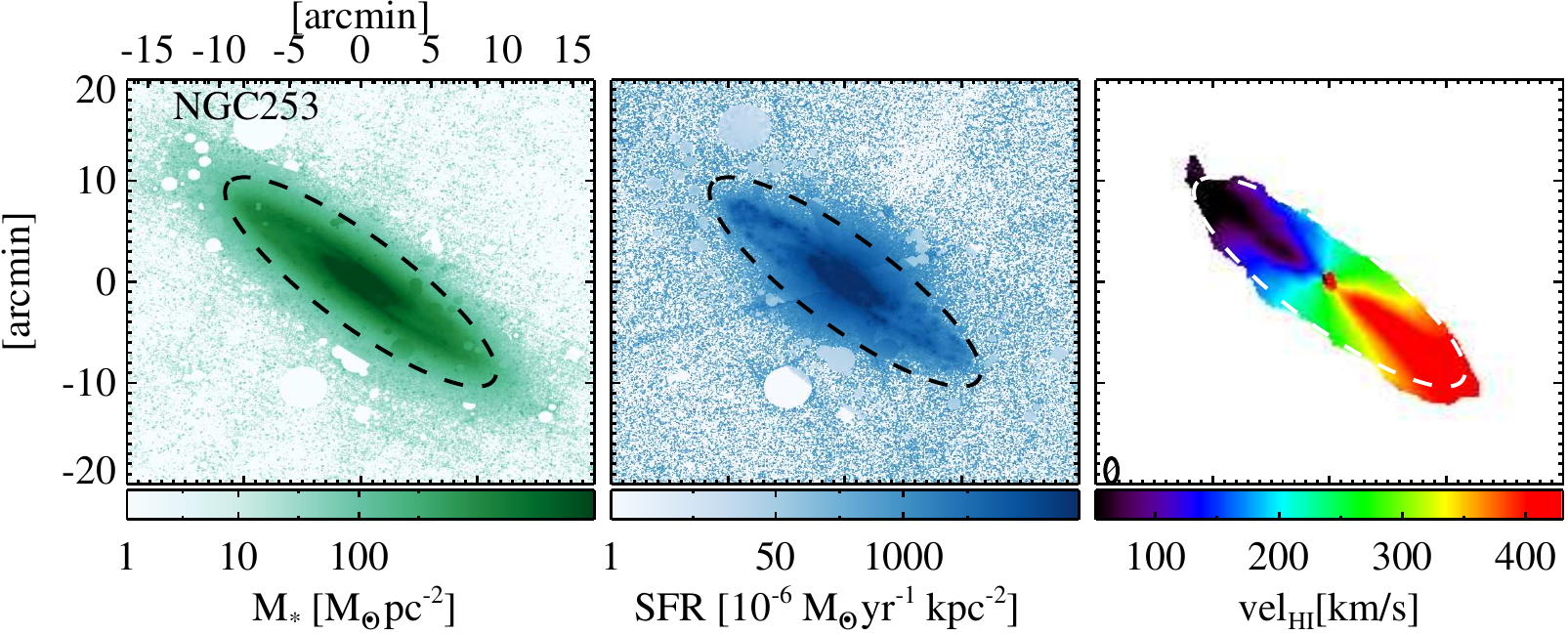}
\includegraphics[width=3cm]{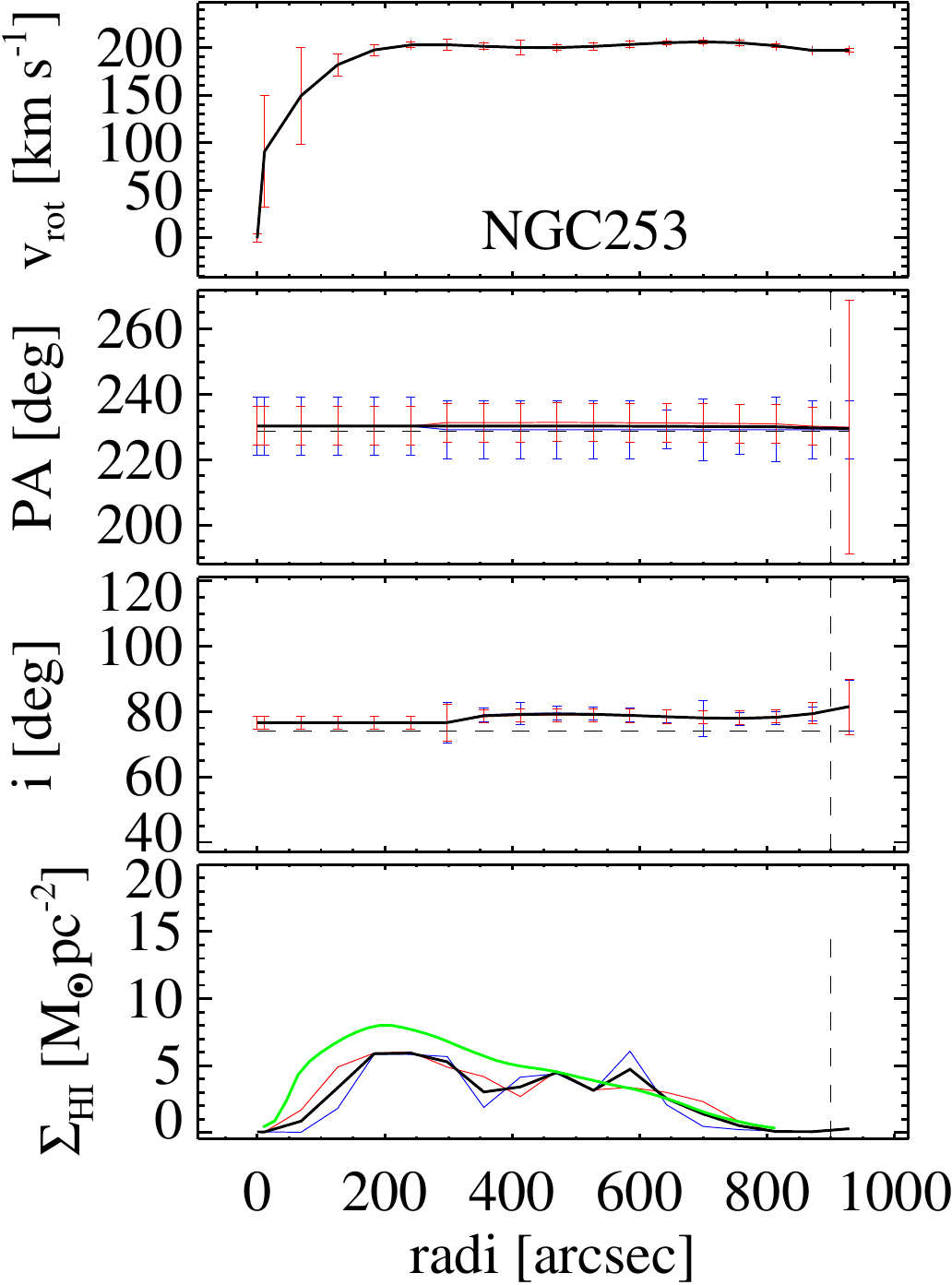}
\includegraphics[width=4cm]{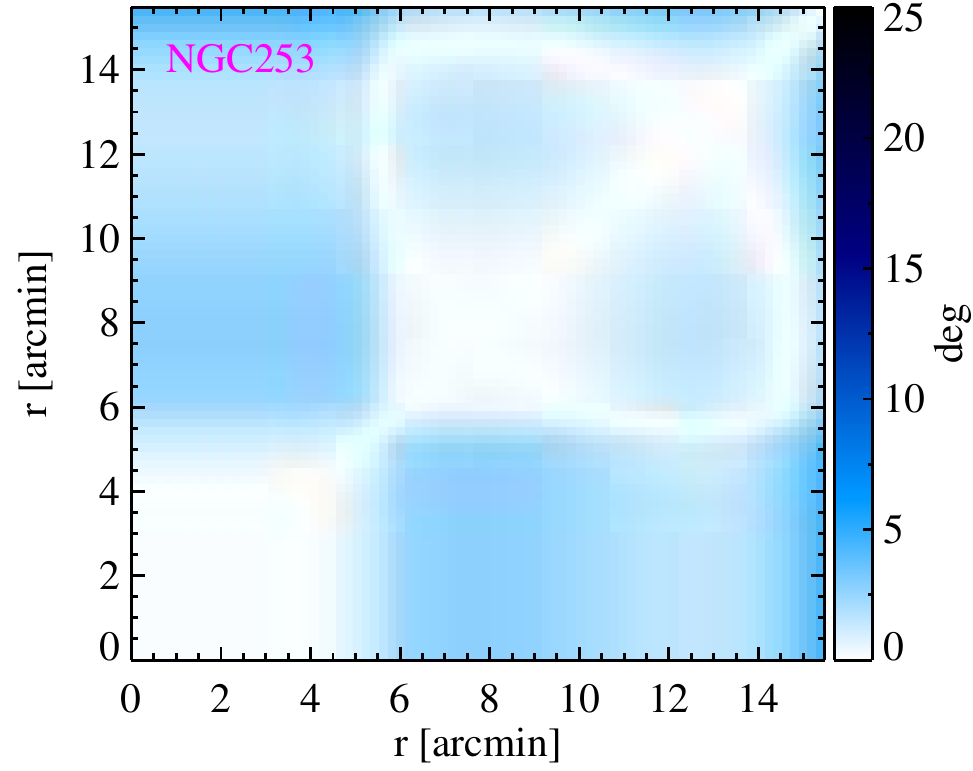}
\vspace{0.2cm}
\includegraphics[width=10cm]{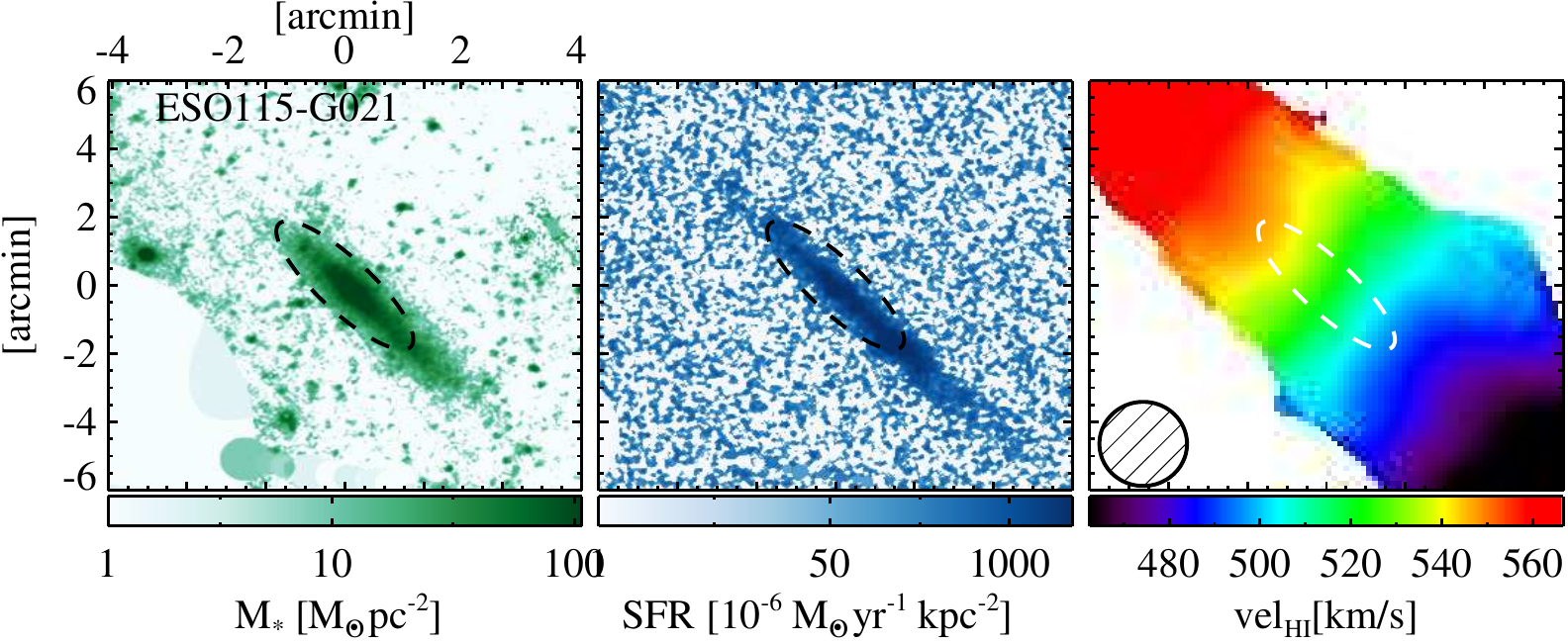}
\includegraphics[width=3cm]{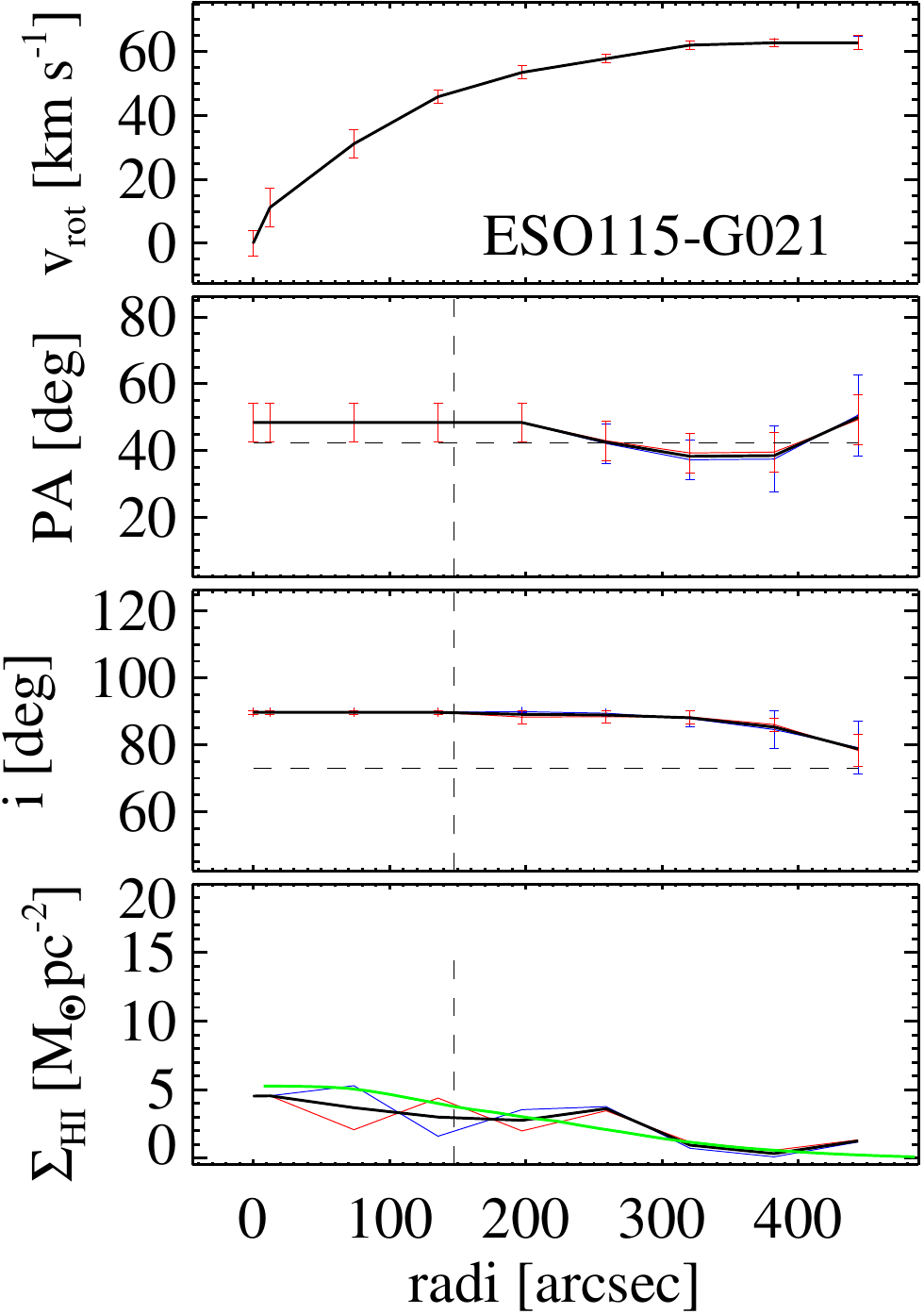}
\includegraphics[width=4cm]{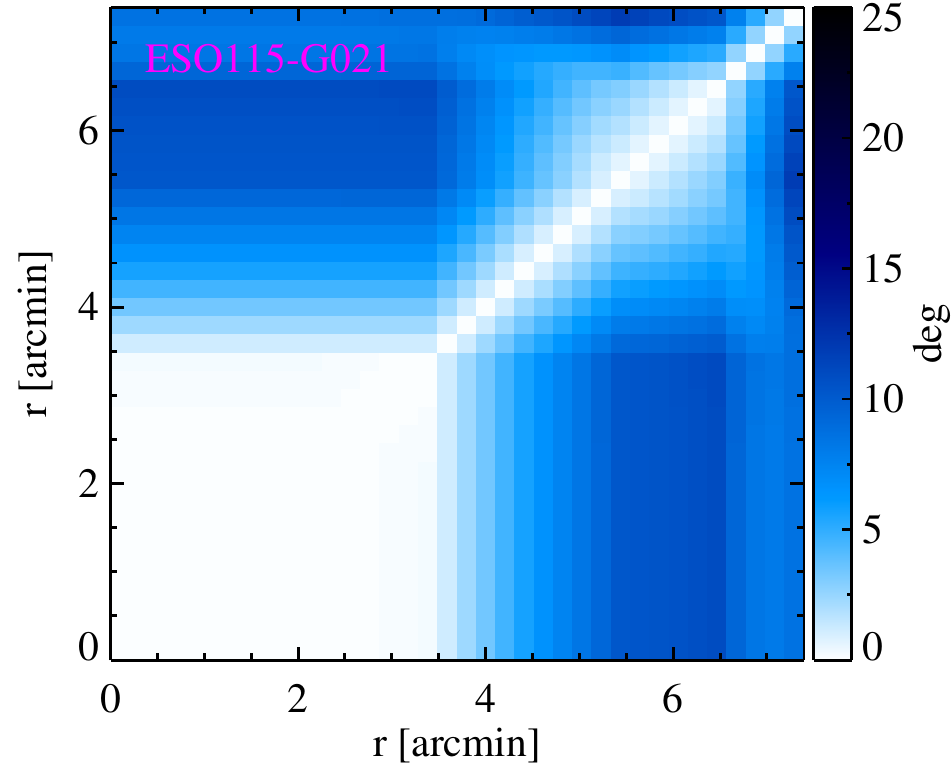}
\vspace{0.2cm}
\includegraphics[width=10cm]{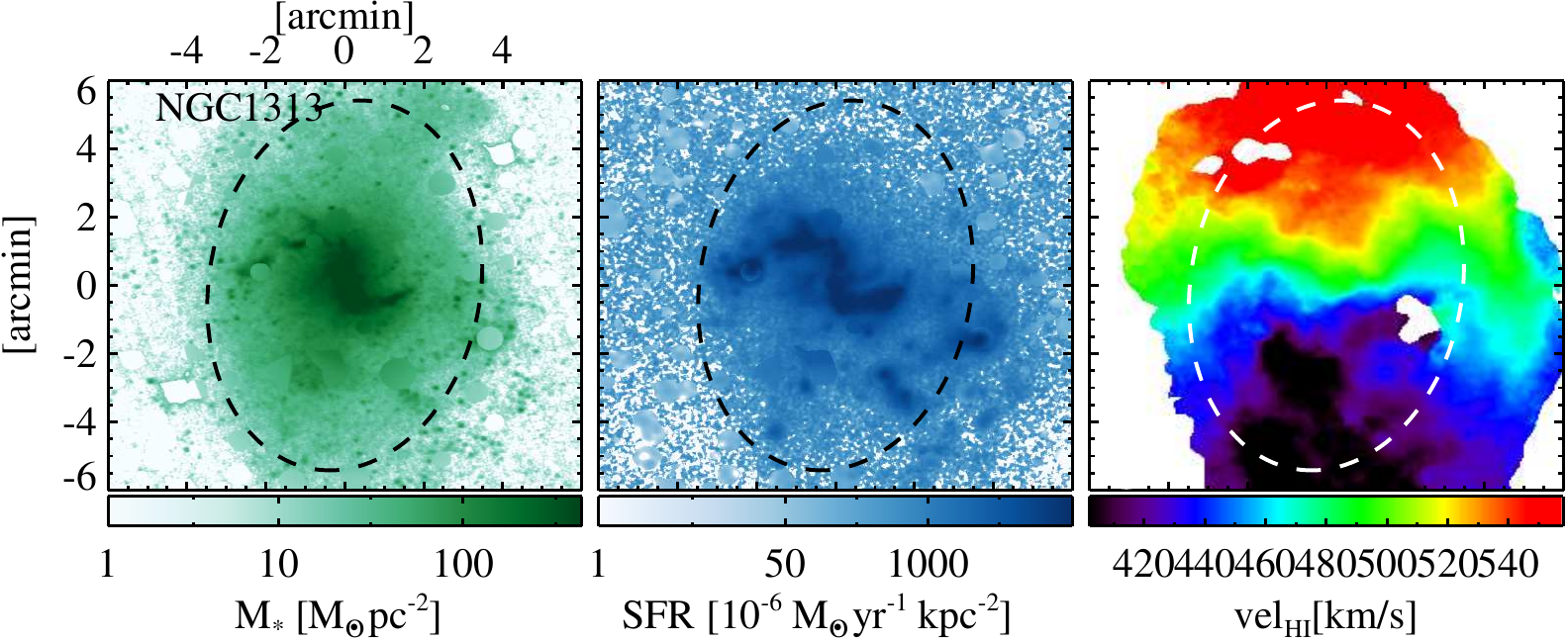}
\includegraphics[width=3cm]{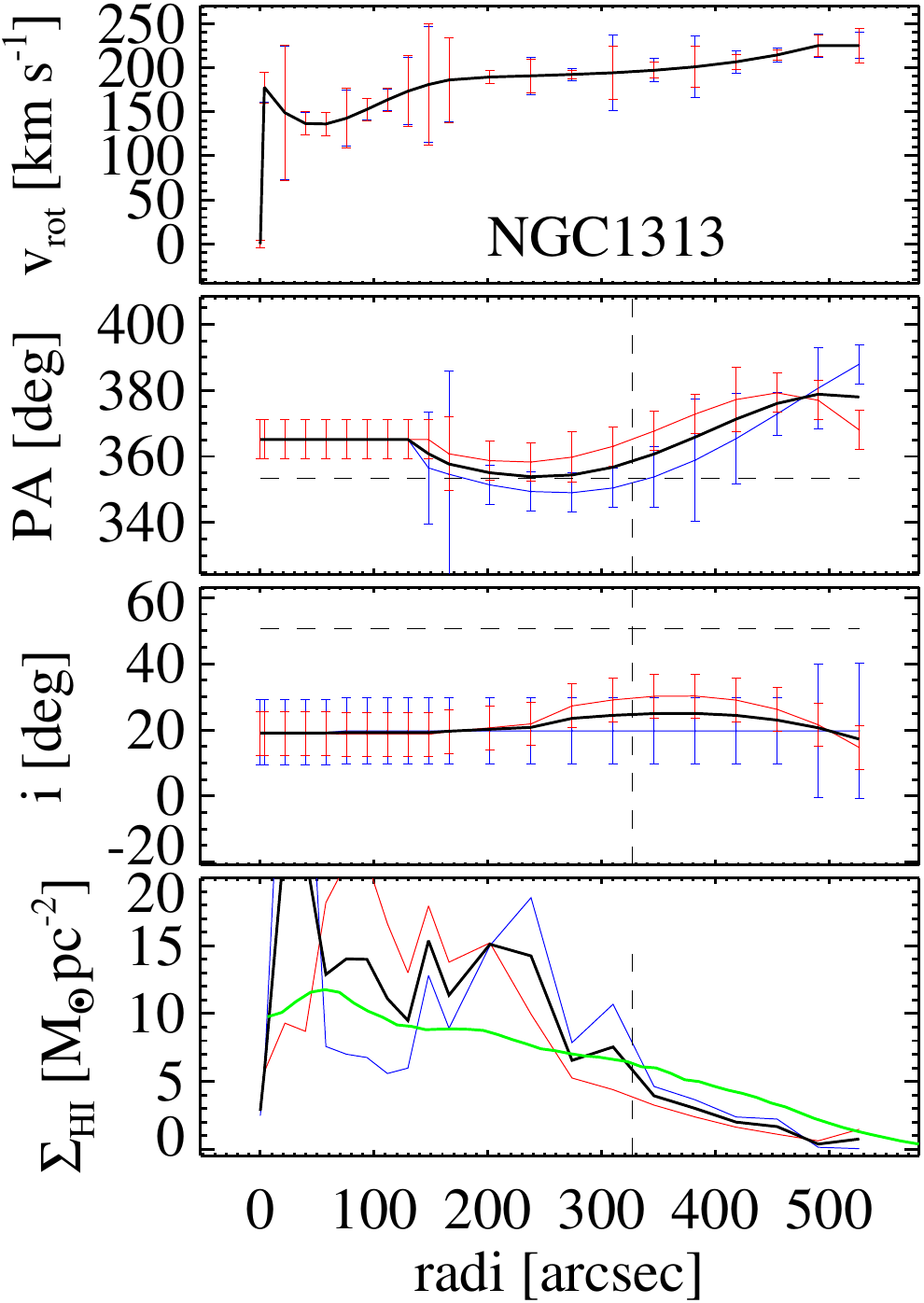}
\includegraphics[width=4cm]{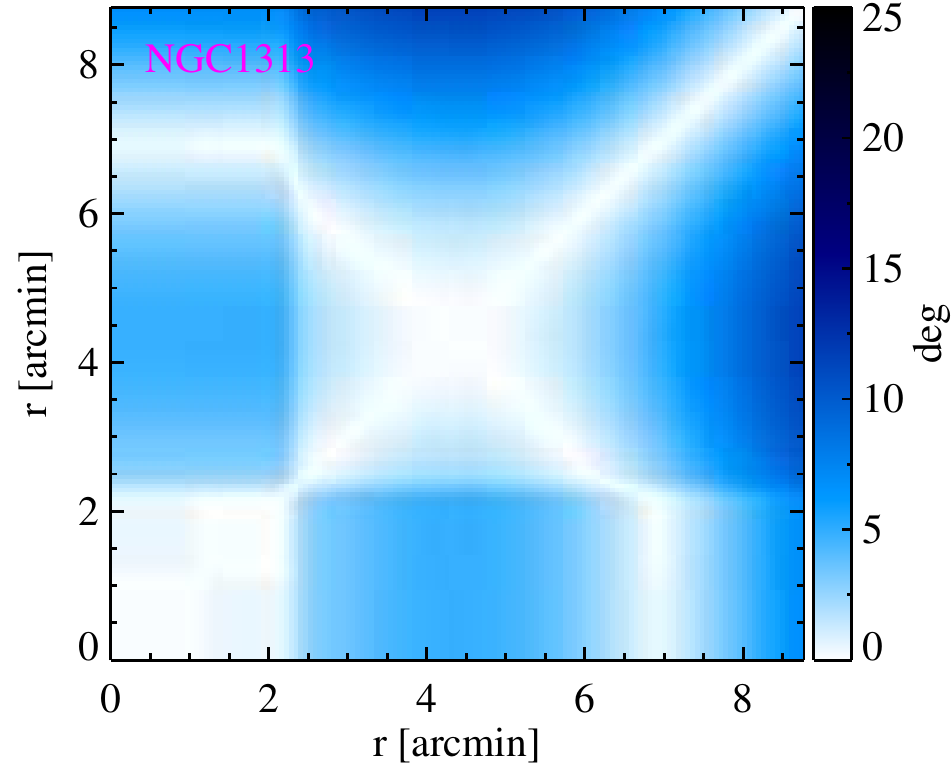}
\vspace{0.2cm}
\caption{From left to right: distributions of $M_*$ and SFR (at the resolution of the WISE W1- and W4-bands, respectively), mean ATCA HI velocity field at the original resolution, tilted ring models and tiltograms (approaching side) for the same ten large LVHIS galaxies as shown in Figure~\ref{fig:atlas_conv} (See Section~\ref{sec:appendix_figure}). To be continued.}
\label{fig:atlas_highrs}
\end{figure*}

\begin{figure*} 
\addtocounter{figure}{-1}

\includegraphics[width=10cm]{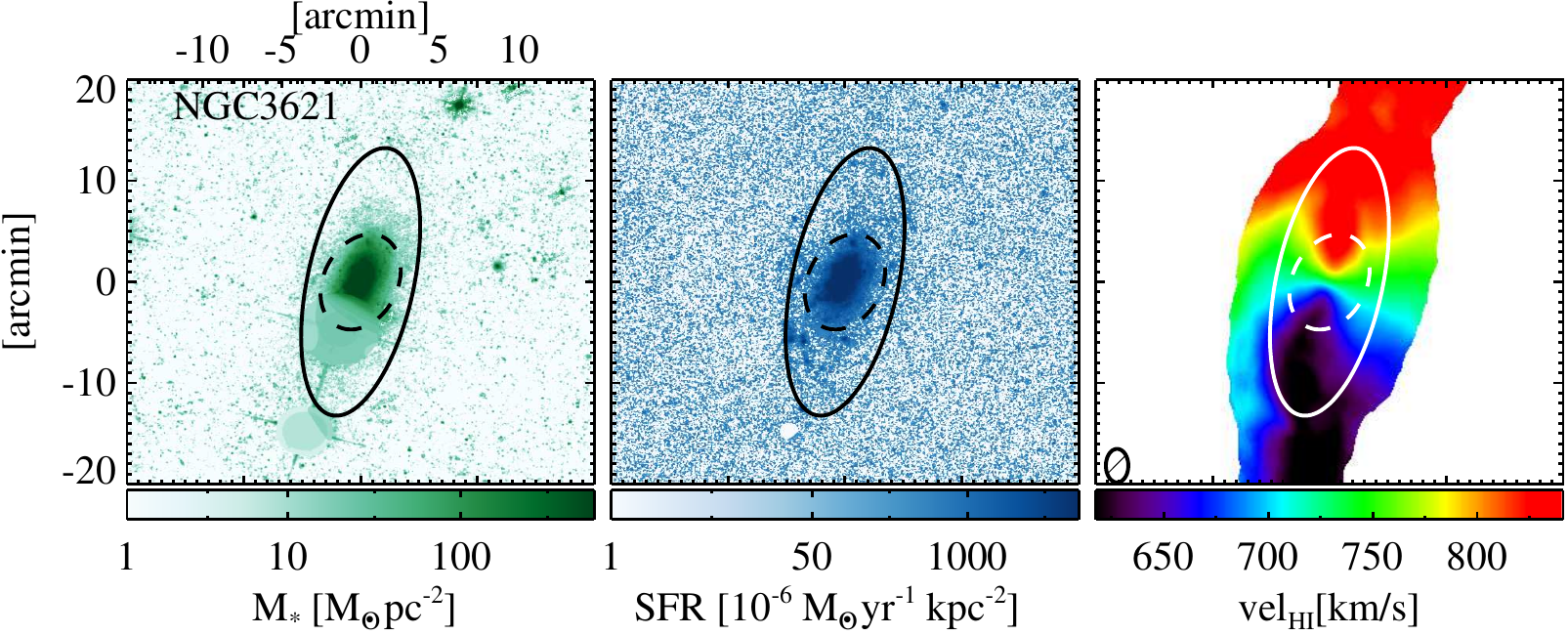}
\includegraphics[width=3cm]{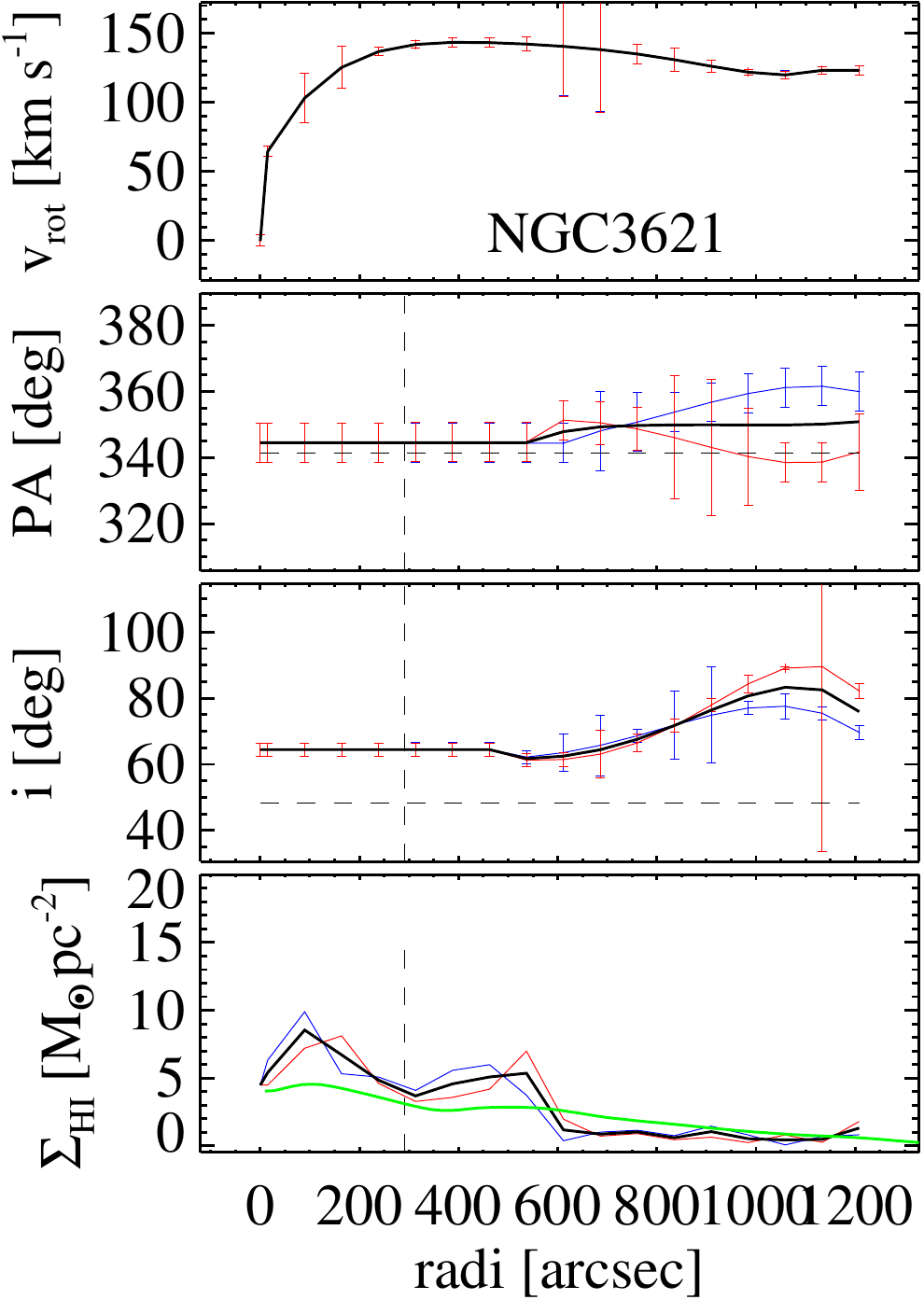}
\includegraphics[width=4cm]{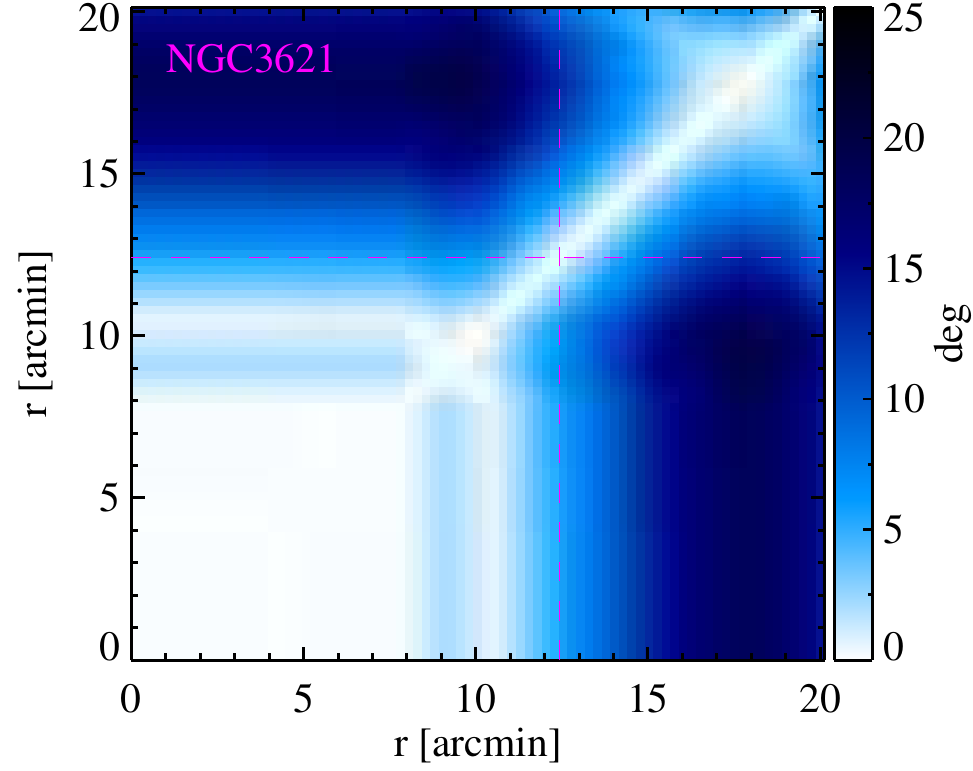}
\vspace{0.2cm}
\includegraphics[width=10cm]{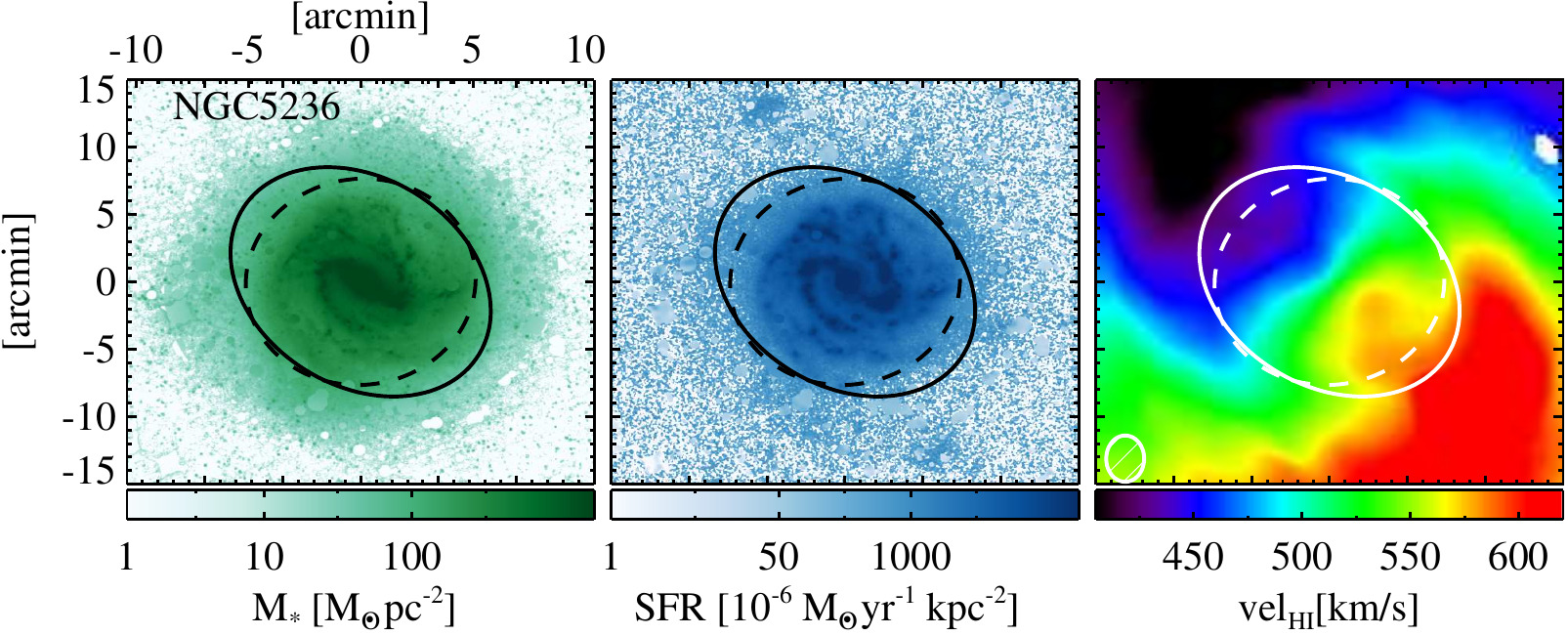}
\includegraphics[width=3cm]{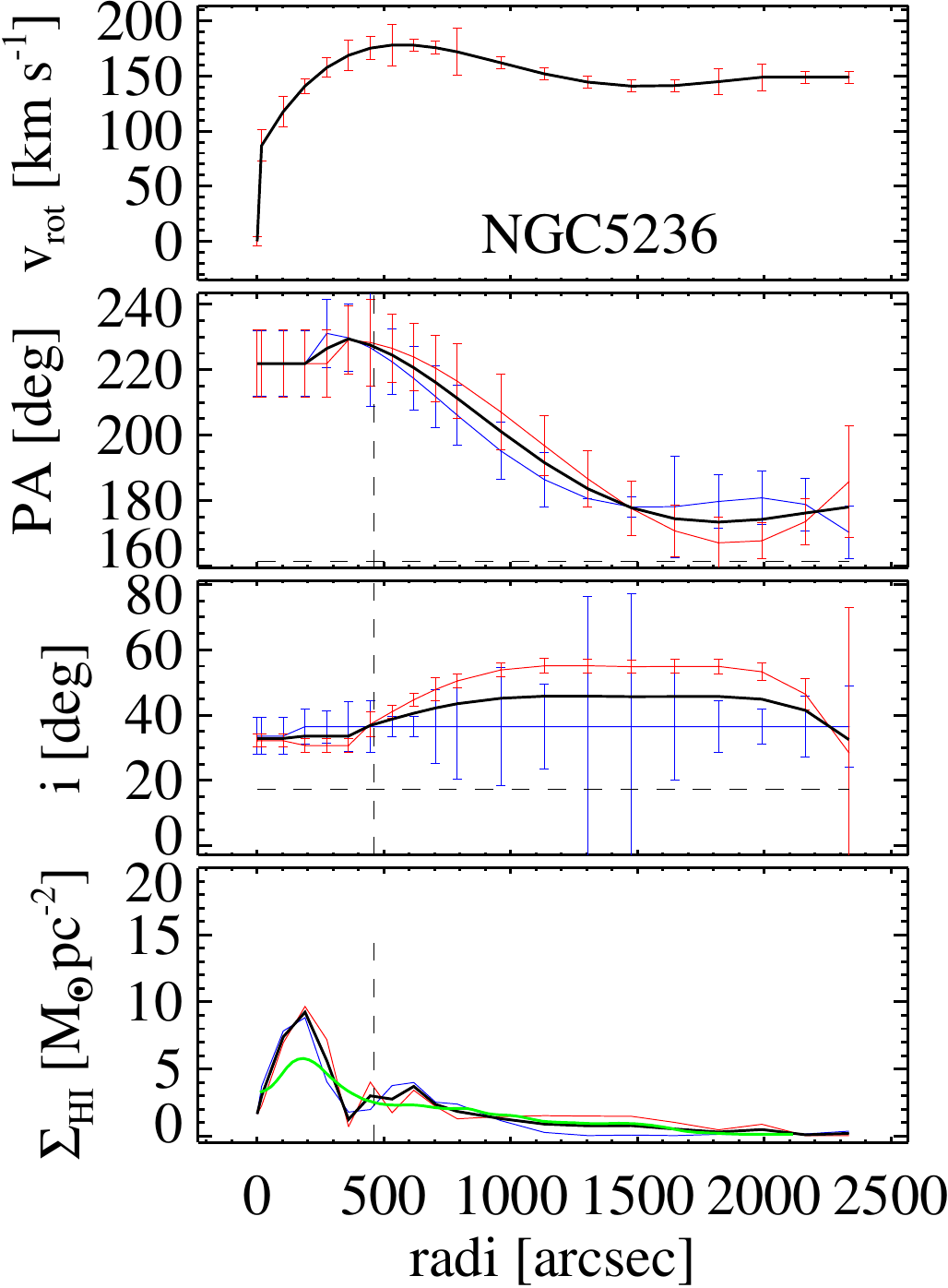}
\includegraphics[width=4cm]{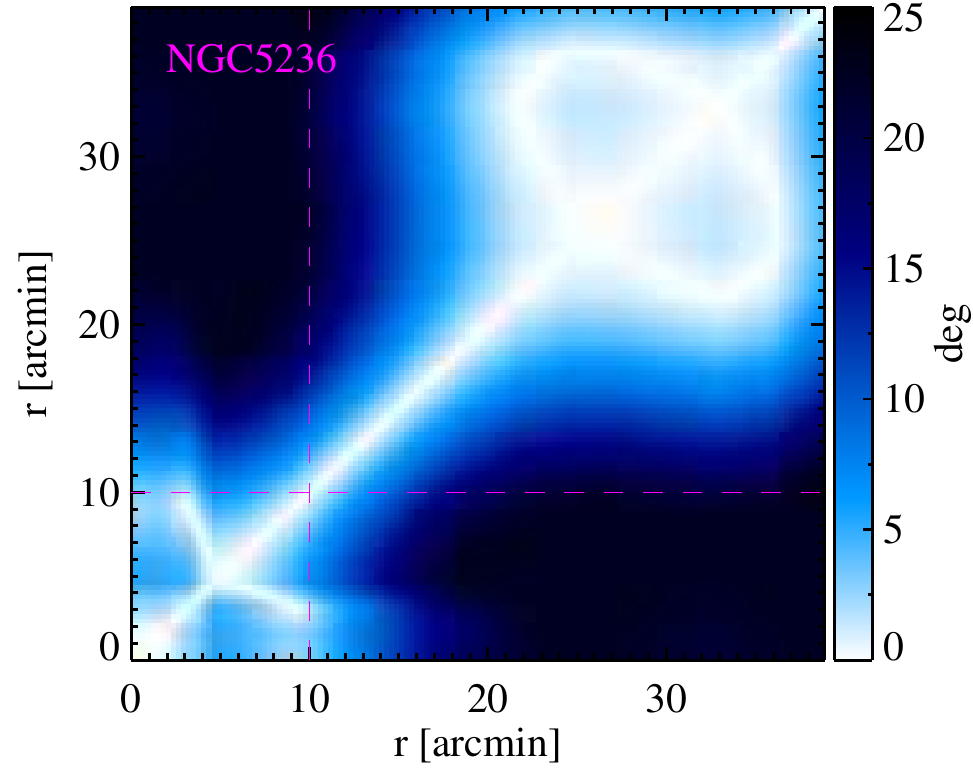}
\vspace{0.2cm}
\includegraphics[width=10cm]{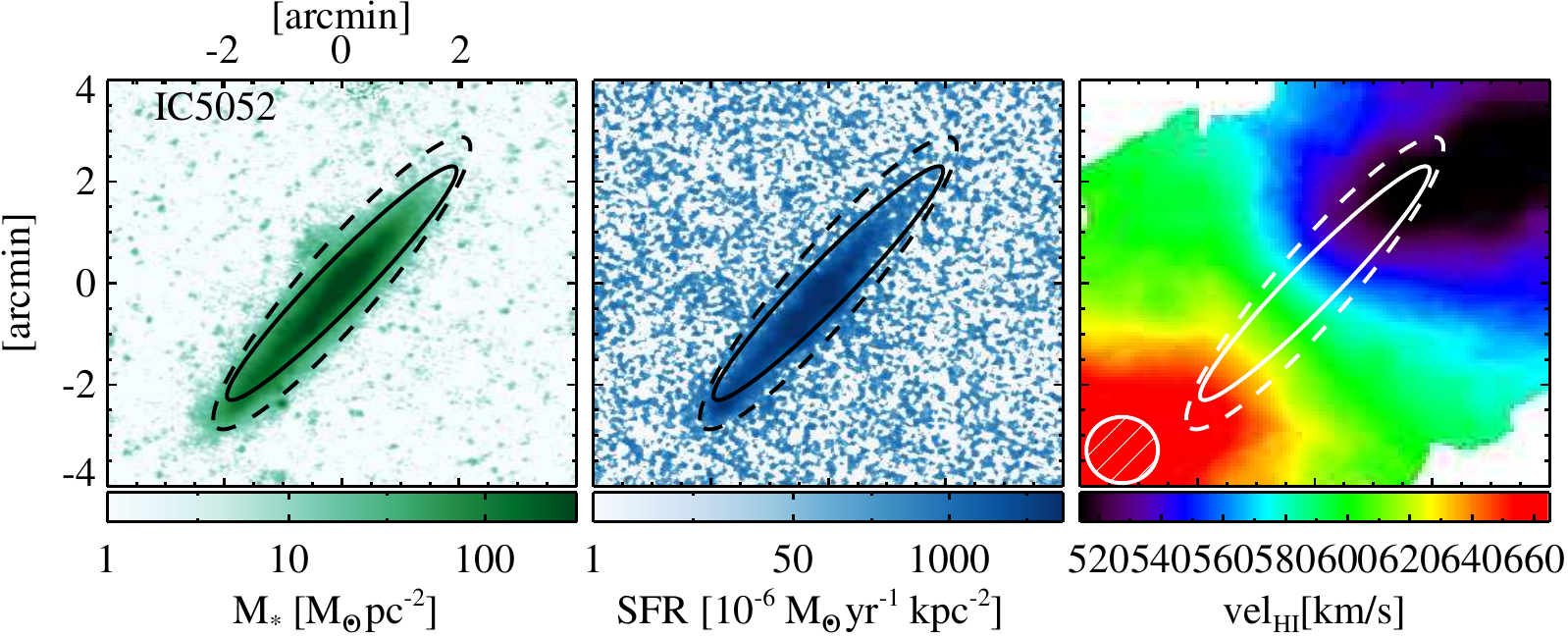}
\includegraphics[width=3cm]{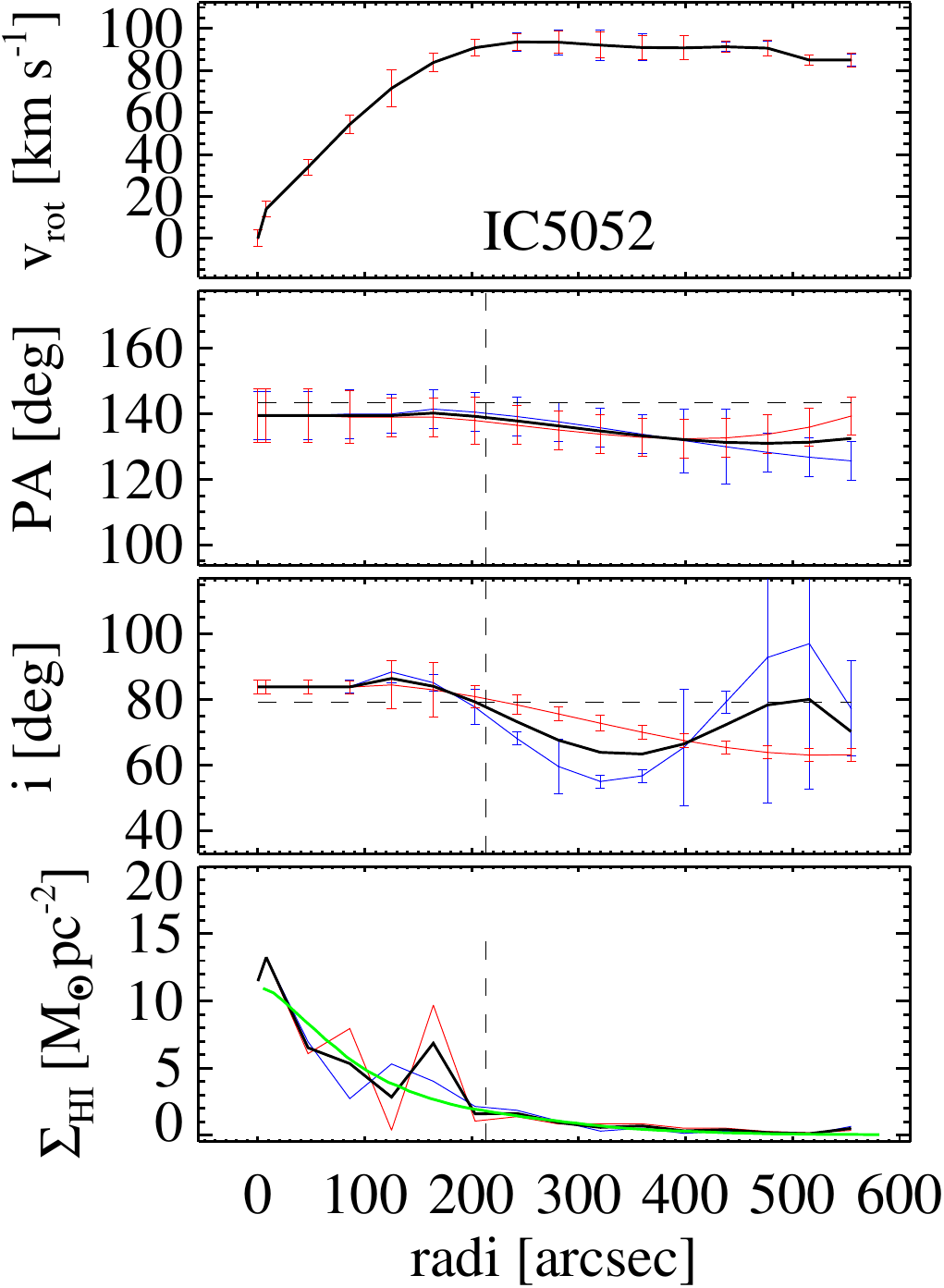}
\includegraphics[width=4cm]{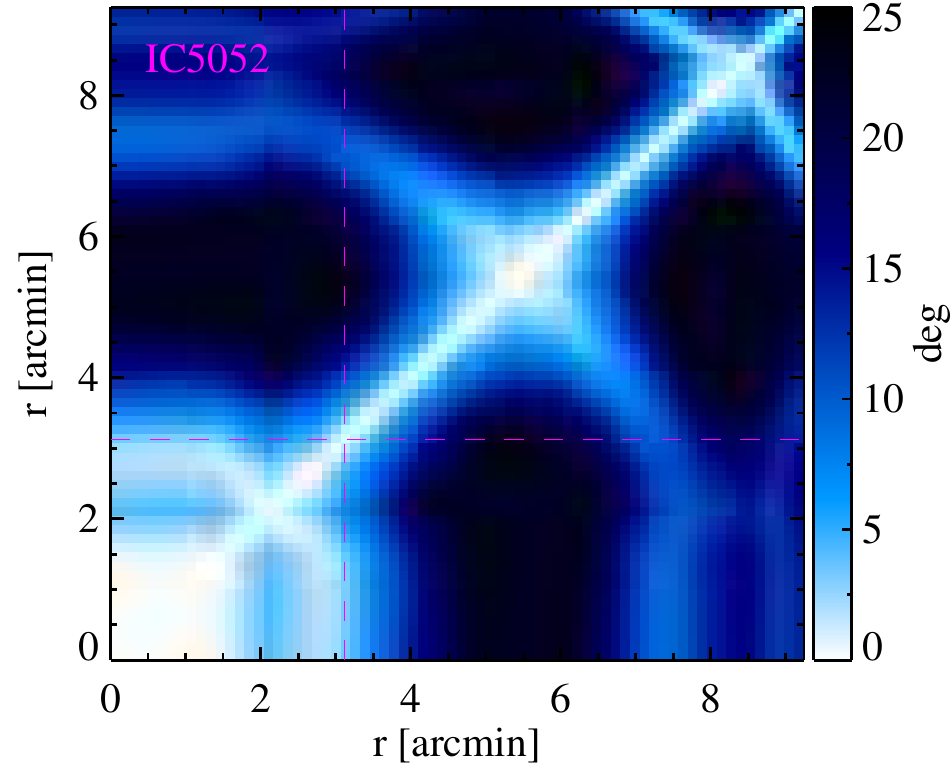}
\vspace{0.2cm}
\includegraphics[width=10cm]{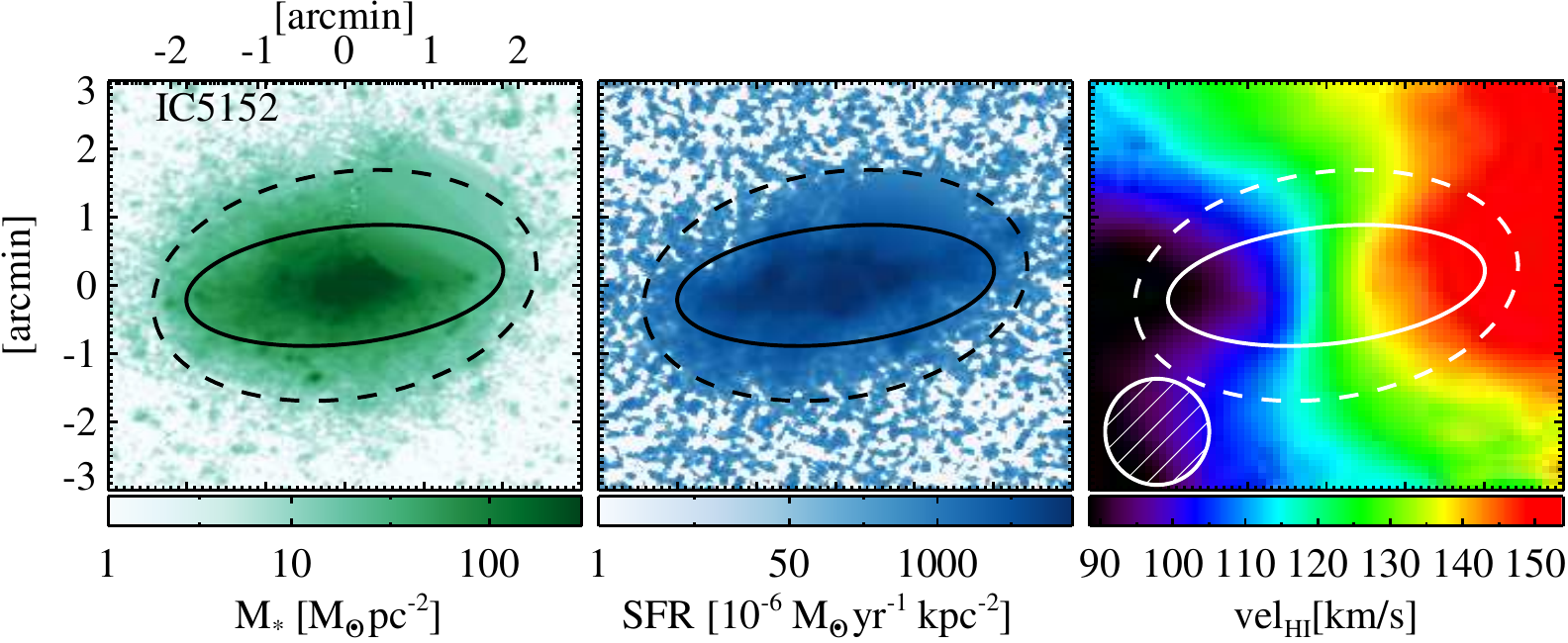}
\includegraphics[width=3cm]{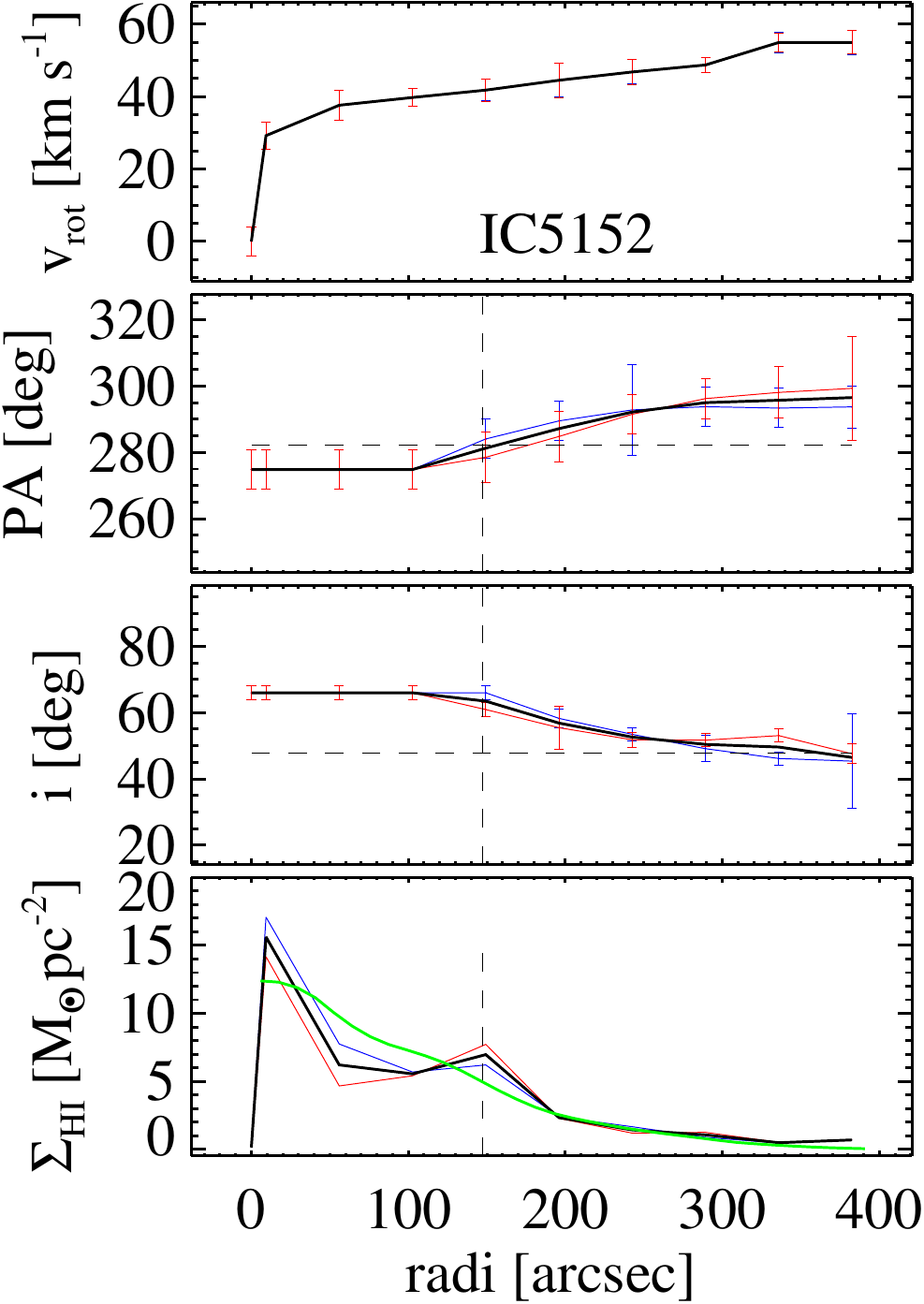}
\includegraphics[width=4cm]{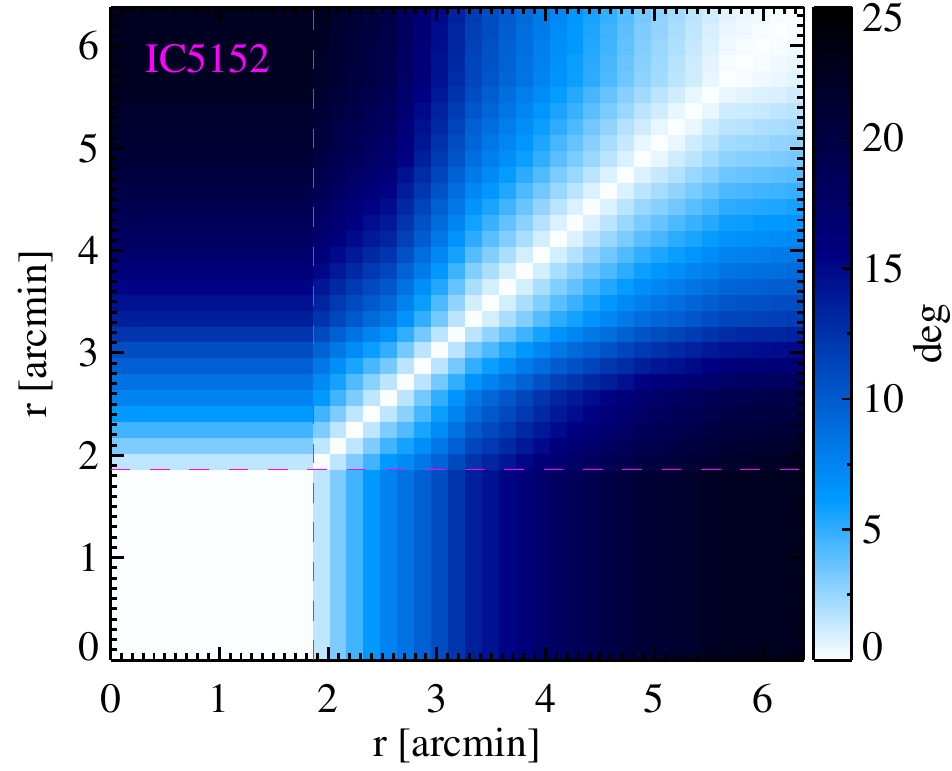}
\vspace{0.2cm}
\includegraphics[width=10cm]{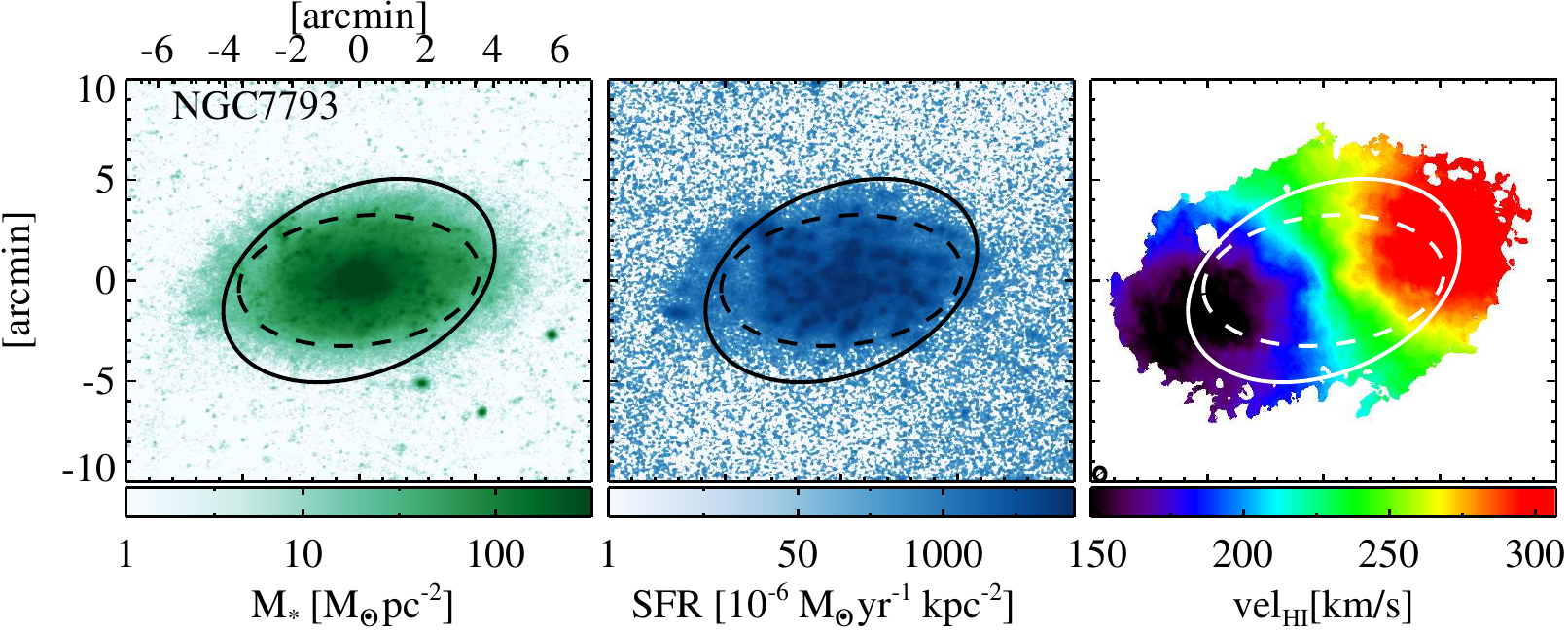}
\includegraphics[width=3cm]{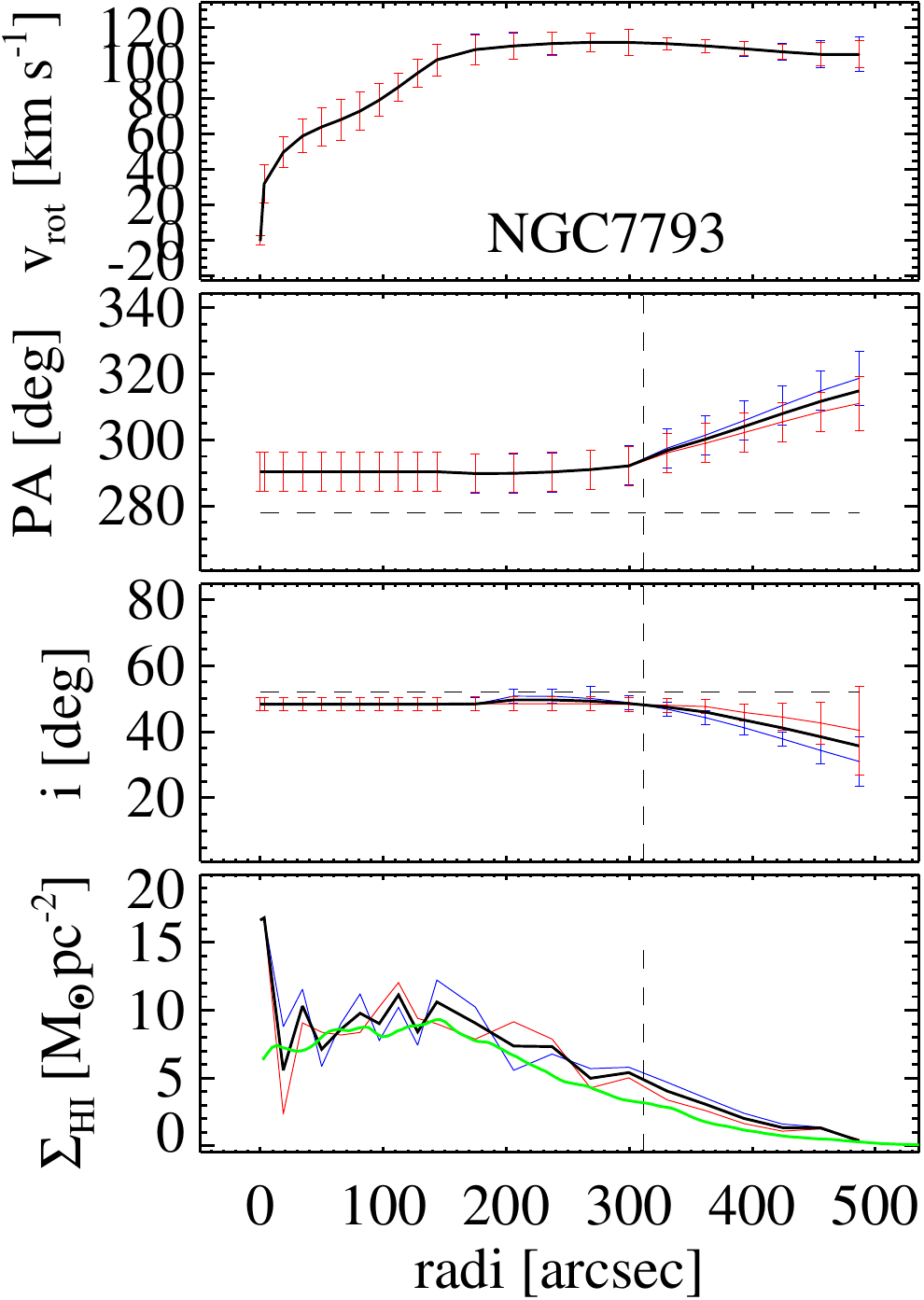}
\includegraphics[width=4cm]{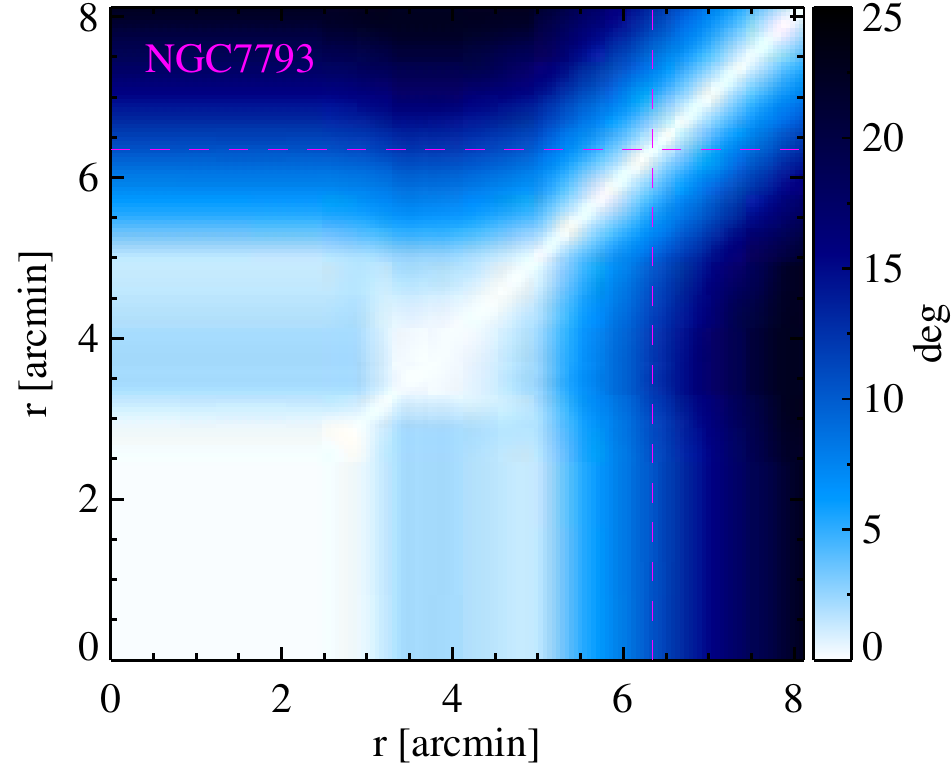}
\vspace{0.2cm}
\caption{Continued. From left to right: distributions of $M_*$ and SFR (at the resolution of the WISE W1- and W4-bands, respectively), mean ATCA HI velocity field at the original resolution, tilted ring models and tiltograms (approaching side) for the same ten large LVHIS galaxies as shown in Figure~\ref{fig:atlas_conv}. }
\label{fig:atlas_highrs}
\end{figure*}

\end{document}